\documentclass[10pt,aps,prc,twocolumn,showpacs,showkeys,nofootinbib,superscriptaddress]{revtex4-2}
\usepackage{graphicx}
\usepackage{amsmath,amsfonts,mathrsfs}

\graphicspath{{./img/}}

\def\vec#1{\boldsymbol{ #1}}

\def\om{\omega}
\def\rmd{{\rm d}}

\def\half{\frac{1}{2}}

\DeclareMathOperator{\Div}{div}
\DeclareMathOperator{\grad}{grad}
\DeclareMathOperator{\rot}{rot}

\newcommand{\lsim}{\stackrel{\scriptstyle <}{\phantom{}_{\sim}}}
\newcommand{\gsim}{\stackrel{\scriptstyle >}{\phantom{}_{\sim}}}

\begin{document}
	
	\title{Helicity and vorticity in heavy-ion collisions at energies available at the JINR Nuclotron-based Ion Collider facility}
	
	\author{N. S. Tsegelnik}
	\email{tsegelnik@jinr.ru}
	\affiliation{Joint Institute for Nuclear Research, RU-141980 Dubna, Russia}
	\author{E. E.  Kolomeitsev}
	\email{kolomei@theor.jinr.ru}
	\affiliation{Joint Institute for Nuclear Research, RU-141980 Dubna, Russia}
	\affiliation{Matej Bel University, SK-97401 Banska Bystrica, Slovakia}
	\author{V. Voronyuk}
	\email{vadimv@jinr.ru}
	\affiliation{Joint Institute for Nuclear Research, RU-141980 Dubna, Russia}
	\affiliation{Bogolyubov Institute for Theoretical Physics, Kiev, Ukraine}

	\begin{abstract}
		Heavy-ion collisions at center-of-mass nucleon collision energies 4.5--11.5\,GeV are analyzed within the parton-hadron-string dynamics (PHSD) transport model. Spectator nucleons are separated, and the transfer of the initial angular momentum of colliding nuclei to the fireball formed by participants is studied. The maximal angular momentum is carried by the fireball in gold-gold collisions with the impact parameter about 5\,fm corresponding to centrality class 10--20\%. The obtained participant distributions were fluidized and the energy and baryon number densities, temperature, and velocity fields are obtained in the Landau frame. 	
		It is shown that the velocity field has dominantly Hubble-like transversal and longitudinal expansion with the vortical motion being only a small correction on top of it. The vorticity field is calculated and illustrated in detail. The formation of two oppositely rotating vortex rings moving in opposite directions along the $z$ axis is demonstrated. Other characteristics of the vortical motion such as the Lamb vector field and the kinematic vorticity number are considered. The magnitude of the latter one
		is found to be smaller than that for the Poiseuille flow and close to the pure shear deformation corresponding to just a flattening of fluid cells. The field of hydrodynamic helicity, which is responsible for the axial vortex effect, is calculated. The separation of positive and negative helicities localized in upper and lower semiplanes with respect to the reaction plane is shown. It is proved that the areas with various helicity signs can be probed by the selection of $\Lambda$ hyperons with positive and negative projections of their momenta orthogonal to the reaction plane.
	\end{abstract}
	
	\maketitle
	
	\tableofcontents
	
	\section{Introduction}\label{sec:level1}
	
	Hyperons, being registered via their weak decays, are ``self-analyzing'' particles, and the asymmetry in momentum distributions of the decay products tells about the averaged spin orientation of the hyperons. The first report about an observation of a nonzero averaged $\Lambda$ polarization in heavy-ion collisions (HICs) is related to early Bevalac experiments with an argon beam colliding with a KCl target at the incident energy 1.8\,GeV per nucleon~\cite{Harris-Lb}. The observed significant polarization of order ($10\pm 5$)\% was obtained on a sample of just 70 $\Lambda$'s. This result was questioned in Refs.~\cite{Anikina}, where the zero results for the $\Lambda$ polarization were obtained for various light-light and light-heavy nucleus collisions at 4.5\,GeV/$c$ momentum per incident nucleon. At the same time significant polarization of produced $\Lambda$'s was routinely observed in proton-proton and proton-nucleus reaction at incident proton momenta from 12\,GeV/$c$ to $\approx 1000$\,GeV/$c$; see~\cite{Castilio-2006,Abe-83,DeGrand-81,Abikim19} and references therein. So, the question remained whether the polarization signal is indeed completely washed out in nucleus-nucleus collisions or some signal survives. With the construction of new high statistic heavy-ion experiments at the BNL Relativistic Heavy Ion Collider (RHIC) facility, the $\Lambda$ polarization can be reliably measured. The STAR Collaboration published in \cite{Adamczyk-Nature} the results of the hyperon polarization measurements demonstrating net $\Lambda$ polarization on the level of 1--2\% in gold-gold collision in the range of center-of-mass energies of two colliding nucleons between $\sqrt{s_{NN}}=7.7$ and 60\,GeV. Thereby the polarization increases with a decrease in the collision energy. At lower collision energy the $\Lambda$ polarization was measured by the HADES Collaboration in Au+Au and Ag+Ag collisions at $\sqrt{s_{NN}} = 2.4$ and 2.55\,GeV~\cite{Kornas-HADES22}, and even larger degrees of polarization were observed,  $\approx 5$ and 3\%, respectively.
	
	Surprisingly, antihyperons $\overline{\Lambda}$ turn out to be also polarized in HICs in contrast to proton-proton and proton-nucleus collisions~\cite{DeGrand-81}. Moreover, the $\overline{\Lambda}$ polarization rises with the lowering of the collision energy much faster than for $\Lambda$, reaching $(7.6\pm 3.3)\%$\footnote{The value of the polarization is recalculated according to the recent measurement of the hyperon decay constant $\alpha_\Lambda$~\cite{Abikim19,PDG2020}, which is about 17\% higher than what was used before.} for $\sqrt{s_{NN}}=7.7$\,GeV.

	The spin polarization of emitted particles is believed to be induced by the coupling of the initial orbital (``mechanical'') angular momentum of two nuclei colliding with a nonvanishing impact parameter and the spin distributed in the matter created in the collision. This is in analogy to the Barnett effect observed more than a century ago~\cite{Barnett} when an electrically neutral unmagnetized metallic object became spontaneously magnetized after being set in rotation. The orbital angular momentum per nucleon in the system of two nuclei $A$ colliding with the impact parameter $b$ and the center-of-mass energy of two nucleons $\sqrt{s_{NN}}$ can be easily estimated as
	\begin{align}
		\vec{l}=\frac{\vec{L}}{A} =\vec{e}_y \frac{b}{2}\sqrt{s_{NN}-4 m_N^2}.
		\label{l-def}
	\end{align}
	(Here and below, we use in equations the system of units with the Planck constant $\hbar$ and the speed of light $c$ taken as unity. However, given the numerical values of physical quantities we will retain these constants for the sake of clarity. Temperature will be measured in the energy units.) The vector $\vec{l}$ is directed along the $y$ axis if the nuclei collide along the $z$ axis in the $xz$ plane; $\vec{e}_y$ is the unit vector in the $y$ direction. For $\sqrt{s_{NN}}=2.5$\,GeV we have $|\vec{l}|\approx 42\hbar(b/10\,{\rm fm})$, for $\sqrt{s_{NN}}=5$\,GeV, $|\vec{l}|\approx 117\hbar(b/10\,{\rm fm})$, and for $\sqrt{s_{NN}}=11$\,GeV we have $|\vec{l}|\approx 275\hbar (b/10\,{\rm fm})$. These numbers are very large, exceeding substantially momenta carried by the highest spin nuclei~\cite{Ward-Fallon}. Several mechanisms of the conversion of this angular momentum to the spin alignment are discussed in the literature.
	
	The general thermodynamic description of the link between the vorticity of the fermionic fluid and its spin polarization was developed in Refs.~\cite{Becattini-Tinti2010,Becattini-Chandra2013,Fang-Pang-Wang2016,Becattini-Karpenko-Lisa2017}.
	The vorticity-induced spin polarization mechanism implemented in hydrodynamic~\cite{Karpenko-Becattini2017,Xie-Wang-Csernai2017,Ivanov-PRC100,Ivanov-PRC102,Ivanov-PRC103,Ivanov-PRC105} and transport models \cite{Li-Pang-Wang-Xia-PRC96,Sun-Ko-PRC96,KTV-PRC97,Wei-Deng-Huang-PRC99,Shi-Li-Liao-PLB788,Vitiuk-BZ2020}
	allowed one to generally reproduce the measured $\Lambda$ polarization.
	However, most of the above-mentioned works were not able to explain the larger polarization of $\overline{\Lambda}$ compared to $\Lambda$. Work~\cite{Vitiuk-BZ2020} argued that the stronger polarization of $\overline{\Lambda}$ could be explained by the different space-time distributions of $\Lambda$ and $\overline{\Lambda}$ and by different freeze-out conditions of both hyperons. An additional mechanism for spin alignment, which distinguishes hyperons and antihyperons, was proposed in Ref.~\cite{Csernai-Kapusta-2019} and is related to the interaction of baryons with vector-meson mean fields, which received magnetic vector components due to vorticity of baryon currents. This mechanism was realized in hydrodynamical codes~\cite{Csernai-Kapusta-2019,Xie-Chen-Csernai-EPJC81,Ivanov-PRC105} that allowed for partial explanation of the experimental splitting in $\Lambda$--$\overline{\Lambda}$ polarizations.

	An alternative approach not related to the equilibrium of spin degrees of freedom is based on the axial vortical effect (AVE) or chiral vortical effect (CVE) \cite{Vilenkin-1,Vilenkin-2,Son-Zhit-2004,Gao-Liang-Pu-2012,Sorin-Teryaev-2017}. In the AVE, the local spin polarization of hyperons (anti-hyperons) is determined by the zero component of the axial current for strange (antistrange) quarks. The latter one is generated by the hydrodynamic helicity, i.e., the projection of the velocity to the vorticity. The AVE was used in Ref.~\cite{BGST-Hseparation} for the first rough estimation of the polarization effect in heavy-ion collisions at energies available at the JINR Nuclotron-based Ion Collider facility (NICA); see also Refs.~\cite{Rogachevsky-ST-2010}. In the CVE the contribution to the axial current is generated by hydrodynamic vorticity. An interesting link between CVE and vortices in the pion superfluid was considered in Ref.~\cite{TZ-2017}. The axial vortical and similar chiral kinetic mechanisms for the $\Lambda$ polarization were realized in Refs.~\cite{BGST-PRC97,Sun-Ko-PRC96} within the quark-gluon string model (QGSM)~\cite{QGSM-1,QGSM-2,QGSM-3} and a multiphase transport model (AMPT)~\cite{AMPT-1,AMPT-2,AMPT-3}. Within the hydrodynamic approach, this mechanism was investigated in Ref.~\cite{Ivanov-PRC102-AVE}.

	Thus, vorticity and helicity are the main hydrodynamic characteristics of the medium created in heavy-ion collisions, and are responsible for the formation of the hyperon polarization signals. The structure of the vorticity field was analyzed in \cite{XTW18} for Au-Au collisions at higher RHIC and LHC (Large Hadron Collider) energies in the framework of AMPT. The circular structure of the transverse vorticity around the beam direction and the quadrupole pattern of the longitudinal vorticity in the transverse plane were found. The other analysis was performed using the QGSM in Refs.~\cite{BGST-Hseparation,BGST-Vsheet} for noncentral (impact parameter 8\,fm) Au + Au collisions at $\sqrt{s_{NN}}=5$\,GeV and using the hadron-string dynamic (HSD) model~\cite{HSD} in Ref.~\cite{Teryaev-Usubov}. It was argued that the vorticity is predominantly localized in a relatively thin layer at the boundary between participants and spectators. Also, noticeable hydrodynamical helicity was observed to manifest specific mirror behavior with respect to the reaction plane. However, there are still open questions concerning the fluidization of particle distribution generated in the transport code: the separation of spectator nucleons and the used definition of the flow velocity; see discussion in~\cite{Deng-Huang-PRC93}. Some of these problems are naturally solved in the three-fluid hydrodynamic approach~\cite{Ivanov-PRC100}, within which a particular structure consisting of two vortex rings is found~\cite{Ivanov-Soldatov-PRC97} in the Au+Au collisions at $\sqrt{s_{NN}}=39$\,GeV.
	
	In this paper, we want to consider in detail the structure and evolution of vorticity and helicity fields created in HICs at various energies in the range accessible for the future NICA collider using the parton-hadron-string
	dynamics (PHSD) model~\cite{PHSD,PHSD-contin}.
	
	In Sec.~\ref{sec:Spec-separ} we discuss the separation of spectator nucleons from the nucleons forming a fireball and the transfer of the angular momentum from two initial nuclei to the fireball medium. Fluidization of the test-particle distributions generated by the PHSD transport code is discussed in Sec.~\ref{sec:Fluid}. The obtained temperature and the particle and energy density fields are described in Sec.~\ref{sec:Tne}. The structure of the velocity field created in collisions at various energies is discussed in Sec.~\ref{sec:V-field}. In Sec.~\ref{sec:vorticity} we calculate the vorticity field. The helicity field is analyzed in Sec.~\ref{sec:helicity}. Conclusions are formulated in Sec.~\ref{sec:conclusion}.
	
	\section{Spectator separation and angular momentum transfer}\label{sec:Spec-separ}
	
	The PHSD model proved to be a reliable tool for the quantitative description of multiplicities and momenta distributions of particles in heavy-ion collisions in a broad energy range from SIS to upper RHIC energies~\cite{PHSD,PHSD-contin}.
	As a transport model it traces momenta and coordinates of all particles at each moment of time. The formal phase-space distribution function for particles (test particles) of type $h$ can be written as
	\begin{align}
		&f_{\rm t.p.}^{(h)}(t,\vec{r}, p_0, \vec{p}) =
		\sum_{i_h} (2\pi)^4 \delta^{(3)}\big(\vec{p} - \vec{p}_{i_h}(t)\big)
		\nonumber\\
		&\quad\times
		\delta\big(p_0-\sqrt{m_h^2+\vec{p\,}^2}\big)\delta^{(3)}\big(\vec{r}-\vec{r}_{i_h}(t)\big) ,
		\label{f-test-p}
	\end{align}
	where $ \vec{r},\vec{p} $ is a point of the coordinate-momentum space, and $ \vec{r}_{i_h}(t) $ and $\vec{p}_{i_h}(t)$ are the coordinate and the momentum of the $i_{h}$th particle that depend on time, $t$. The $\delta$-function with $p_0$ keeps the particle on mass shell specified by a hadron mass $m_h$. The code is able to treat particles with continuum mass spectra (broad resonances) where the mass-shell $\delta$ function is replaced with a dynamically varying spectral function; see, e.g.,~\cite{PHSD-contin,Leupold-resonances}. In practice, each resonance particle is now represented by an ensemble of particles with various masses populated and interacting according to the spectral function weight. The spectral functions are used also for the description of partons in the deconfined phase. The relative volume occupied by the partonic phase is small at energies $\sqrt{s_{NN}}\lsim 12$\,GeV. For instance, at $\sqrt{s_{NN}}=11.5$\,GeV in Au+Au collisions at $b=2$\,fm, the fraction of the deconfined phase in the full volume does not exceed 20\% for times at the maximum overlap~\cite{Moreau:2021clr,Moreau:2019aux}. However, for the most central collisions the parton fraction can reach $\approx 40$\% in the mid rapidity region. After the maximal overlap the fireball expands and the partonic fraction decreases rapidly and is insignificant for later times. For larger impact parameters the partonic fraction decreases also. We apply the version with particles moving freely between two successive collisions without influence of mean fields; however, the chiral symmetry breaking effects introduced in Ref.~\cite{Palmese2016} are included to provide the correct strange particle multiplicities.
	
	As in many transport codes, PHSD uses the parallel ensemble method, that consists of the parallel simulation of $N$ collision events. This allows computing with good accuracy collective quantities, e.g., energy and particle densities, since the statistical fluctuations are reduced by averaging over $N$ events. This ensemble average we will not indicate explicitly, assuming that all physical quantities calculated with the particles distributions (\ref{f-test-p}) are ensemble averaged. In our calculation we use $N=100$ for energies $\sqrt{s_{NN}}>5$\,GeV and $N=200$ for
	$\sqrt{s_{NN}}\lsim 5$\,GeV. The code was re-initialized 200--250 time so that, finally, statistics with $\approx (2\mbox{--}5)\times 10^4$ collisions are collected for each impact parameter, collision energy, and other varied parameter.

	Particles of colliding nuclei are usually divided in spectators and participants where the former ones do not suffer violent collisions; therefore their rapidities do not differ much from the initial rapidity of colliding nuclear beams, $y_{\rm b}= \frac{1}{2} \ln \frac{\sqrt{s_{NN}} +  \sqrt{s_{NN} - 4 m^2_N}}{\sqrt{s_{NN} -  \sqrt{s_{NN} - 4  m^2_N}}} $. Rapidities of participants, on the contrary, decrease fast due to collisions and form quickly a thermal distribution centered at the midrapidity ($y=0$) with the width $\sim \sqrt{2T/m_N}\approx 0.6 \sqrt{T/150\,{\rm MeV}}$. Thus, typical rapidities of participants, $|y_{\rm part}|\simeq 0+0.6$, are several times smaller than the beam rapidities, $y_{\rm b}=1.5$ for collisions at $\sqrt{s_{NN}}=4.5$\,GeV and $y_{\rm b}=2.5$ at $\sqrt{s_{NN}}=11.5$\,GeV.  Using this criterion we separate the spectator part in the distribution function (\ref{f-test-p}), defined as
	\begin{align}
		&f_{\rm t.p.}^{(h,{\rm spec})}(t,\vec{r}, p_0, \vec{p})=\sum_{q=\pm}
		f_{\rm t.p.}^{(h)}(t,\vec{r}, p_0, \vec{p})
		\theta\big(\Delta y_{\rm b}-|q\,y-y_{\rm b}|\big),
		\label{spec-part}
	\end{align}
	and count the remaining particles,
	\begin{align}
		&f_{\rm t.p.}^{(h,{\rm part})}(t,\vec{r}, p_0, \vec{p})=f_{\rm t.p.}^{(h)}(t,\vec{r}, p_0, \vec{p})-
		f_{\rm t.p.}^{(h,{\rm spec})}(t,\vec{r}, p_0, \vec{p}),
	\end{align}
	as the participants. The rapidity width of the spectator distribution is controlled by the parameter $\Delta y_{\rm b}=0.27$, which takes into account the Fermi motion of nucleons in the nucleus ($p_F=0.25$\,GeV/$c$ in rest frame of nuclei).
	
\begin{figure}
		\includegraphics[width=0.49\textwidth]{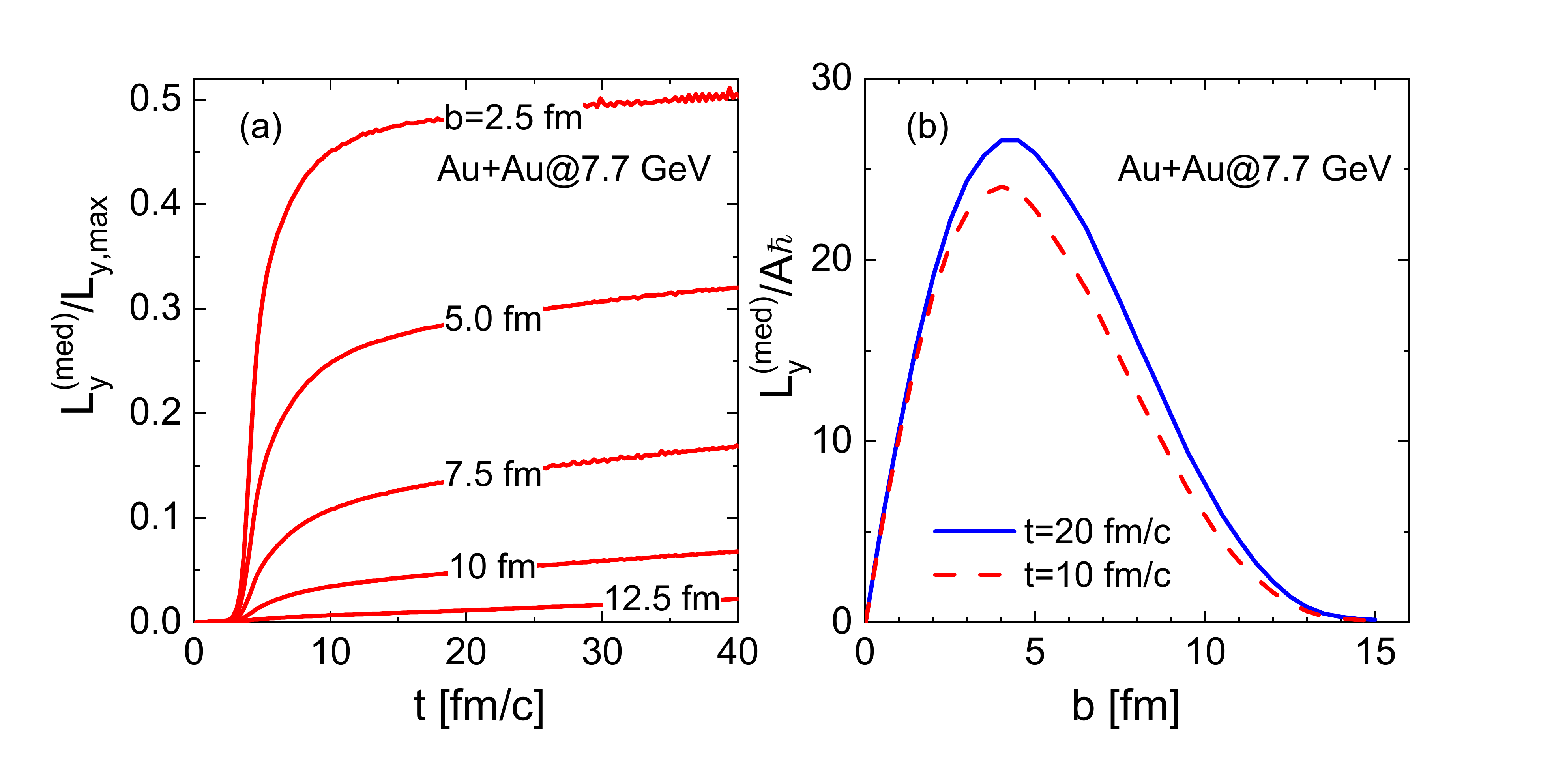}
		\caption{(a) The projection of the angular momentum orthogonal to the reaction plane, which is transferred to the medium, as a function of time for Au+Au collisions at $\sqrt{s_{NN}}=7.7$\,GeV and various impact parameters indicated by labels. The momentum is normalized to the maximal value corresponding to the given impact parameter (\ref{l-def}), $L_{y,{\rm max}}/A=189\hbar\,(b/10\,{\rm fm})$.
		(b) The $y$ component of the angular momentum stored in the medium as a function of the impact parameter for two moments of time.
		}
		\label{fig:L}
	\end{figure}

	The total angular momentum carried by the particles can be calculated as
	\begin{align}
		\vec{L}(t)=\sum_{h}\intop\rmd ^3 r \frac{\rmd^4 p}{(2\pi)^4} [\vec{r}\times \vec{p}]
		f_{\rm t.p.}^{(h)}(t,\vec{r}, p_0, \vec{p}),
		\label{L-tot}
	\end{align}
	where the sum over $h$ goes over all particle (hadron) types.
	We verified that for all considered collision energies and impact parameters the value of $\vec{L}$ calculated by this expression coincides with that given by Eq.~(\ref{l-def}) with the precision $\lsim 1\%$ and stays constant during the whole duration of the collision up to times $50\,{\rm fm}/c$.
	The question now is, which part of this total angular momentum, $\vec{L}^{\rm (med)}$ , is transferred to the medium?
	To calculate this quantity we replace the distribution function in (\ref{L-tot}) as $f_{\rm t.p.}^{(h)}\to f_{\rm t.p.}^{(h),{\rm part}}$.
	The evolution of $\vec{L}_y^{\rm (med)}$ for a fixed impact parameter is shown in Fig.~\ref{fig:L}(a) for Au+Au collisions at $\sqrt{s_{NN}}=7.7$\,GeV. Other components of the vector $\vec{L}^{\rm (med)}$ are proved to fluctuate strongly from event to event and are very small on average, $|\vec{L}_{x,z}^{\rm  (med)}|/|\vec{L}_y^{\rm (med)}|<10^{-3}$. On the time axis the zero time corresponds to the initialization of nuclei in the PHSD model before their collision. The touching time of nuclei is $\simeq 2.2\,{\rm fm}/c$ for $\sqrt{s_{NN}}=7.7$\,GeV and $b=7$\,fm. The overlap time interval for two nuclei of radius $R$ is
	\begin{align}
		\delta t_{\rm over}=\frac{2R}{\gamma_b}=\frac{4R m_N}{\sqrt{s_{NN}}}\,,
	\end{align}
	where $\gamma_b=\sqrt{s_{NN}}/2m_N$ is the Lorentz factor of colliding nuclei,
	that makes $\delta t_{\rm over}\simeq 3.6$\,fm/$c$ in our case. So, during the first $4\mbox{--}6$\,fm/$c$, when the nuclei approach and overlap, the momentum distribution of nucleons is represented by two counterstreaming flows of spectator nucleons. Then stating from  $t \approx 3\mbox{--}4$\,fm/$c$ the number of participants starts rapidly growing, and so does the angular momentum of the medium.
	
\begin{figure}
		\centering
		\includegraphics[width=5cm]{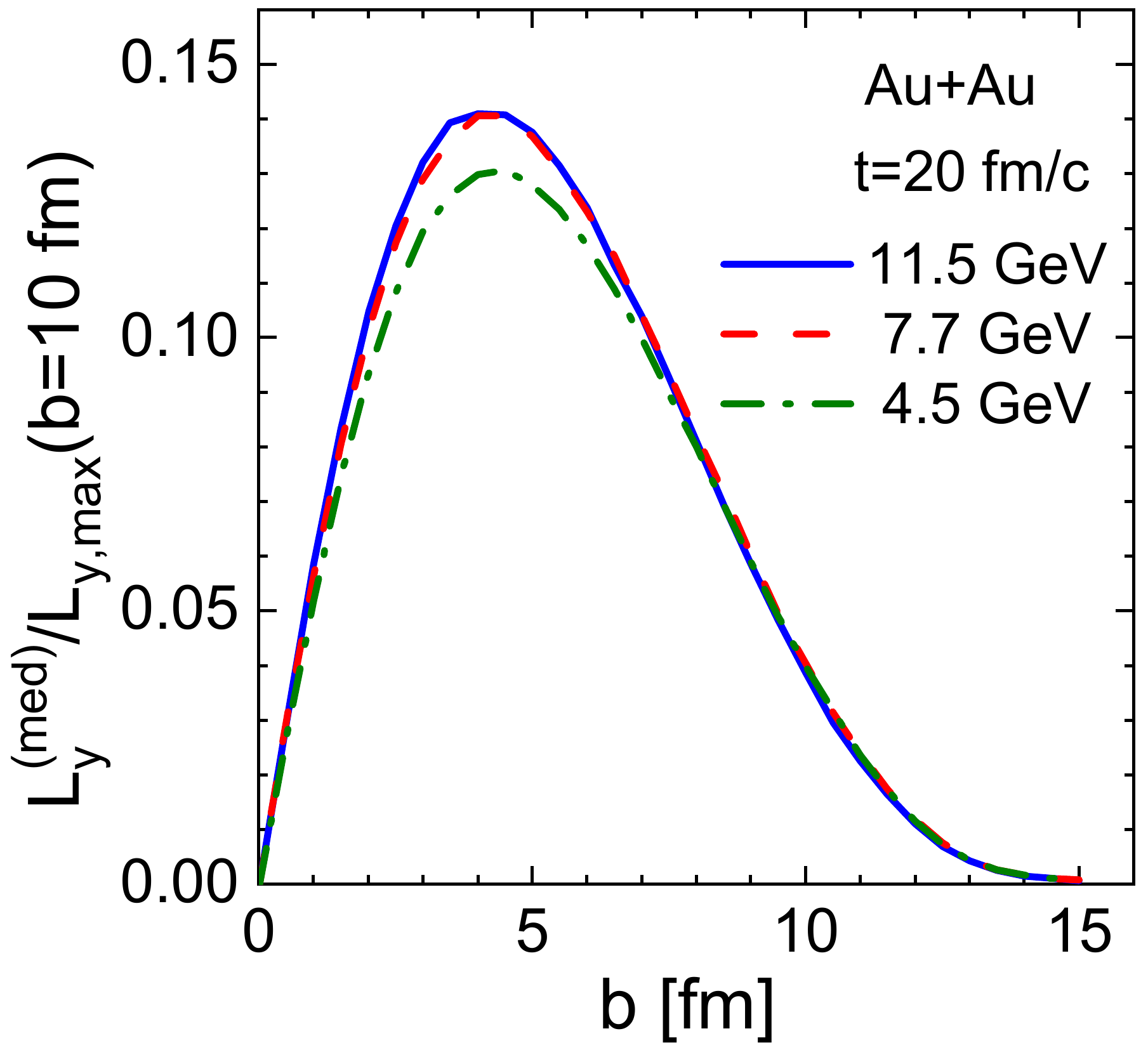}
		\caption{The transferred angular momentum at time $t=20$\,fm/$c$ normalized by the initial angular momentum of colliding nuclei (for $b=10$\,fm) at various colliding energies.}
		\label{fig:L-b-diffs}
	\end{figure}

	In Fig.~\ref{fig:L}(a) we see that the transfer of the angular momentum to the medium occurs over the timescale of $\approx (5\mbox{--}10)\,{\rm fm}/c$, and $\vec{L}^{\rm (med)}$ does not change significantly, while slightly growing, at $t\gsim 10$\,fm/$c$. Also, we observe that only a small part of the total angular momentum is, actually, transferred to the medium. So, for impact parameters $b>2.5$\,fm it less than 50\% and decreases with the increase of $b$ since the overlap of colliding nuclei decreases. The dependence of the transferred momentum, $L_y^{\rm (med)}$, on the impact parameter is shown in Fig.~\ref{fig:L}(b). It essentially differs from the linear proportionality with $b$ and shows a clear maximum for $b\approx 5$\,fm. Similar behavior of the transferred angular momentum was obtained in Ref.~\cite{Becattini-Piccinini-Rizzo} within the Glauber model. We see also that after $t=10\,{\rm fm}/c$ the $b$ dependence does not change much, not more than be 10--15\% [compare solid and dashed curves in Fig.~\ref{fig:L}(b)], and saturates for $t\gsim 20\,{\rm fm}/c$ [see Fig.~\ref{fig:L}(a)]. The dependence $\vec{L}^{\rm (med)}(b)$ weakly varies with the collision energy. Figure~\ref{fig:L-b-diffs} shows the function $\vec{L}^{\rm (med)}(b)$ normalized by the maximum available angular momentum of colliding nuclei, Eq.~(\ref{l-def}), at the impact parameter $b=10$\,fm for several colliding energies. We see that the result for $\sqrt{s_{NN}}=11.5$\,GeV (solid line) almost perfectly coincides with the result for 7.7\,GeV (dashed line) and, at a lower energy,
	$\sqrt{s_{NN}}=4.5$\,GeV, the function (dash-dotted line) is smaller by 7\% around maximum at $b\approx 5$\,fm but coincides with the results for other energies at $b<2$\,fm and $b>8$\,fm.

\begin{figure}
		\centering
		\includegraphics[width=6cm]{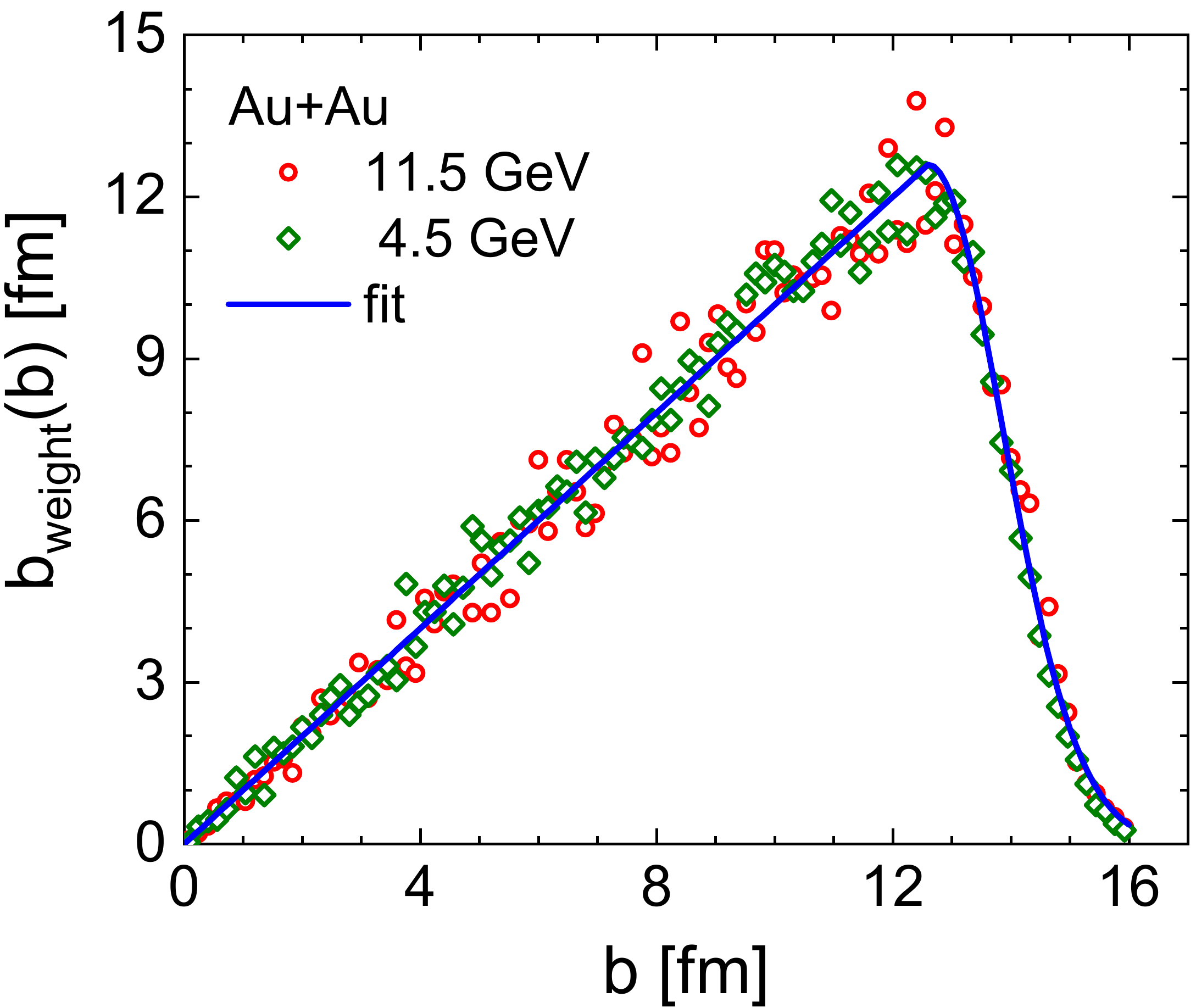}
		\caption{The impact parameter weighted with the probability that the nucleus-nucleus interaction occurs at impact parameter $b$ [see Eq.~(\ref{p-event})], $b_{\rm weight}(b)=b\,p_{\rm even}(b)$, as a function of $b$ for two collision energies. The solid line shows the parametrization~(\ref{b-weight}).  }
		\label{fig:b-weighted}
	\end{figure}

\begin{figure}
		\centering
		\includegraphics[width=8.5cm]{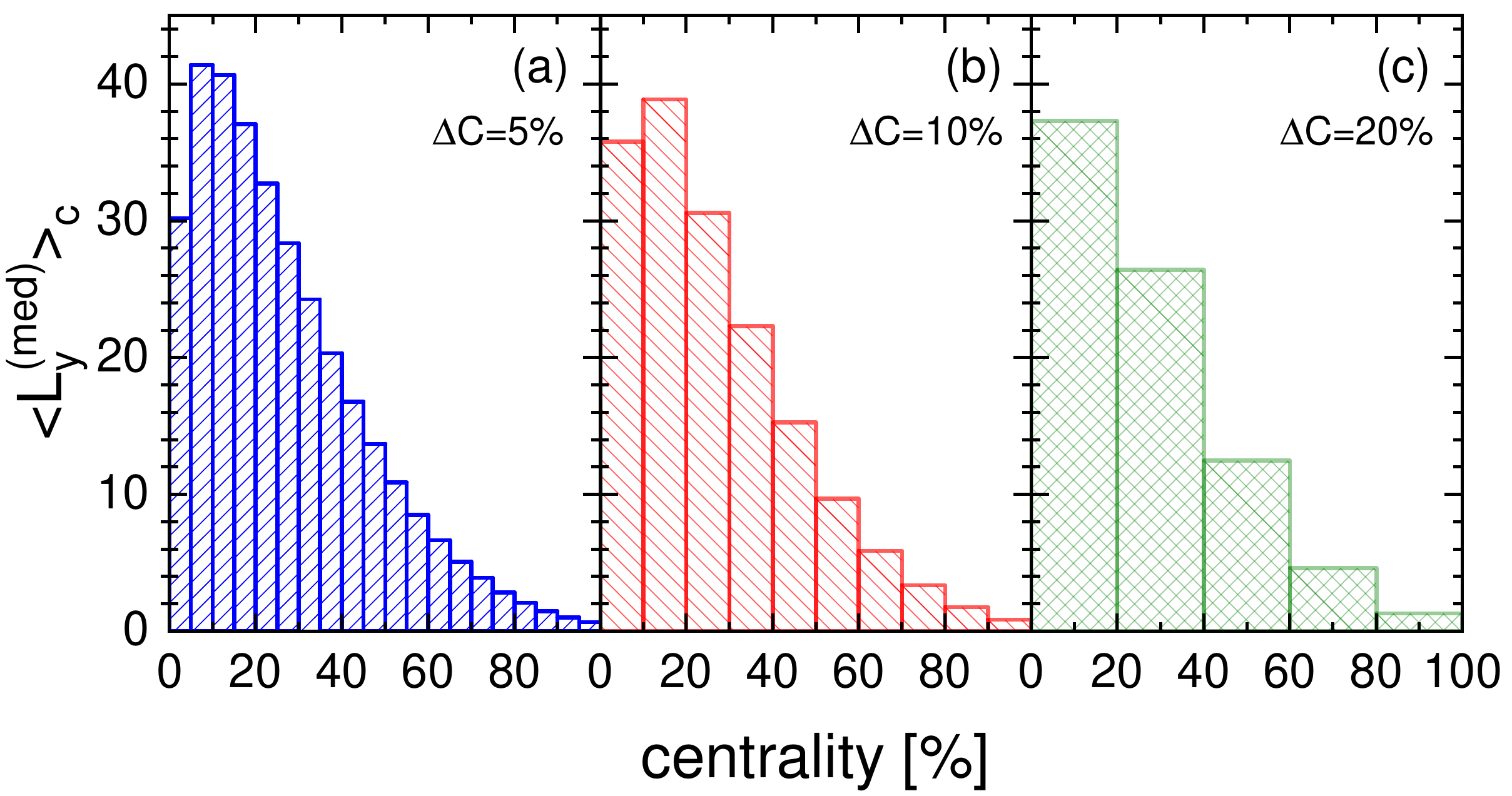}
		\caption{Averaged angular momentum carried by the medium as a function of centrality for the Au+Au collision at $\sqrt{s_{NN}}=7.7$\,GeV. Panels (a), (b), and (c) show the results for various centrality binnings: $\Delta C=$5\%, 10\%, and 20\%, respectively.    }
		\label{fig:L-aver}
	\end{figure}

	Physical quantities measured in heavy-ion experiments are averaged within some centrality class.
	To get a feeling about the transferred angular momentum in the collision with specific centrality selection we have to average
	$\vec{L}^{\rm (med)}(b)$ over the impact parameter with the weight $2\pi b\, p_{\rm event}(b)$. Here $p_{\rm event}(b)$ stands for
	the probability density that the nucleus-nucleus interaction has occurred at a given impact parameter, which can be expressed through the differential number of collision events, $\rmd N_{\rm event}$, that occurred for given $b$ normalized to the total number of events, $N_{\rm event}$:
	\begin{align}
		p_{\rm event}(b){\rmd b}=\frac{\rmd N_{\rm event}}{N_{\rm even}}.
		\label{p-event}
	\end{align}
	The distributions of the ``weighted impact parameter'' $b_{\rm weight}(b)=b\, p_{\rm event}(b)$ are calculated in PHSD for two collision energies, $\sqrt{s_{NN}}=4.5$ and 11.5\,GeV, and shown in Fig.~\ref{fig:b-weighted}. As we see, the distributions weakly depend on the collision energy and can be parametrized by the expression
	\begin{align}
		&b_{\rm weight}(b) 
		=\left\{
		\begin{array}{ll}
			b\,,& b\le b_{\rm d},\\
			12.6\,{\rm fm} \times e^{0.31(b/{\rm fm}-12.6)^2}, & b_{\rm d} < b < b_{\rm max}.
		\end{array}
		\right.
		\label{b-weight}
	\end{align}
	Here the maximum impact parameter for the Au+Au collisions is $b_{\rm max}=16$\,fm and $b_{\rm d}=12.6$\,fm. The rapid smooth dropoff of the function $b_{\rm weight}(b)$ for $b>b_{\rm d}$ is determined by the diffuseness of the density distribution in the nucleus.
	Now the averaged transferred angular momentum corresponding to the impact factor range $b_1<b<b_2$ is then defined as
	\begin{align}
		\langle \vec{L}^{\rm (med)}\rangle_{b_1}^{b_2} =\frac{
			\intop_{b_1}^{b_2}\vec{L}^{\rm (med)}(b') b_{\rm weight}(b')\rmd b'
		}
		{
			\intop_{b_1}^{b_2} b_{\rm weight}(b')\rmd b'
		}\,.
		\label{Laver}
	\end{align}
	The relation between the centrality of collision and the impact parameter is
	\begin{align}
		C(b)=\frac{\intop_{0}^{b} b_{\rm weight}(b') \rmd b'}
		{\intop_{0}^{b_{\rm max}} b_{\rm weight}(b') \rmd b'}.
	\end{align}
	With this definition the most central collisions (small $b$) correspond to  small values of $C$.
	In Fig.~\ref{fig:L-aver} we show the averaged transferred angular momentum as a function of centrality for various centrality binnings. For a fine centrality binning, $\Delta C\lsim 10\%$, one can resolve a maximum in the angular momentum transfer at $C\sim 10\mbox{--}20\%$; see panels (a) and (b) in Fig.~\ref{fig:L-aver}. For a coarser binning, see Fig.~\ref{fig:L-aver}(c), the maximum disappears and the magnitude of the averaged angular momentum is slightly reduced for the smallest centrality bin.
	From calculations shown in Fig.~\ref{fig:L-aver} we see that the transferred angular momentum decreases with an increase of parameter $C$, i.e., with increase of the impact parameter, when $C>10\mbox{--}20\%$ (depending on the binning step). Interestingly, the centrality dependence of hyperon polarization shows the opposite trend and increases with the $C$ increase. This is observed both at high collision energies~\cite{Adam-PRC98} and at low ones~\cite{Abdallah-PRC104}. This means that the formation of polarization signal of hyperons has more complicated nature than a direct transformation of the initial angular momentum and depends on the production mechanism of hyperons and, thereby, on their phase-space distributions.

	\section{Fluidization}\label{sec:Fluid}
	
	Our next task is to obtain the hydrodynamical characteristics of the medium (fluid) created in the heavy-ion collision. In other words, we have to fluidize the test-particle distributions generated by the transport code and determine local energy and baryon densities and velocities of the fluid. The flow velocity, $u_\mu$, we define in the Landau frame where the 4-velocity is the eigenvector of the full energy-momentum tensor, $T^{\mu \nu}$,
	\begin{equation}\label{eq:hydro:rel:Tmunu-ideal-eigen}
		T^{\mu \nu}\, u_{\mu} = \varepsilon\, u^{\nu},
	\end{equation}
	and the corresponding the eigenvalue, $\varepsilon$, is then local energy density. The four-velocity is normalized as $u_\mu u^\mu=1$ and can be written as
	$u_\mu=\gamma (1,\vec{v})$  through the three-velocity $\vec{v}$ and $\gamma=(1-\vec{v}^2)^{-1/2}$.
	When the flow velocity is determined, the local baryon density can be computed from the baryon current, $J^\mu_B$, as $n_B=u_\mu J_B^\mu$. To determine the local temperature one has use the equation of state (EoS) of the medium and solve the equation $\varepsilon(n_B,T)=\varepsilon$. We use the EoS of~\cite{SDM09}, which includes all known hadrons with masses up to 2\,GeV/$c^2$ in the zero-width approximation. The equation of state of the hadron resonance gas at finite temperature and baryon density is calculated thermodynamically, taking into account a density-dependent mean field that guarantees the nuclear matter saturation. This EoS was used in the hydrodynamic calculations~\cite{KKT-Hydro,KT-HysHSD}. For the detailed description of the EoS, see Ref.~\cite{KKT-Hydro}.
	
	Now we specify how we calculate the energy-momentum tensor $T_{\mu\nu}$ and the baryon current $J_B^\mu$ at each space-time point $(t,\vec{r})$. First, in order to make the transition from discrete particles to continuous medium we introduce a smearing function $\Phi(\vec{r}, \vec{r}_i(t))$ instead of the spatial $\delta$ function in (\ref{f-test-p}):
	\begin{align}\label{eq:fluidization:distrib-func}
		f_{\rm t.p.}(t,\vec{r},p_0, \vec{p}) &= \sum_{h,i_h} \frac{(2\pi)^4}{\mathcal{N}} \delta^{(3)}(\vec{p} - \vec{p}_{i_h}(t))
		\nonumber\\
		&\times \delta\big(p_0-\sqrt{m_h^2+\vec{p\,}^2}\big) \Phi(\vec{r}, \vec{r}_{i_h}(t)),
	\end{align}
	where $\mathcal{N} =\int d^3r\, \Phi(\vec{r}, \vec{r}_{i_h}(t))$ is the normalization factor. Then, the energy-momentum tensor looks as follows:
	\begin{align}
		\label{eq:Tmunu}
		T^{\mu\nu}(t,\vec{r}) &= \int \frac{d^4p}{(2\pi)^3} \frac{p^\mu p^\nu}{p^0 }f_{\rm t.p.}(t,\vec{r},p_0, \vec{p})
		\nonumber \\
		&= \frac{1}{\mathcal{N}} \sum_{h,i_h} \frac{p^\mu_{i_h}(t) p^\nu_{i_h}(t)}{p^0_{i_h}(t) }\, \Phi(\vec{r}, \vec{r}_{i_h}(t))
	\end{align}
	where $p^{\mu}_{i_h}=(p^{0}_{i_h}, \vec{p}_{i_h})$ -- four-momentum of particle $i_h$ of type $h$. Similarly, the baryon current is given by
	\begin{equation} \label{eq:baryoncurrent}
		J_B^\mu(t,\vec{r}) =   \frac{1}{\mathcal{N}} \sum_{h,i_h} B_{i_h}\,
		\frac{p^\mu_{i_h}(t) }{p^0_{i_h}(t) }\, \Phi(\vec{r}, \vec{r}_{i_h}(t)),
	\end{equation}
	where $B_{i_h}$ is the baryon charge of particle $i_h$.
	
	The smearing kernel $\Phi(\vec{x}, \vec{x}_i(t))$ is often taken in a Gaussian form. We will follow the particle-in-cell (PIC) method well known in hydrodynamics and plasma physics. Namely, we will use a square-law spline kernel (the cloud-in-cell method) as a smearing kernel~\cite{Birdsall1997}. It is fast and provides continues distributions of considered quantities. In contrast to the Gaussian kernel it corresponds to particles with a finite size. For a one-dimensional grid in the $x$ direction with step $\Delta_x$, for any coordinate $x$ the nearest grid point is $x_a=a_x\,\Delta_x$, where $a_x=[x/\Delta_x]$ (here $[x]$ is the floor of $x$). Thereby, the contribution of each particle to three nearest grid points $x_a$ and $x_{a\pm1}$ is defined by the following functions:
	\begin{align}
		\Phi(x_a,x_i) &= \frac{1}{\Delta_x} W_0 (x_a/\Delta_x - [x_i/\Delta_x]),
		\nonumber\\
		\Phi(x_{a\pm1},x_i) &= \frac{1}{\Delta_x} W_{\pm1} (x_a/\Delta_x - [x_i/\Delta_x]),
	\end{align}
	where
	\begin{equation}\label{eq:fluid:wight}
		W_k(x) =
		\begin{cases}
			\displaystyle \frac{3}{4}-x^2, & k=0,\\[1em]
			\displaystyle \half \, \Big( \half\pm x \Big)^2, & k=\pm 1.
		\end{cases}
	\end{equation}
	Note, that $ W_{0}(x) + W_{+1}(x) + W_{-1}(x) = 1 $. In the three-dimensional case, the grid points are $\vec{r}_a=(a_x\Delta_x,a_y\Delta_y,a_z\Delta_z)$, and the full smearing function is just a product of the one-dimensional functions for each space direction:
	\begin{align}
		\Phi(\vec{r}_{a},\vec{r}_i) = \Phi(x_{a},x_i)\, \Phi(y_{a},y_i)\, \Phi(z_{a},z_i).
		\label{Phi-3D}
	\end{align}
	
	All numerical calculations are done on the space grid with $\{\Delta_x,\Delta_y,\Delta_z\}=\{1,1,1/\gamma_b\}$\,fm. For each cell of the grid we calculate contributions to the $T^{\mu\nu}$ tensor and the $J_B^{\mu}$ vector from 27 neighboring cells.
	Then one can analytically solve Eq.~(\ref{eq:hydro:rel:Tmunu-ideal-eigen}) and obtain velocity, energy density, and temperature of fluid for each central point of the cell.
	When further spatial derivatives has to be calculated, e.g., for vorticity, we use the formula
	\begin{align}
		\label{eq:NumericalPartials}
		\partial_i u^j_{\vec{r}_a}\approx \frac{u^j_{\vec{r}_{a+i}}-u^j_{\vec{r}_{a-i}}}{2 \, \Delta_i}\,,\quad i,j=x,y,z\,,
	\end{align}
	where we denote the position of neighboring cells as $\vec{r}_{a\pm i}=\vec{r}_a\pm(\delta_{ix}\Delta_x+\delta_{iy}\Delta_y+\delta_{iz}\Delta_z)$.

	After all necessary quantities are defined on the grid (let us denote them generically  $A_{\vec{r}_a}$), the same smearing function (\ref{Phi-3D}) is used for continuous interpolation of the considered quantities at any point inside a grid. So, the value $A(x)$ at any point $\vec{r}$ in the vicinity of the nearest grid knot $\vec{r}_a$ is given by
	\begin{equation} \label{interpolation}
		A(\vec{r}) = \sum\limits_{i,j,k=\pm 1,0} A_{\vec{r}_a + (i\Delta_x,j\Delta_y,k\Delta_z)} \, \Phi(\vec{r}, \vec{r}_{a}) \Delta_x\, \Delta_y\, \Delta_z.
	\end{equation}
	
	As the result of the high collision statistics and the interpolation procedure described in this section, we obtain very smooth distributions of the velocity, the temperature, and the energy and particle density fields.

	\section{$T$, $n$, $\varepsilon$ profiles}\label{sec:Tne}
	
	\begin{figure} 
		\centering
		\includegraphics[width=0.49\textwidth]{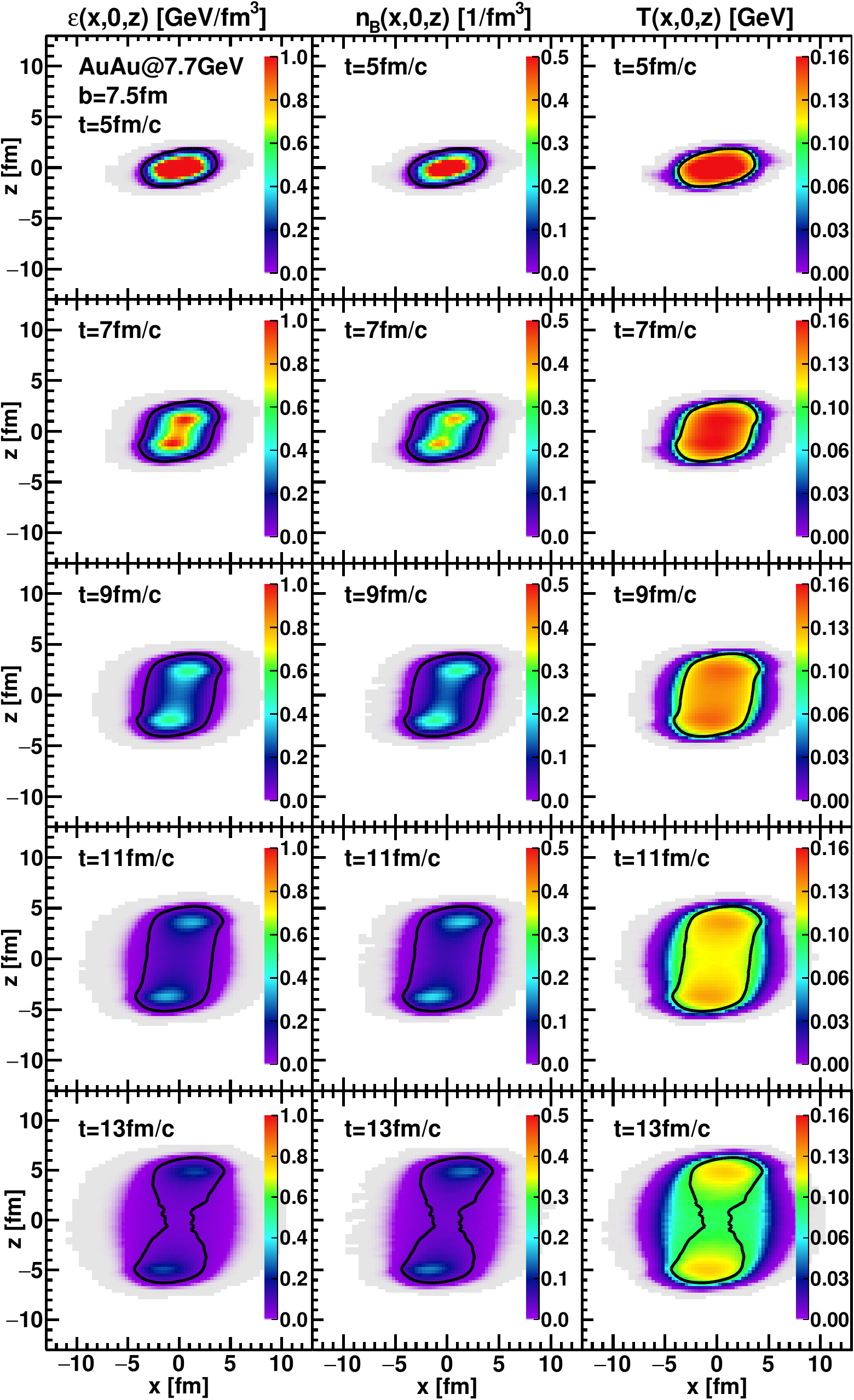}
		\caption{Time dependence of energy density, baryon density, and temperature for  Au+Au collisions at $\sqrt{s_{NN}}=7.7$\,GeV at impact parameter $b=7.5$\,fm. Solid lines indicate the contour of the condition $\epsilon_c=0.05\,{\rm GeV/fm^3}$. Light grey fields show cells containing at least one particle in one of the simulated collision events.}
		\label{fig:ENT-contour-Au77-b75}
	\end{figure}
	
	The evolutions of energy density, baryon density, and corresponding temperature fields in the $y=0$ plane are shown in Fig.~\ref{fig:ENT-contour-Au77-b75} for Au+Au collisions at $\sqrt{s_{NN}}=7.7$\,GeV and impact parameter $b=7.5$\,fm. Only participants are shown here, while the spectator nucleons are separated as discussed in Sec.~\ref{sec:Spec-separ}. The earliest time corresponds to the maximum overlap moment $t\simeq 5$\,fm/$c$. The energy and baryon densities and the temperature have maximal values and the hot fluidized zone of the fireball has the shape of a slightly tilted pill with radius $\approx 4$\,fm in the $xy$ plane and thickness 4\,fm in the $z$ direction. It starts longitudinal and transversal expansion, forming after roughly 7\,fm/$c$ a tilted cylinder similar to the Bjorken expansion model. After 9\,fm/$c$ the densities and temperature fields form two maxima moving in opposite directions and corresponding to the excited fragments of nuclei that passed through each other.
	Black solid lines on the plots show the contour in the $xz$ plane outside of which the energy density is smaller than $0.05\,{\rm GeV/fm^3}$. This was suggested in Ref.~\cite{Huovinen-Petersen} as a criterion of applicability of the hydrodynamics description. We see that at time 13\,fm/$c$ the central part of the fireball is substantially disintegrated (freeze-out stage). The full disintegration of the fireball fluid occurs at $\approx 15\mbox{--}16$\,fm/$c$.

\begin{figure}
		\centering
		\includegraphics[width=8.5cm]{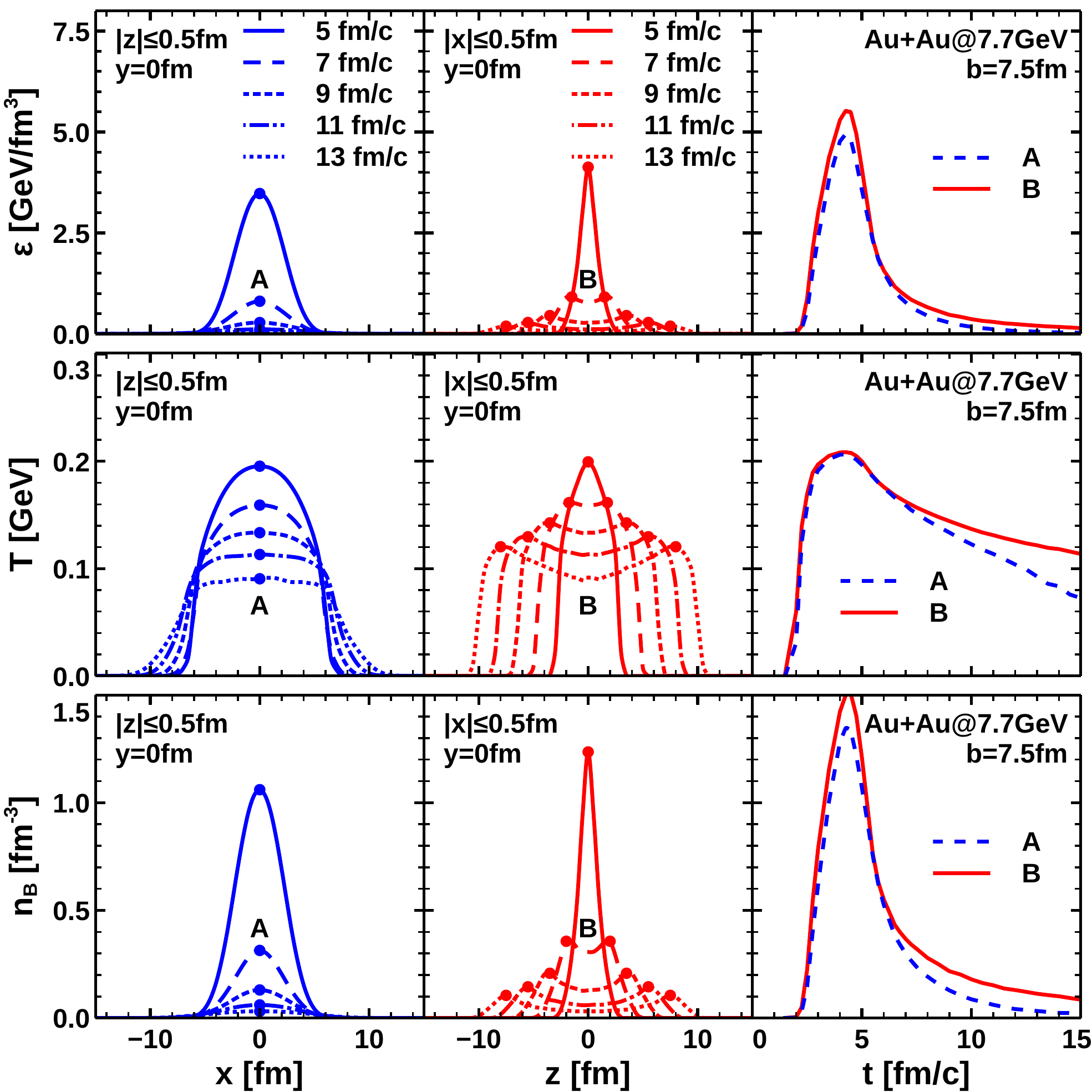}
		\caption{
			Profiles of the energy density (upper row), the temperature (middle row), and the baryon density (lower row) plotted for $y=0$\,fm along the $x$ axis and integrated over the interval $|z|<0.5$ (first column) and plotted along the $z$ axis and integrated for $|x|<0.5$ (second column). The third column shows the maxima of the $x$ and $z$ profiles by points $A$ and $B$ correspondingly. Calculations are done for Au+Au collisions at $\sqrt{s_{NN}}=7.7$\,GeV, impact parameter $b=7.5$\,fm, and five moments of time. Time of the maximum overlap is about 4.9\,fm/$c$.}
		\label{fig:profilesXZ-7.7}
	\end{figure}

	For better visualisation of the distribution of the thermodynamic quantities shown in Fig.~\ref{fig:ENT-contour-Au77-b75}, in Fig.~\ref{fig:profilesXZ-7.7} we present profiles of this quantities along the $x$ and $z$ axes (shown in the first and second columns) in the plane $y=0$\,fm. After the nucleus overlapping is completed at 5\,fm/$c$, all profiles have maxima for the center cell at $x=z=0$\,fm. Note that the some differences in the maxima of the $x$ and $z$ profiles occur because they are obtained after the summation over the final intervals of coordinates $|z|<0.5$\,fm  and $|x|<0.5$\,fm for the first and second profiles, respectively. The difference is more pronounced if one profile is much sharper than the other one; compare distributions for the temperature and for the energy and baryon densities. If one integrates the $x$ profile for $|x|<0.5$\,fm and the $z$ profile for $|z|<0.5$\,fm one obtain exactly the same value.
	In the next 2\,fm/$c$ the hight of the $\varepsilon$ and $n$ profiles drops by factor 4. Thereby, the profiles of the energy and baryon densities (upper and lower rows) broaden slowly in the $x$ direction and fast in the $z$ directions, in which their widths become almost twice larger.
	At the later times $z$ profiles exhibit two symmetric maxima for positive and negative $z$ moving away with the speed close to $c$.
	The temperature profiles decreases slower than those for $\varepsilon$ and $n$ and broaden similarly in both $x$ and $z$ directions.

	\begin{figure}
		\centering
		\includegraphics[width=8.5cm]{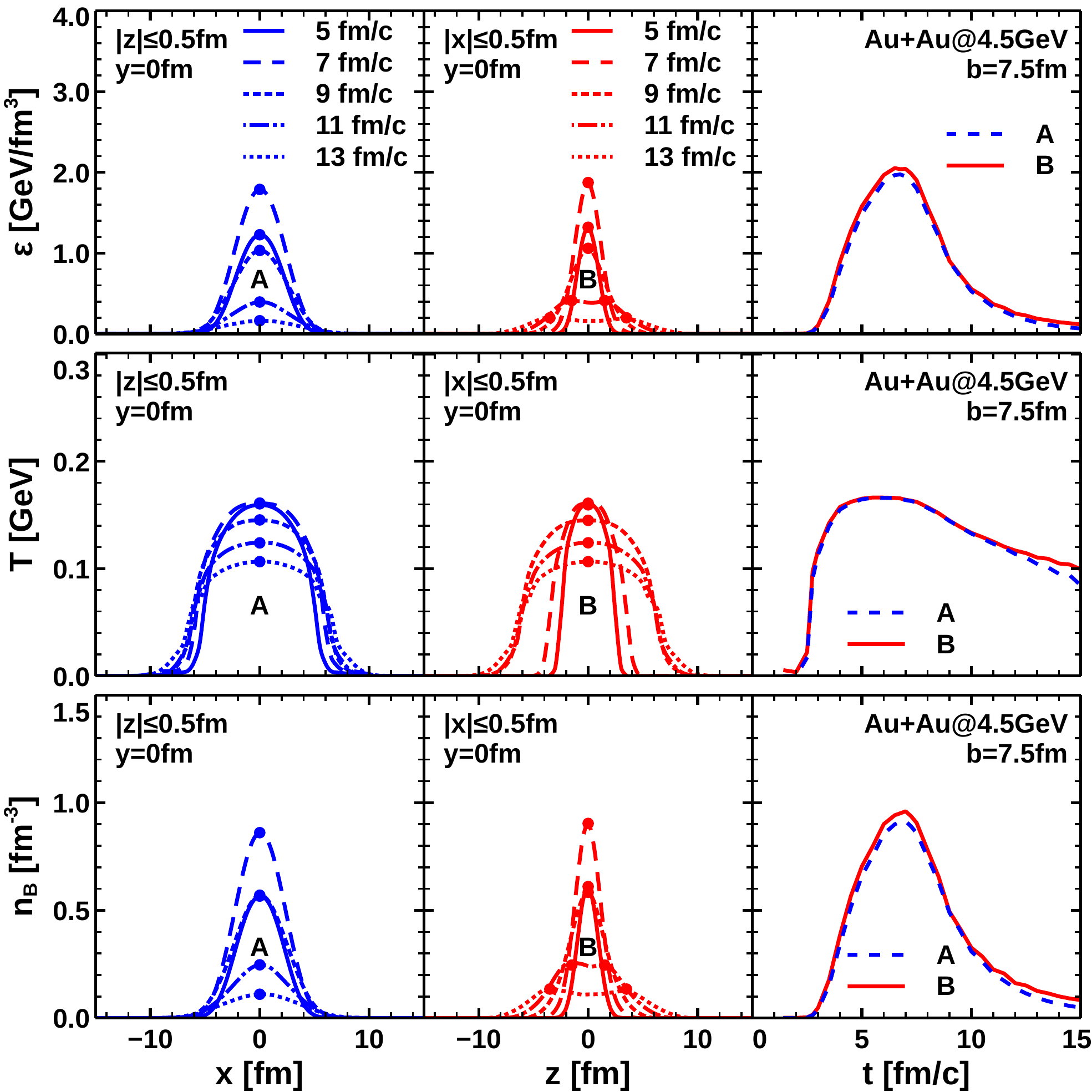}
		\caption{The same as in Fig.~\ref{fig:profilesXZ-7.7} but for collision energy $4.5\,$GeV.
		The time of the maximum overlap is about 7.5\,fm/$c$.}
		\label{fig:profilesXZ-4.5}
	\end{figure}
	
	\begin{figure}
		\centering
		\includegraphics[width=8.5cm]{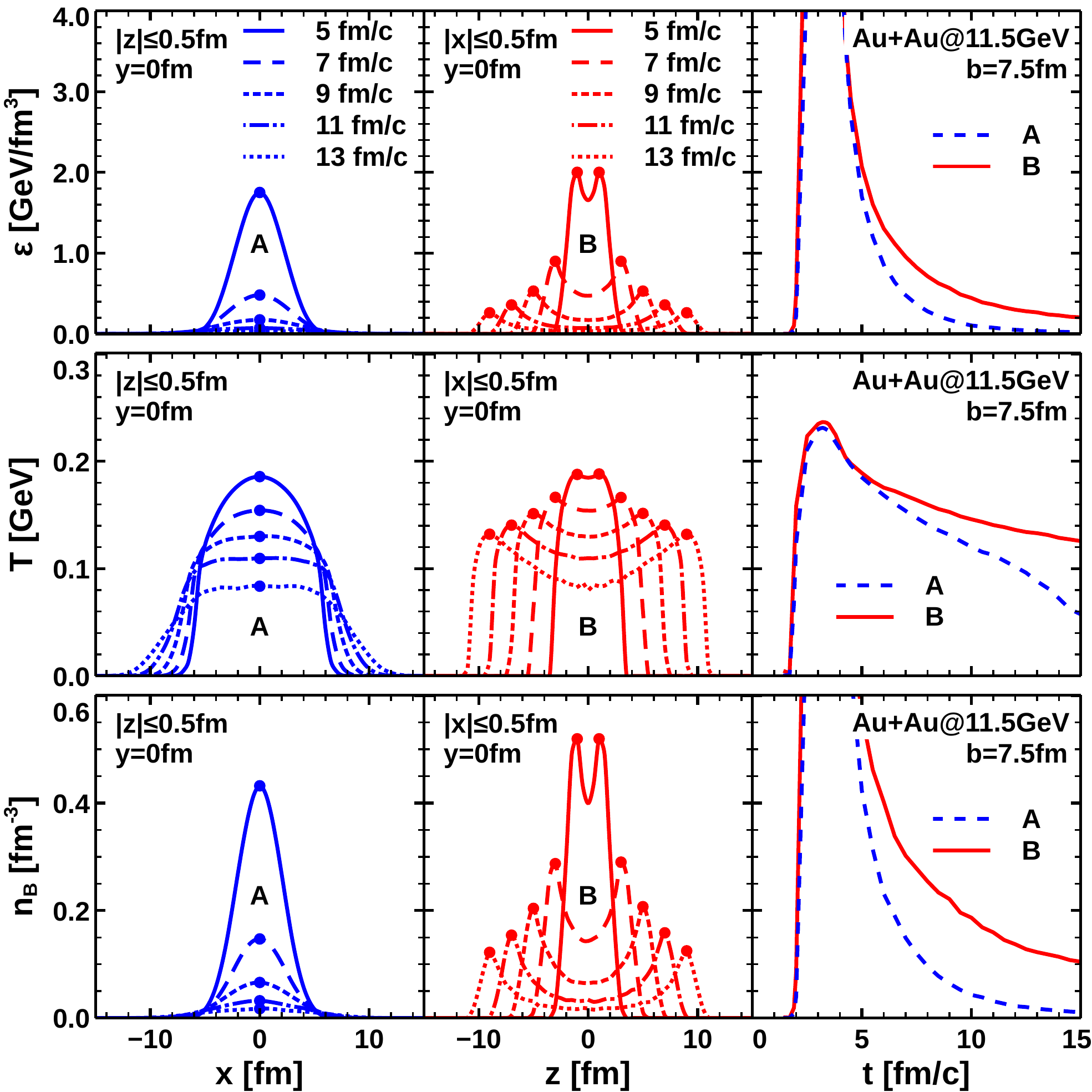}
		\caption{The same as in Fig.~\ref{fig:profilesXZ-7.7} but for collision energy $11.5\,$GeV.
		The time of the maximum overlap is about 3.8\,fm/$c$. The maximum of the energy density is 12\,GeV/fm$^{3}$ at $t=3.3$\,fm/$c$. The maximum of the baryon number density is 2.1/fm$^{3}$ at $t=3.4$\,fm/$c$.  }
		\label{fig:profilesXZ-11.5}
	\end{figure}

	Next we consider how the thermodynamical characteristics of the fireball changes with the variation of the collision energy. In Fig.~\ref{fig:profilesXZ-4.5} we show the $x$ and $y$ profiles of $\varepsilon$, $T$, and $n_B$ for the Au+Au collisions at$\sqrt{s_{NN}}=4.5$\,GeV and $b=7.5$\,fm. The striking difference seen in the profiles of all quantities is the slow-down of the evolution. The compression phase lasts now till $\approx 7$\,fm/$c$, and the expansion phase from 7\,fm/$c$ to 13\,fm/$c$. The double-hump structures in the $z$ profiles of $\varepsilon$ and $n$ clearly seen in Fig.~\ref{fig:profilesXZ-7.7} are barely seen in Fig.~\ref{fig:profilesXZ-4.5}. Maxima of all thermodynamical quantities are smaller for collisions at $\sqrt{s_{NN}}=4.5$\,GeV than for collisions at $\sqrt{s_{NN}}=7.7$\,GeV.
	
	\begin{figure*}
		\centering
		\includegraphics[width=1\textwidth]{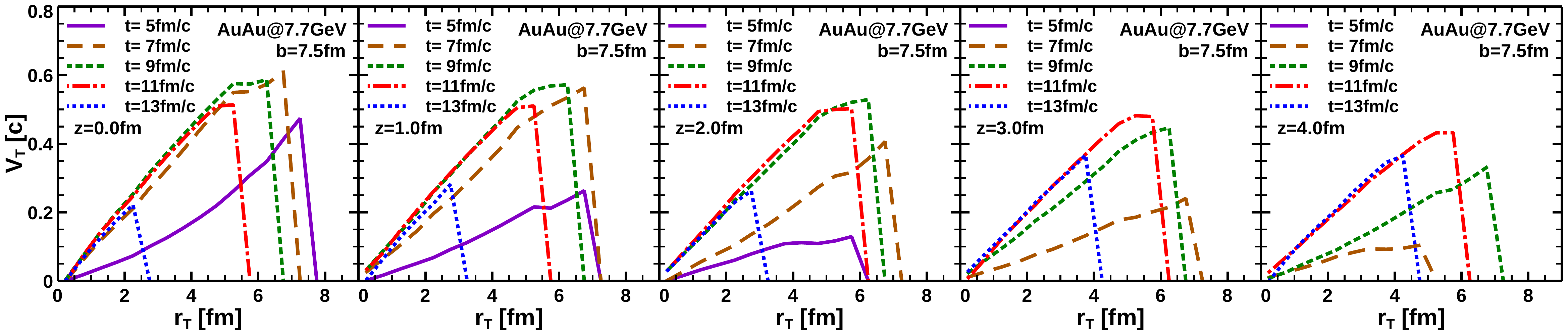}
		\caption{Transverse velocity (\ref{vT-def}) as a function of transverse radius  for various values of $z$ and various moments of time. Cut $\varepsilon>0.05\,{\rm GeV/fm^3}$ is applied.}
		\label{fig:vT}
	\end{figure*}

	Consider now higher energies. In Fig.~\ref{fig:profilesXZ-11.5} we show profiles for collisions at $\sqrt{s_{NN}}=11.5$\,GeV. The evolution of the system in the transversal direction is similar to that for 7.7\,GeV. In the $z$ direction the system expands very rapidly and the double-hump structure appears also.

	\section{Velocity field }\label{sec:V-field}
	
	We turn now to the velocity field created in the collisions. We concentrate on the collision energy of 7.7\,GeV. In Fig.~\ref{fig:vT} we show the transverse velocity
	\begin{align}
		v_T=\sqrt{v_x^2+v_y^2}
		\label{vT-def}
	\end{align}
	as a function of the transverse radius $r_T=\sqrt{x^2+y^2}$ for various $z$ slices ($z\ge 0$) at various moments of time. We see that after $\approx 9$\,fm/$c$ the profile $v_T\propto r_T$, almost independent of $z$, is formed for $z\le 2$\,fm, and after $\approx 11$\,fm/$c$ this dependence gets extended for $z\lsim 4$\,fm. At earlier times, the Hubble-like flow is not yet formed completely and the transverse velocity has a steeper dependence on $r_T$ and is not  extended to $z>0.5$\,fm, decreasing with a $z$ increase. At times $>11$\,fm/$c$ the outer regions start freezing out and the region of the collective flow shrinks as we do not take into account fluids with the energy density $\varepsilon<0.05\,{\rm GeV/fm^3}$.
	
	We have to note that the Hubble-like expansion does not necessarily happen around the central cell. Only for $z=0$ the radial expansion of the fluid occurs around the point $(x_0=y_0=z_0=0)$, whereas for $z\gsim 1$\,fm the center of the expansion is shifted to $x_0>0$, but $y_0=0$ because we consider noncentral collisions. The position of the center changes with time from $x_0\simeq 0.5$ and 1.5\,fm for $z=1$ and 2\,fm, respectively, at $t=5$\,fm/$c$, to smaller values $x_0\simeq 0.25$\,fm for $z<4$\,fm and 0.5\,fm for $z \approx 4$\,fm at $t=13$\,fm/$c$. The coordinate of the center is a asymmetric function of $z$.

\begin{figure} 
		\centering
		\includegraphics[width=5cm]{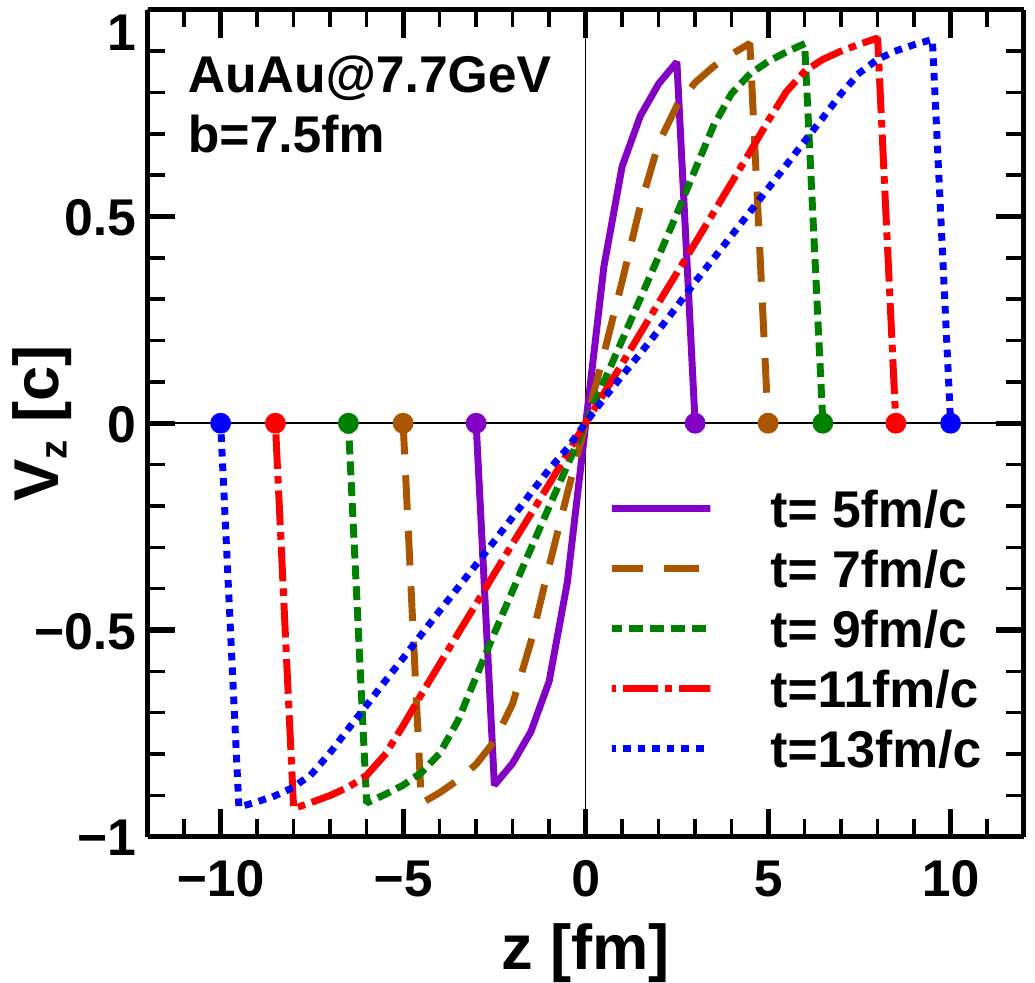}
		\caption{Profiles of the longitudinal velocity $v_z$ as a function of $z$ for $r_T=0$ and several moments of time. Cut $\epsilon>0.05\,{\rm GeV/fm^3}$ is applied.}
		\label{fig:vZ-profiles}
	\end{figure}

	The profile of the longitudinal direction, $v_z$, is shown in Fig.~\ref{fig:vZ-profiles}. The Hubble-like behavior, $v_z= \alpha_{\|} z$ is established already at earlier times.
	
\begin{figure} 
		\parbox{4.3cm}{\includegraphics[width=4.25cm]{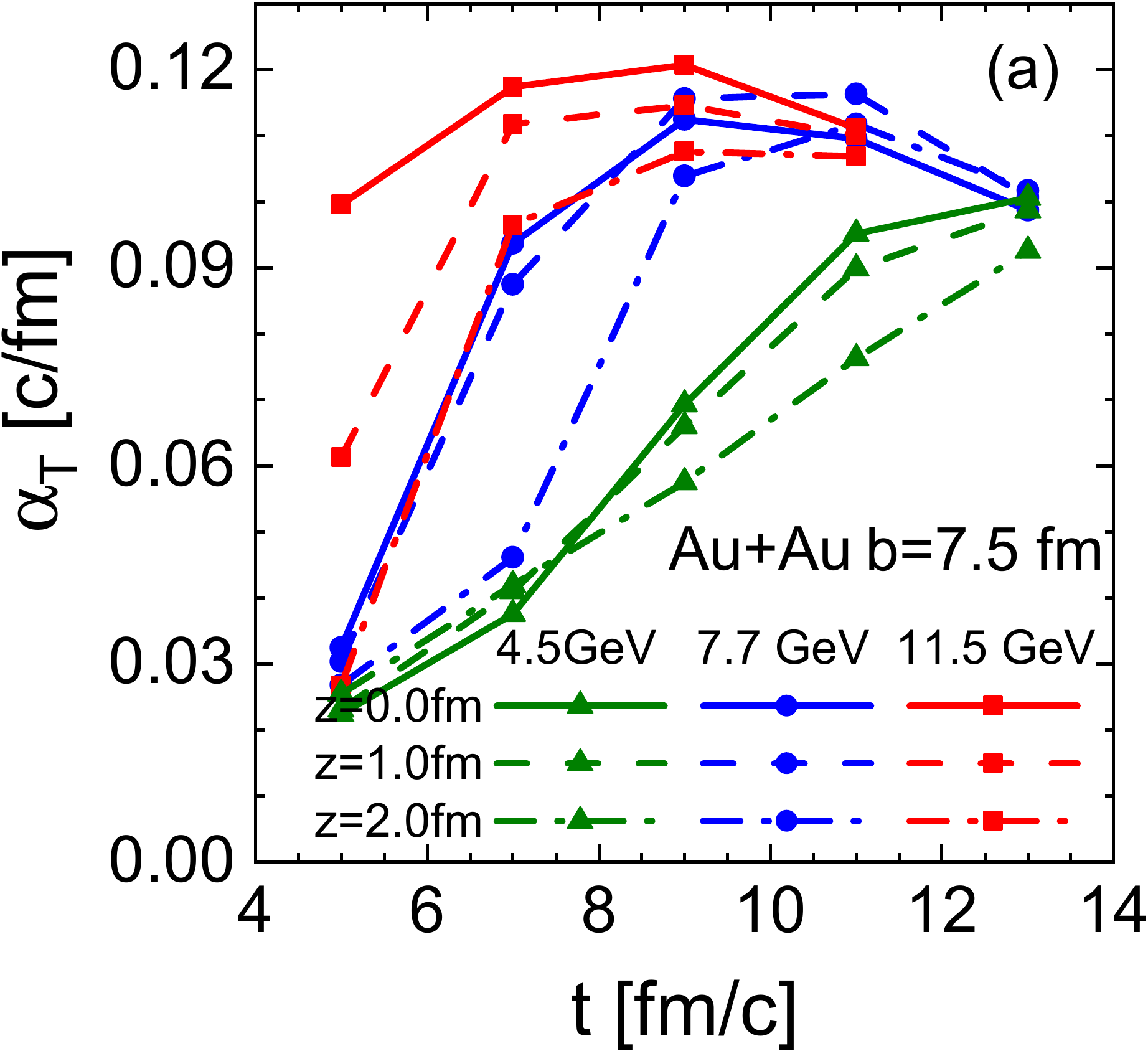}}
		\parbox{4.2cm}{\includegraphics[width=4.1cm]{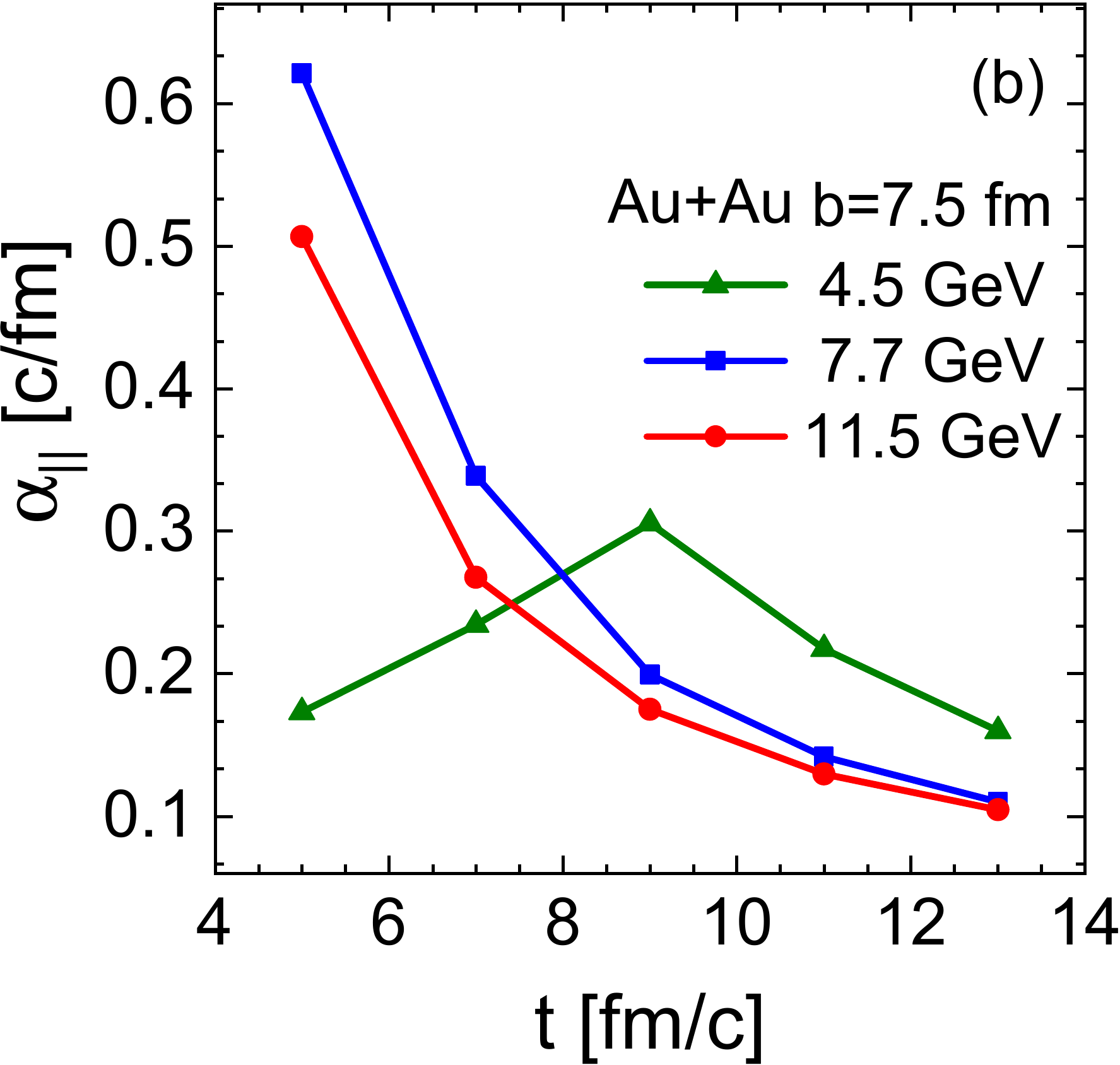}}
		\caption{Evolution of the parameters of the Hubble-like parametrization of the velocity field (\ref{V-Hubble}) with $\beta_T=\beta_\|=1$ for various collision energies. Panel (a) shows the transverse flow parameter $\alpha_T$ for the various $z$ slices. Panel (b) shows the longitudinal flow parameter $\alpha_\|$, which is almost $z$-independent.  }
		\label{fig:Habble-param}
	\end{figure}

	Thus, the structure of the velocity field after the fluidization of the particle distributions obtained in the PHSD transport model has mainly the Hubble-like structure for each moment of time,
	\begin{align}
		\vec{v}_{\rm H} &=\alpha_T\,r_T^{\beta_T} \vec{e}_T + \alpha_{\|}\,z^{\beta_{\|}}\,\vec{e}_z\,,\quad
		\label{V-Hubble}
	\end{align}
	where $\alpha_T$, $\beta_T$ and $\alpha_{\|}$, $\beta_\|$ do not depend on coordinates (but, maybe, on time) and we introduced the unit vectors in the transverse and longitudinal directions, $\vec{e}_T =\vec{r}_T/r_T= (x,y,0)/r_T$ and $\vec{e}_z=(0,0,1)$\,. Here and below we do not indicate the time dependence of velocity components explicitly. From Figs.~\ref{fig:vT} and \ref{fig:vZ-profiles} we conclude that $\beta_\| \approx 1$ and $\beta_T \approx 2$ for earlier times and $\beta_T \approx 1$ for later times. The dependence of the parameters of transverse and longitudinal expansions in Eq.~(\ref{V-Hubble}) on time and the collision energy is illustrated in Fig.~\ref{fig:Habble-param}.
	For the collision energy $\sqrt{s}=7.7$\,GeV,  the parameter $\alpha_T$ [see panel (a) in Fig.~\ref{fig:Habble-param}] increases with time between $t\simeq 5$ and 11\,fm/$c$ and the transverse flow propagates from lower $|z|$ to larger $|z|$ so that a common (weakly $z$ independent) transverse motion is formed for $|z|\lsim 3$\,fm at $t\simeq 10\mbox{--}11$\,fm/$c$. At later times the transverse flow starts decelerating.
	The coefficient of the longitudinal expansion, $\alpha_\|$, is shown in Fig.~\ref{fig:Habble-param}(b). It depends very weakly on $z$ and decreases with a time increase.
	
	For higher energy, $\sqrt{s}_{NN}=11.5$\,GeV, the picture is qualitatively similar, only the transverse flow parameter, $\alpha_T$, reaches slightly higher values and varies on a smaller timescale. The longitudinal flow parameter, $\alpha_\|$, is smaller than for the collision energy 7.7\,GeV.
	
	The situation is qualitatively different for lower collision energy, $\sqrt{s}_{NN}=4.5$\,GeV. The transverse flow parameter increases with time for much longer, between $t=5$\,fm/$c$ and $t\simeq 13$\,fm/$c$. The longitudinal flow parameter first increases reaching values higher than for
	$\sqrt{s}_{NN}=7.7$ and 11.5\,GeV at $t\simeq 9$\,fm/$c$, and then decreases for later times.
	One could expect that the drastic changes in the flow pattern between $\sqrt{s_{NN}}=4.5$\,GeV and 7.7\,GeV can manifest in the changes of particle flow pattern observed experimentally.
	
	There are two types of corrections to Eq.~(\ref{V-Hubble}):
	\begin{align}
		\vec{v} &= \vec{v}_{\rm H} + \delta \vec{v} +\delta \vec{v}_{\rm asym}.
		\label{Vtot}
	\end{align}
	One is axially symmetric but mixes $r_T$ and $z$ dependence of transverse and longitudinal components of velocities in (\ref{V-Hubble}):
	\begin{align}
		\delta \vec{v} 
		=\delta\alpha_{T}(r_T,z) \vec{e}_T + \delta\alpha_{\|}(r_T,z) \vec{e}_z.
		\label{dV-sym}
	\end{align}
	The other type of correction is responsible for violation of the axial symmetry. There is the term related to the shift of the center of the Hublle-like expansion: $-\alpha_T (x_0(z),0,0)$. We put $\beta_T=1$. Additionally, one can add a term responsible for an elliptic flow (the term linear in $x$ and $y$). So the symmetry breaking term can be written as
	$\delta \vec{v}_{\rm asym} = -\alpha_T (x_0(z),0,0)+\delta\alpha_{T,\rm as}(r_T,z)( x, - y,0)/2$. In terms of the unit vectors $\vec{e}_T$ and $\vec{e}_z$ we have
	\begin{align}
		\delta \vec{v}_{\rm asym}
		&= -\alpha_T x_0(z)\Big(\frac{x}{r_T}\vec{e}_T+\frac{y}{r_T}[\vec{e}_T\times \vec{e}_z]\Big)
		\nonumber\\
		&+ \delta\alpha_{T,\rm as}(r_T,z)\Big(
		\frac{x^2-y^2}{2r_T}\vec{e}_T
		\nonumber\\
		&+\frac{xy}{r_T}[\vec{e}_T\times \vec{e}_z]
		\Big).
		\label{dV-asym}
	\end{align}
	The hydrodynamic directed, $v_1^{\rm (hydro)}$, and elliptic, $v_2^{\rm (hydro)}$, flows can be expressed for such a parametrization as follows:
	\begin{align}
		v_1^{\rm (hydro)} &= \intop_0^{2\pi}\frac{\rmd \phi}{2\pi} (\vec{v})_x = -\alpha_T x_0(z),
		\nonumber\\
		v_2^{\rm (hydro)} &= \intop_0^{2\pi}\frac{\rmd \phi}{2\pi} \frac{(\vec{v})^2_x - (\vec{v})^2_y}{\sqrt{(\vec{v})^2_x + (\vec{v})^2_y}}
		\nonumber\\
		& \approx
		r_T \delta\alpha_{T,\rm as}(r_T,z)\Big(1+\frac{\delta\alpha_{T}}{\alpha_T\, r_T}\Big),\,\, x_0\ll r_T.
		\label{hydro-flow}
	\end{align}
	
	The analysis of the velocity fields generated with the fluidized PHSD model shows that on average the correction terms are numerically much smaller than the Hubble-like term:
	\begin{align}
		|\delta \vec{v} + \delta \vec{v}_{\rm asym}|\ll |\vec{v}_{\rm H}|.
	\end{align}
	
	Any vector field can be decomposed into irrotational and solenoidal components (the Helmholtz decomposition) and one can write the velocity field in terms of the scalar and vector potentials, $\vec{v}= \grad \phi -\rot \vec{\psi}$, where only two components of the vector potential are independent and they can be fixed by the gauge condition $\Div\vec{\psi}=0$. The potentials obey the Poisson equations $\Delta \phi=\Div\vec{v}$ and $\Delta \vec{\psi}=-\rot\vec{v}$. The quantity $\theta=\Div\vec{v}$ is called the dilation of the velocity field, which measures the isotropic expansion or compression of the fluid. The other quantity, which is of our primary interest, is the vorticity of the fluid,
	\begin{align}
		\vec{\om}=\rot \vec{v},
	\end{align}
	which measures the rotation of fluid particles. Thus, the vorticity defines the solenoidal part of the velocity field.
	
	The local variation of the velocity field in the vicinity of point $\vec{r}_0$ can be written for $\vec{r}\sim \vec{r}_0$ as
	\begin{align}
		v_i(\vec{r})=v_i(\vec{r}_0)+ (\vec{r}-\vec{r}_0)_j\partial_jv_i(\vec{r}_0)
		+ O(|\vec{r}-\vec{r}_0|^2)\,.
		\label{v-change}
	\end{align}
	The gradient tensor can be decomposed into symmetric and antisymmetric tensors as
	\begin{align}
		\partial_i v_j &=\xi_{+,ij}+\xi_{-,ij}\,,
		\nonumber\\
		\xi_{+,ij}&=\frac12(\partial_i v_j + \partial_{j}v_i)
		\,,\,\,
		\xi_{-,ij} =\frac12(\partial_i v_j - \partial_{j}v_i).
		\label{xi-decomp}
	\end{align}
	The symmetric one is the strain rate tensor which characterizes isotropic expansion as well as stretching and shearing deformations of the fluid. Its trace is the dilation scalar, $\xi_{+,ii}=\theta$. The antisymmetric one, which can be expressed through the vorticity vector $\xi_{-,ij}=\frac12\epsilon_{ijk}\om_k$, describes the rigid-body rotation of the fluid element. It indicates both the direction and rate of rotation of the fluid at a point. Finally, Eq.~(\ref{v-change}) can be written as
	\begin{align}
		\vec{v}(\vec{r})&\approx \vec{v}(\vec{r}_0)+{\textstyle\frac12}\grad D(\vec{r},\vec{r}_0)
		+{\textstyle\frac12} [\vec{\om}(\vec{r}_0)\times (\vec{r}-\vec{r}_0)],
	\end{align}
	where we introduce the deformation scalar $
	D(\vec{r},\vec{r}_0)=\sum_{ij}(\vec{r}-\vec{r}_0)_i \xi_{+,ij}(\vec{r}_0)
	(\vec{r}-\vec{r}_0)_j\,.$

	The Hubble-like flow (\ref{V-Hubble}) is irrotational, $\rot \vec{v}_{\rm H}\equiv 0$. Hence the vorticity is determined by the symmetry violating terms as, e.g., given in Eqs.~(\ref{dV-sym}) and (\ref{dV-asym}) above. Therefore only a small fraction of the velocity flow generated in heavy-ion collision possesses a nonvanishing vorticity:
	\begin{align}
		\vec{\om} &= \rot(\delta\vec{v} +  \delta \vec{v}_{\rm asym})
		\nonumber\\
		&=\Big(\frac{\partial \delta\alpha_{\|}}{\partial r_T} z
		-\frac{\partial\delta\alpha_{T}}{\partial z} r_T\Big)[\vec{e}_T\times \vec{e}_z ]
		\nonumber\\
		&-\frac{\partial\delta\alpha_{T,\rm as}}{\partial z}\frac{x^2-y^2}{2r_T}
		[\vec{e}_T\times \vec{e}_z ]
		\nonumber\\
		&+\frac{xy}{r_T}\Big(\frac{\partial\delta\alpha_{T,\rm as}}{\partial z}
		\vec{e}_T
		-\frac{\partial\delta\alpha_{T,\rm as}}{\partial r_T}\vec{e}_z
		\Big)
		\nonumber\\
		&-\alpha_T\frac{\partial x_0(z)}{\partial z}\Big(\frac{y}{r_T}\vec{e}_T
		-\frac{x}{r_T}[\vec{e}_T\times \vec{e}_z ]
		\Big).
		\label{vort-exp}
	\end{align}
	Here we put $\beta_T=\beta_\|=1$ in Eq.~(\ref{V-Hubble}).
	We see that the axially symmetric part (\ref{dV-sym}) produces vorticity directed only in the azimuthal direction (terms $\propto [\vec{e}_T\times \vec{e}_z]$). The asymmetric part (\ref{dV-asym}) induces the dependence of the azimuthal component of vorticity on the azimuthal angle and the transverse and longitudinal components of vorticity.
	

	\section{Vorticity field}\label{sec:vorticity}
	
	\subsection{Vorticity, Lamb vector, and helicity}

	As demonstrated in Refs.~\cite{Becattini-Tinti2010,Becattini-Chandra2013,Fang-Pang-Wang2016} within a statistical approach, the particle with mass $m$, spin $s$, and four-momentum $p^\mu$ acquires in the presence of the thermal vorticity $\varpi_{\mu\nu} = \frac{1}{2} \big(\partial_{\nu} (u_{\mu}/T) - \partial_{\mu} (u_{\nu}/T)\big)$  an average spin polarization characterized by the spin four-vector
	\begin{align}
		\label{eq:becattini:S-def}
		S^{\mu}(x,p)=-\textstyle{\frac16} s\,(s+1) \varepsilon^{\mu\nu\lambda\delta} \varpi_{\nu\lambda} p_\delta/m\,.
	\end{align}
	If we neglect the gradient of the temperature, which has weaker spatial and time dependence than energy density and velocity (see
	Figs.~\ref{fig:profilesXZ-7.7}, \ref{fig:profilesXZ-4.5}, and \ref{fig:profilesXZ-11.5}),
	the thermal vorticity can be expressed as $\varpi_{\mu\nu}\approx \om_{\mu\nu}/2T$ through the kinematic vorticity tensor $\omega_{\mu \nu} = (\partial_{\nu} u_{\mu} - \partial_{\mu} u_{\nu})$. The latter tensor provides a natural relativistic generalization\footnote{It should be mentioned that neither thermal nor kinematic vorticities enjoy the conservation properties as in, e.g., the Helmholtz-Kelvin theorem even for an ideal barotropic fluid. Alternative definitions of relativistic vorticities are discussed in Ref.~\cite{Deng-Huang-PRC93}. } for the nonrelativistic vorticity $\vec{\om}$:
	\begin{align}
		\bar{\omega}^{\mu} = \epsilon^{\mu \nu \rho \sigma} u_{\nu} \partial_{\rho} u_{\sigma}
		= - \frac{1}{2} \epsilon^{\mu \nu \rho \sigma} u_{\nu} \omega_{\rho \sigma},
	\end{align}
	or, inversely,
	\begin{align}
		\om^{\mu\nu}=-\epsilon^{\mu\nu\rho\sigma}u_\rho \bar{\om}_\sigma\,.
		\label{om-munu-om}
	\end{align}
	The components of the vorticity vector are
	\begin{align}
		\label{eq:vorticity:rel-vector-parts}
		\bar{\omega}^\mu = \gamma^{2} \big((\vec{v}\vec{\om}),\vec{\omega} + [\vec{v} \times \partial_{t} \vec{v}]\big).
	\end{align}
	Thus, in the nonrelativistic limit we have  $\bar{\omega}^\mu \approx \big((\vec{v}\vec{\om}),\vec{\om}\big)$, where the relativistic subleading term is a pseudoscalar,
	\begin{align}
		h=(\vec{\om}\vec{v}),
		\label{h-def}
	\end{align}
	called the helicity density of the flow~\cite{Moreau-61,Moffat-h-def}.
	The kinetic vorticity tensor $\om_{\mu\nu}$ contains also information about the acceleration of the fluid:
	\begin{align}
		A^\mu &=\om^{\mu\nu}u_\nu= u_\nu\partial^{\nu} u^{\mu}= \gamma(\partial_t -  (\vec{v}\vec{\nabla})) u^\mu\,.
	\end{align}
	
	Since the bulk of the fluid in the fireball moves with velocities smaller than 0.5$c$--0.6\,$c$ (see Figs.~\ref{fig:vZ-profiles} and \ref{fig:Habble-param}), we will consider the nonrelativistic hydrodynamics.
	In the nonrelativistic limit we have $A^\mu=(0,\vec{a})$, where the acceleration $\vec{a}=\partial_t \vec{v} +(\vec{v} \nabla)\vec{v}$ can be written with the help of Eq.~(\ref{eq:hydro:id-1}) as follows:
	\begin{align}
		\vec{a}=\frac{\partial \vec{v}}{\partial t} +\vec{\lambda}_\om +
		\frac12\grad \vec{v}^2,
		\label{a-def}
	\end{align}
	where
	\begin{align}
		\vec{\lambda}_\om=[\vec{\om}\times \vec{v}]
		\label{lamb-def}
	\end{align}
	is the Lamb vector, also known as the vortex force  transverse to the fluid motion. It is a measure of the Coriolis acceleration of a velocity field under the effect of its own rotation.
	
	Substituting Eq.~(\ref{om-munu-om}) in Eq.~(\ref{eq:becattini:S-def}) and neglecting the temperature gradients we find
	\begin{align}
		S^\mu&\approx\frac{s(s+1)}{6mT}\big(\bar{\om}^\mu (u\cdot p) -u^\mu (\bar{\om}\cdot p) \big)
		\nonumber\\
		&\approx \frac{s(s+1)}{6mT}\big((\vec{\om}\vec{p}), E\vec{\om}-[\vec{p}\times \vec{\lambda}_\om]\big)+O(v^2)\,.
	\end{align}
	In the rest frame of the particle, which is used for experimental identification of the fermion polarization, this four-vector becomes $S^{*\mu}=(0, \vec{S}^*)$, where in the  nonrelativistic limit of the fermion ($\vec{p}\ll m$)  we have~\cite{KTV-PRC97}
	\begin{align}
		\vec{S}^* & \approx \vec{S}-\frac{\vec{p}}{2m}S^0
		\approx\frac{s(s+1)}{6}\Big(\frac{\vec{\om}}{T} - \big[\frac{\vec{p}}{m}\times \frac{\vec{\lambda}_\om}{T}\big]
		\Big)
		\nonumber\\
		&+O(\vec{v}^2,\vec{p}^2/m^2) \,.
	\end{align}
	We see that the Lamb vector is responsible for coupling of the particle velocity with the polarization. The helicity provides only relativistic correction to the polarization (\ref{eq:becattini:S-def}).
	
	The helicity density and the Lamb vector are of crucial importance in vorticity dynamics. For example, they determine the decomposition of the velocity in two orthogonal components,
	\begin{align}
		\vec{\om}^2\vec{v} &= h\,\vec{\om} - [\vec{\om}\times \vec{\lambda}_\om].
	\end{align}
	Multiplying this relation by $\vec{v}$ we find $\vec{\om}^2\vec{v}^2 =h^2+ \vec{\lambda}_\om^2$. Hence, if a local $\vec{\om}$ and $\vec{v}$ are very slowly varying then an increase of the angle between $\vec{\om}$ and $\vec{v}$ implies an increase of the Lamb vector and decrease of the helicity density. The Lamb vector is maximized for a flow confined to the plane with $(\vec{\om}\cdot \vec{v})=0$, implying a maximum local transverse force for fixed $\vec{u}^2$ and $\vec{\om}$; see Eq.~(\ref{a-def}). In the alternative situation, if the Lamb vector vanishes, $\vec{\lambda}_\om=0$ but $h\neq 0$, we deal with the situation when the vorticity is parallel to the flow velocity, $\vec{\om}=h\vec{v}$. This means that $\vec{v}$ is an eigenvector of the $\rot$ operator with eigenvalue $h$, $\rot \vec{v} =h \vec{v}$. This type of flow is called the Beltrami flow or helical flow. It is a stationary flow with finite extensive helicity, i.e., scaling with the volume of the system. The Beltrami flows play an important role in the study of turbulent and chaotic flows in hydrodynamics. Reference \cite{Pelz-Yakhot} suggested that, in various regions of space, turbulent flows organize into a coherent hierarchy of weakly interacting superimposed approximate Beltrami flows; see also Refs.~\cite{Constantin-Majda}. The properties of the Beltrami flow were investigated, e.g., in Refs.~\cite{Changchun,Dritschel,Fre-Sorin}.
	
	\begin{figure*} 
		\centering
		\includegraphics[width=0.99\textwidth]{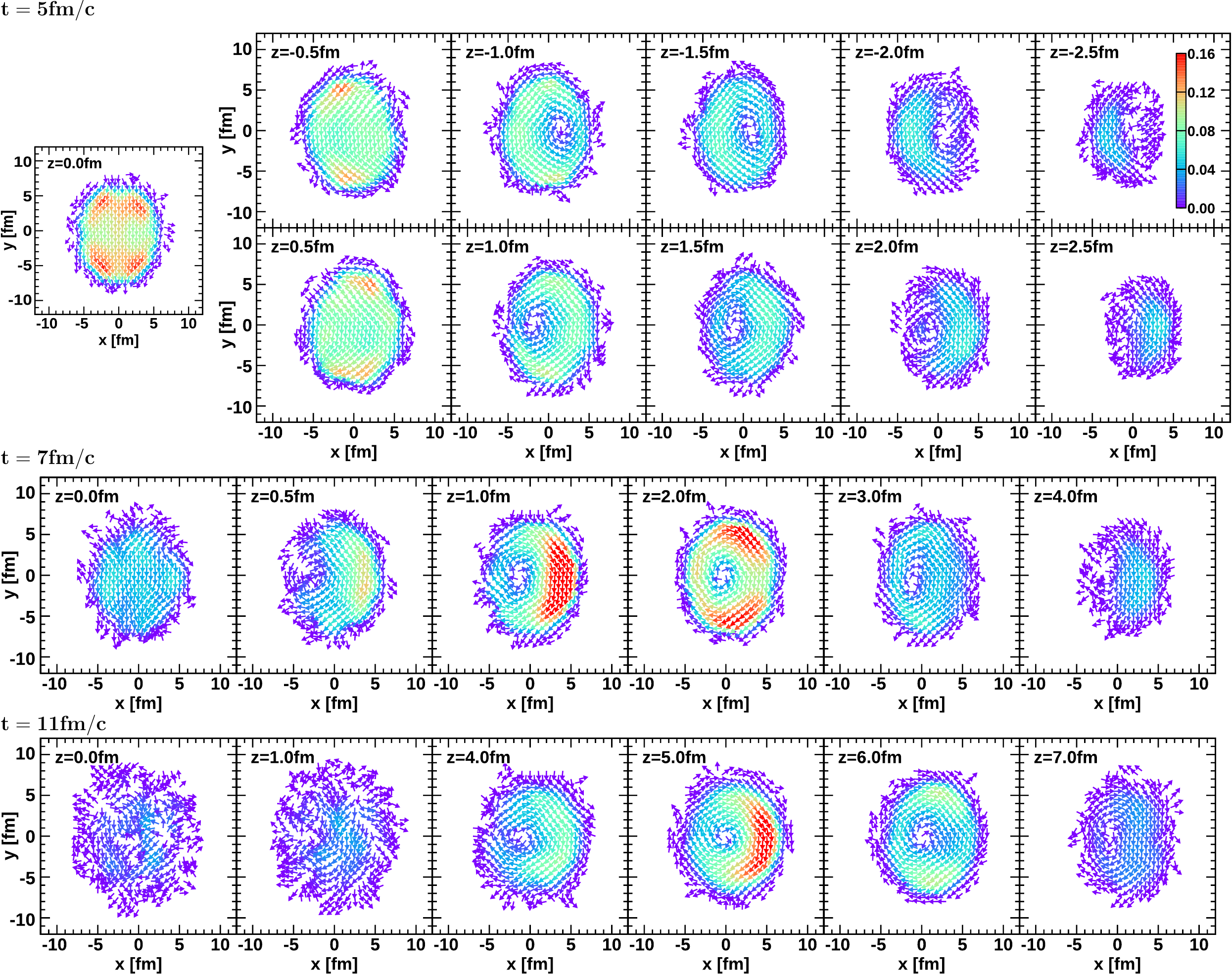}
		\caption{Vorticity field ($\om_x,\om_y$) in the $x$-$y$ plane at various $z$ slices for Au+Au collisions at $\sqrt{s_{NN}}=7.7$\,GeV with the impact parameter $b=7.5$\,fm at times $t=5$, 7, and $11\,{\rm fm}/c$. All arrows are of equal length. The magnitude of the vorticity is indicated by the colored scale in units $c$/fm. For $t=5$\,fm/$c$ (the moment of the maximum overlap of colliding nuclei) we show both positive and negative $z$ to visualized the symmetry of the vorticity field. For later times only $z\ge 0$ are plotted. }
		\label{fig:W-Au77-b75}
	\end{figure*}

	The concept of the integral helicity of a fluid volume $V$
	\begin{align}
		H=\intop_{V} (\vec{\om}\cdot \vec{v})\rmd^3 x
		\label{H-int}
	\end{align}
	has gained interest since 1961, when Moreau showed in Ref.~\cite{Moreau-61} that helicity is an invariant of Euler equations of ideal fluid motion. A similar conserving quantity was found also in magnetohydrodynamics~\cite{Woltjer-58}, where the role of vorticity is played by magnetic field $\vec{B}$ and velocity is replaced by the corresponding vector potential $\vec{A}$, $\vec{B}=\rot\vec{A}$. In contrast to other conservative quantities, like momentum and energy, helicity does not correspond to any space–time symmetry. Rather, as shown in Ref.~\cite{Moffat-h-def}, it is related to the topology of the flow. This quantity measures the state of ``knottedness'' of vortex filaments.

	The Lamb vector characterizes not only the essential nonlinearity of the convective fluid acceleration, Eq.~(\ref{a-def}), in the hydrodynamic equations, but also was found to be instrumental in the analysis of coherent motion in fluids (e.g., large stable vortical structures as the Great Red Spot of Jupiter)~\cite{Hammam}.
	
	In the simplest case of an incompressible fluid, the hydrodynamic equation describing the evolution of velocity field is (see the Appendix~\ref{app:vorticity-eq})
	\begin{align}
		\frac{\partial \vec{v}}{\partial t} & + \vec{\lambda}_\om  = -\vec{\nabla}\Big(\frac{p}{\rho} + \frac{\vec{v}^2}{2}\Big)+\nu \nabla^2 \vec{v}.
		\label{NS-eq-text}
	\end{align}
	Here $p(\vec{r},t)$ is the pressure, $\nu$ is the kinematic shear viscosity (\ref{kin-viscos}), and $\rho$ stands for the mass density, $\rho=\epsilon/c^2$.
	whereby $\nu$ and $\rho$ are constant. Then, the continuity equation (\ref{eq:hydro:euler}) implies $\Div\vec{v}=0$.
	Applying the divergence operator to Eq.~(\ref{NS-eq}), we obtain that the divergency of the Lamb vector is the source term in a Poisson equation for the Bernoulli function $\Psi=\frac{p}{\rho} + \frac{\vec{v}^2}{2}$:
	\begin{align}
		\Div\vec{\lambda}_\om = -\nabla^2 \Psi.
		\label{divL}
	\end{align}
	In derivation we use that $\Div \vec{v}=0$.
	The divergency of the Lamb vector is called the hydrodynamic charge~\cite{Rousseaux}, $q_H = \Div\vec{\lambda}_\om$.
	From Eq.~(\ref{divL}) we see that regions with $q_H>0$ correspond to regions where $\Psi$ is concentrated, while in regions with $q_H<0$ function $\Psi$ is depleted~\cite{Hammam}.
	In Refs.~\cite{Marmanis,Sridhar} it was shown that one can construct the hydrodynamic analog of the Maxwell equations, where the hydrodynamic charge plays a role of the electric charge.
	
	Using the Helmholtz theorem the Lamb vector can be decomposed as  $\vec{\lambda}_\om=\vec{\lambda}_{\om,\perp} + \vec{\lambda}_{\om,\|}$ in irrorational, $\vec{\lambda}_{\om,\perp}$, and solenoidal, $\vec{\lambda}_{\om,\|}$ parts, with $\rot\vec{\lambda}_{\om,\perp} =0 =\Div\vec{\lambda}_{\om,\|}$. From Eqs.~(\ref{divL}) and (\ref{NS-eq-text}) we find
	\begin{align}
		\vec{\lambda}_{\om,\perp}=-\nabla \Psi, \quad \vec{\lambda}_{\om,\|}=-\frac{\partial\vec{v}}{\partial t} -\nu \nabla^2\vec{v}\,.
	\end{align}
	This decomposition shows that, in the particular case of a stationary flow and when the viscous effects are negligible, i.e., $\vec{\lambda}_{\om,\|}=0$, the Lamb vector
	$\vec{\lambda}_{\om}= -\nabla \Psi$ constitutes the directional normal to the surface of constant $\Psi$, which is called the Lamb surface and is formed by streamlines and vortex lines at in each point velocity and vorticity are orthogonal to this normal, $(\vec{v} \nabla \Psi) = (\vec{v}\,\vec{\lambda}_\om)\equiv 0$ and $(\vec{v} \nabla \Psi) = (\vec{v}\,\vec{\lambda}_\om)\equiv 0$. These properties of the Lamb vector allowed Ref.~\cite{Rousseaux} to use the Lamb vector and the hydrodynamic charge to locate and characterize coherent structures such as vortices in experimental data.

	Applying the circulation operation to Eq.~(\ref{NS-eq-text}) and taking into account that
	for incompressible or weakly compressible flow when the entropy gradient can be neglected and the fluid is nearly barotropic\footnote{For the barotropic equation of state, the pressure is a function of only density $\epsilon$ or matter density $\rho=\epsilon/c^2$. In general pressure as well es the entropy are functions of both density and temperature, $p=p(\rho,T)$ and $s=s(\rho, T)$. The constancy of the entropy implies the connection of the temperature and density.
		\label{foot:barotropic}} we obtain the Helmholtz equation for vorticity
	\begin{align}
		\frac{\partial \vec{\om}}{\partial t} + [\vec{\nabla} \times \vec{\lambda}_\om]= \nu \Delta \vec{\om}.
		\label{HH-eq}
	\end{align}
	Here the solenoidal part of the Lamb vector is responsible for the nonlinear coupling of the vorticity with the velocity field in the system. It describes the torque exerted by the Coriolis force. The viscous term on the right-hand side is responsible for the decay of the vorticity due to diffusion. Note that in the approximation of the incompressible fluid the Helmholtz equation has the trivial solution $\vec{\om}=0$, since in (\ref{HH-eq}) there is no term responsible for the generation of vorticity.
	Taking into account the density gradients (but still assuming that $\nu$ is constant) we obtain the extended vorticity equation [see Eq.~(\ref{NS-eq-rot})]
	\begin{align}
		\frac{\partial \vec{\om}}{\partial t} + [\vec{\nabla} \times \vec{\lambda}_\om] =  \frac{1}{\rho^2} [\vec{\nabla} \rho \times \vec{\nabla} p] + \nu \Delta \vec{\om},
		\label{Biermann}
	\end{align}
	where the first term on the right-hand side is the vorticity source term. This term is called the Biermann battery following Ref.~\cite{Biermann}, where a similar term was considered as a source of magnetics field in stars.
	The equation for the vorticity becomes more involved if the density dependence of the kinematic vorticity is taken into account; see the derivation in the Appendix~\ref{app:vorticity-eq}. We cast it here in the compact form
	\begin{align}
		&\Big(\frac{\partial \vec{\om}}{\partial t} + [\vec{\nabla} \times \vec{\lambda}_\om] -\frac{1}{\rho^2} [\vec{\nabla} \rho \times \vec{\nabla} p] - \nu \Delta \vec{\om}\Big)_i
		\nonumber\\
		&\qquad = -\nu\Gamma^{[\nu\rho]}_{ij}\om_j - \nu D^{[\nu\rho]}_{ijk}\nabla_j \om_k +S^{[\nu\rho]}_{i}.
	\end{align}
	We see that there appear new structures that control evolution of vorticity in time: the ``width'' term, $\Gamma^{[\nu\rho]}_{ij}$ given in Eq.~(\ref{app:Gamma}); and in space: the diffusion tensor, $D^{[\nu\rho]}_{ijk}$, given in Eq.~(\ref{app:D}), and the new source terms, $\vec{S}^{[\nu\rho]}$ given in Eq.~(\ref{app:S}).

	\subsection{Vorticity in heavy-ion collisions}
	
	\begin{figure}
		\parbox{4.2cm}{\includegraphics[width=4.2cm]{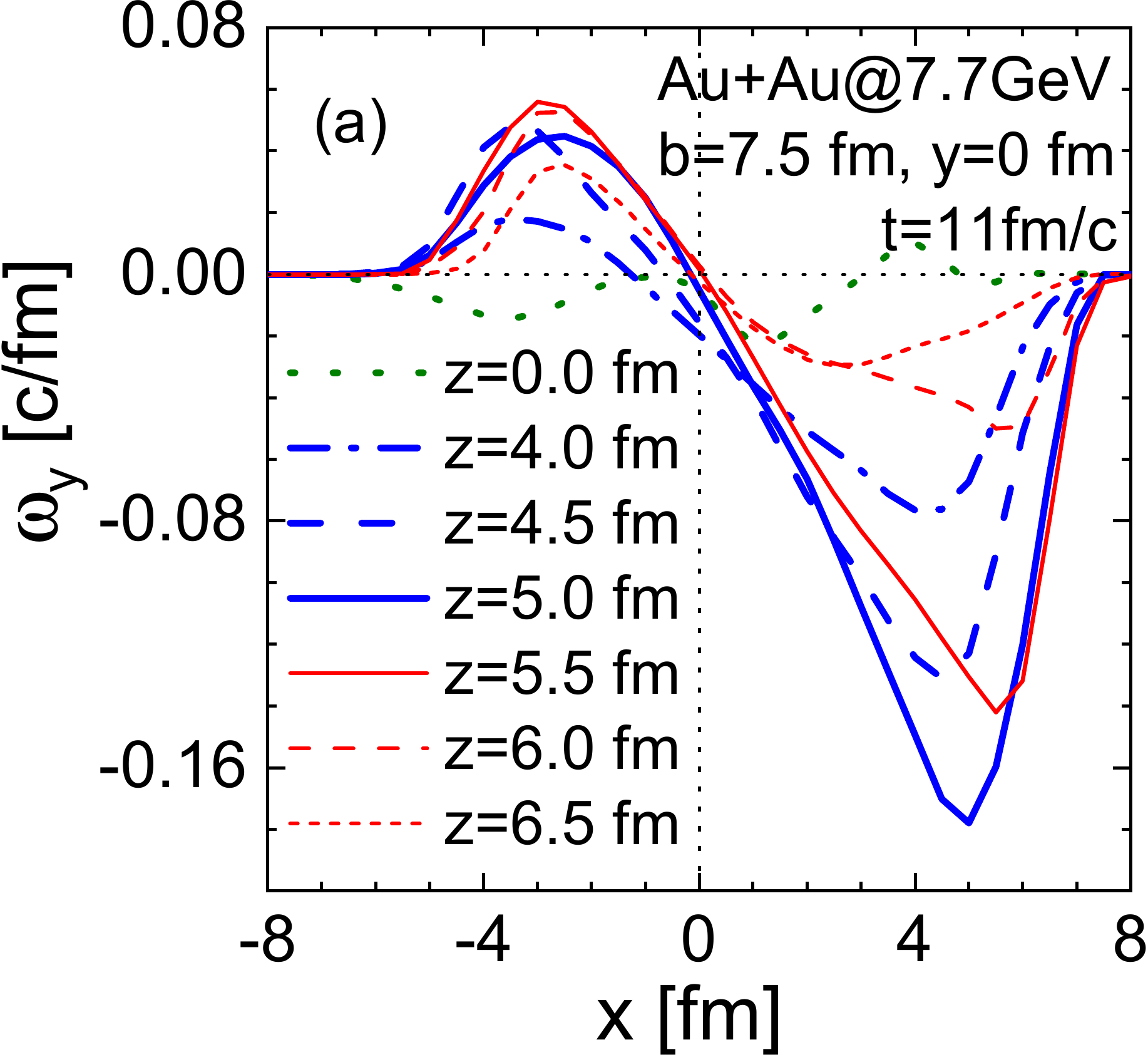}}
		\parbox{4.2cm}{\includegraphics[width=4.2cm]{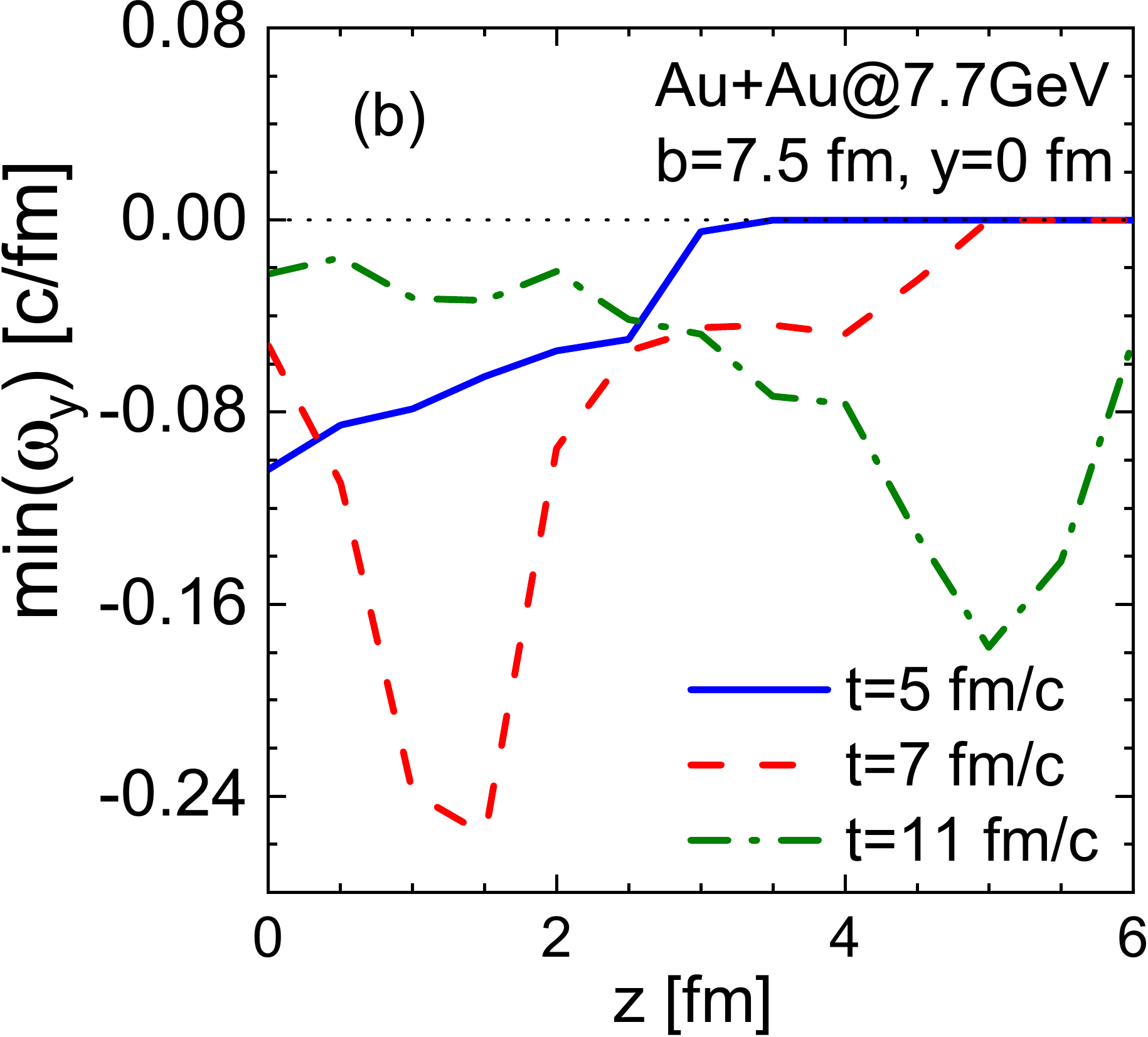}}\\
		\parbox{4.2cm}{\includegraphics[width=4.2cm]{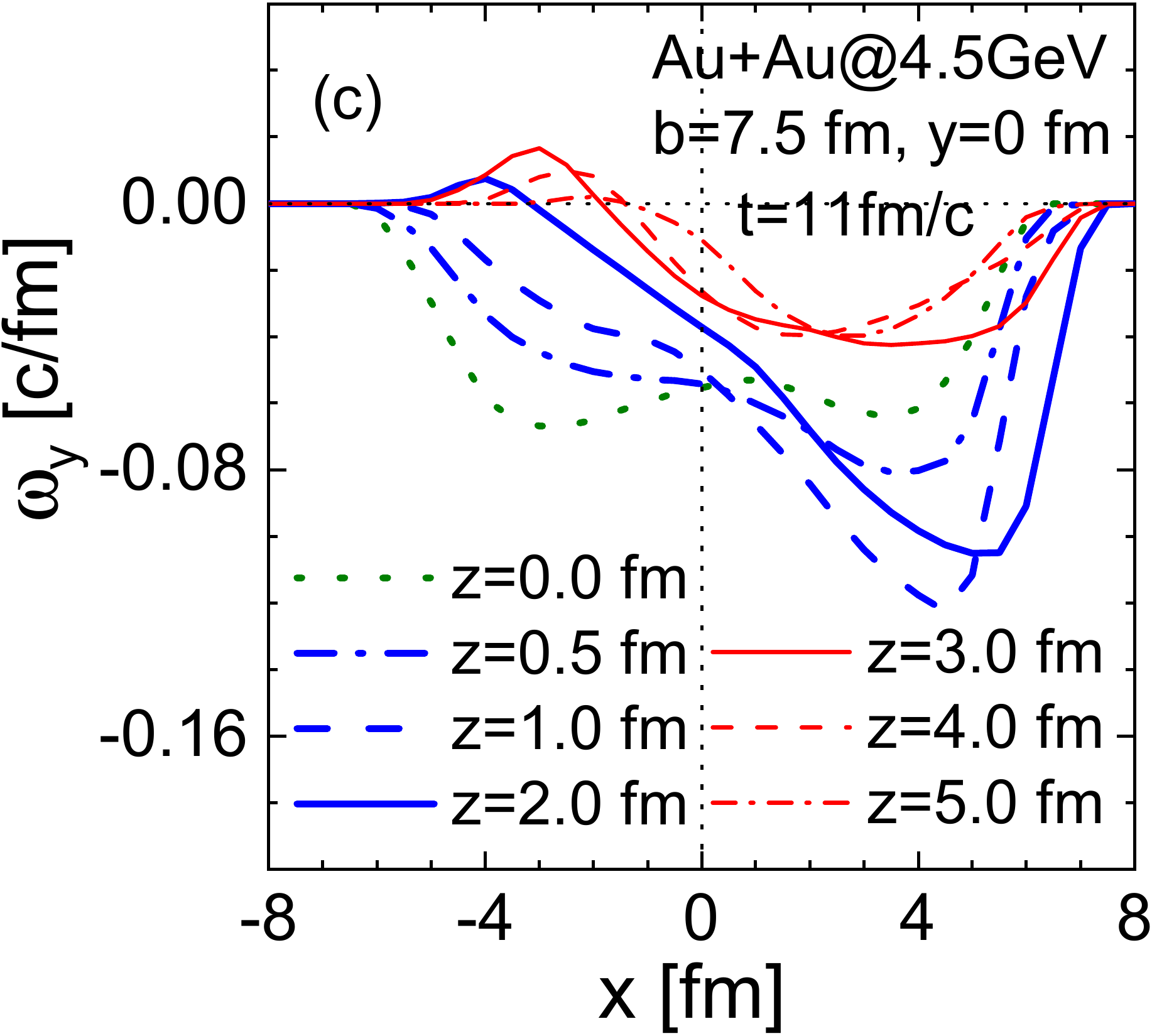}}
		\parbox{4.2cm}{\includegraphics[width=4.2cm]{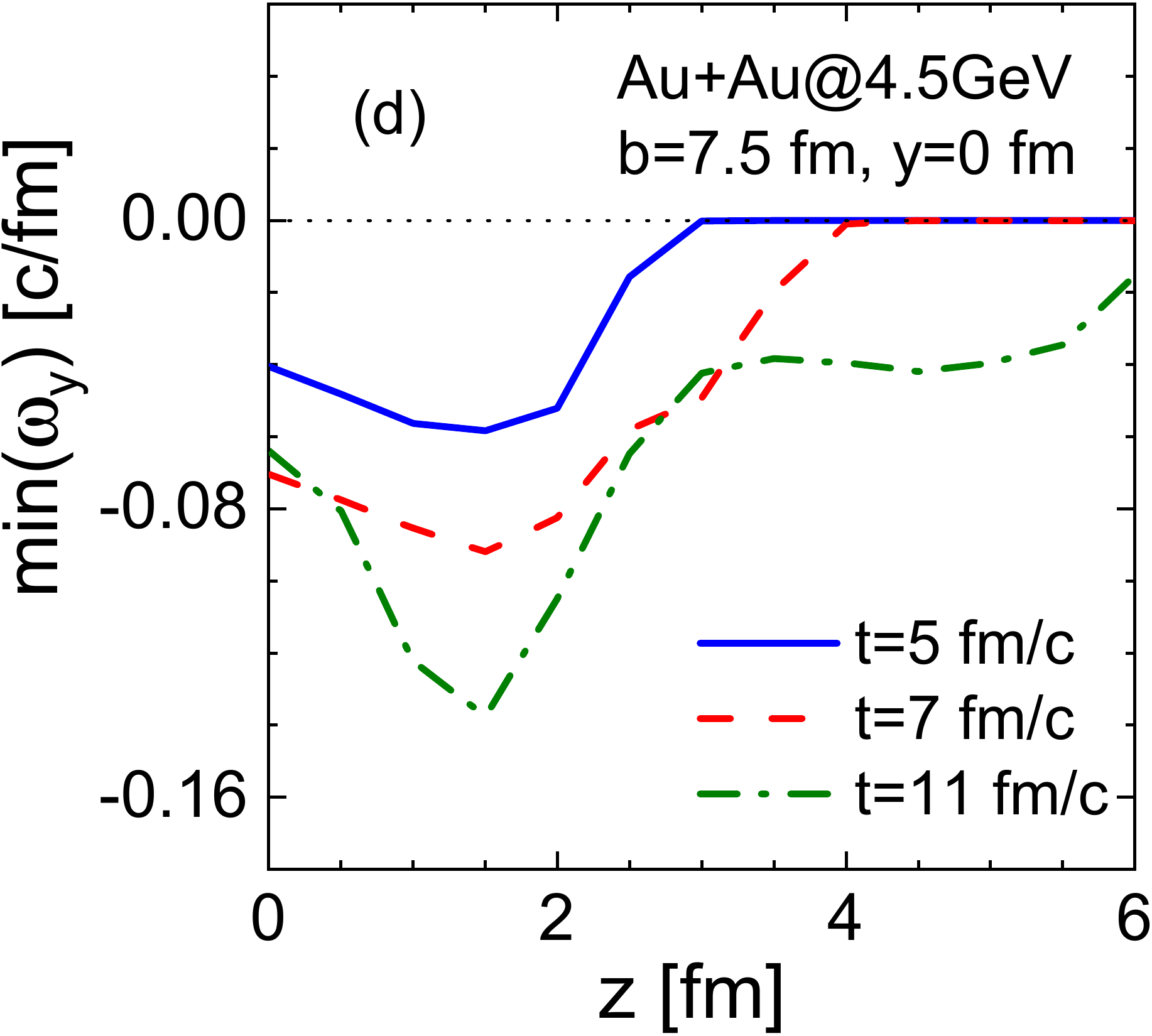}}\\
		\parbox{4.2cm}{\includegraphics[width=4.2cm]{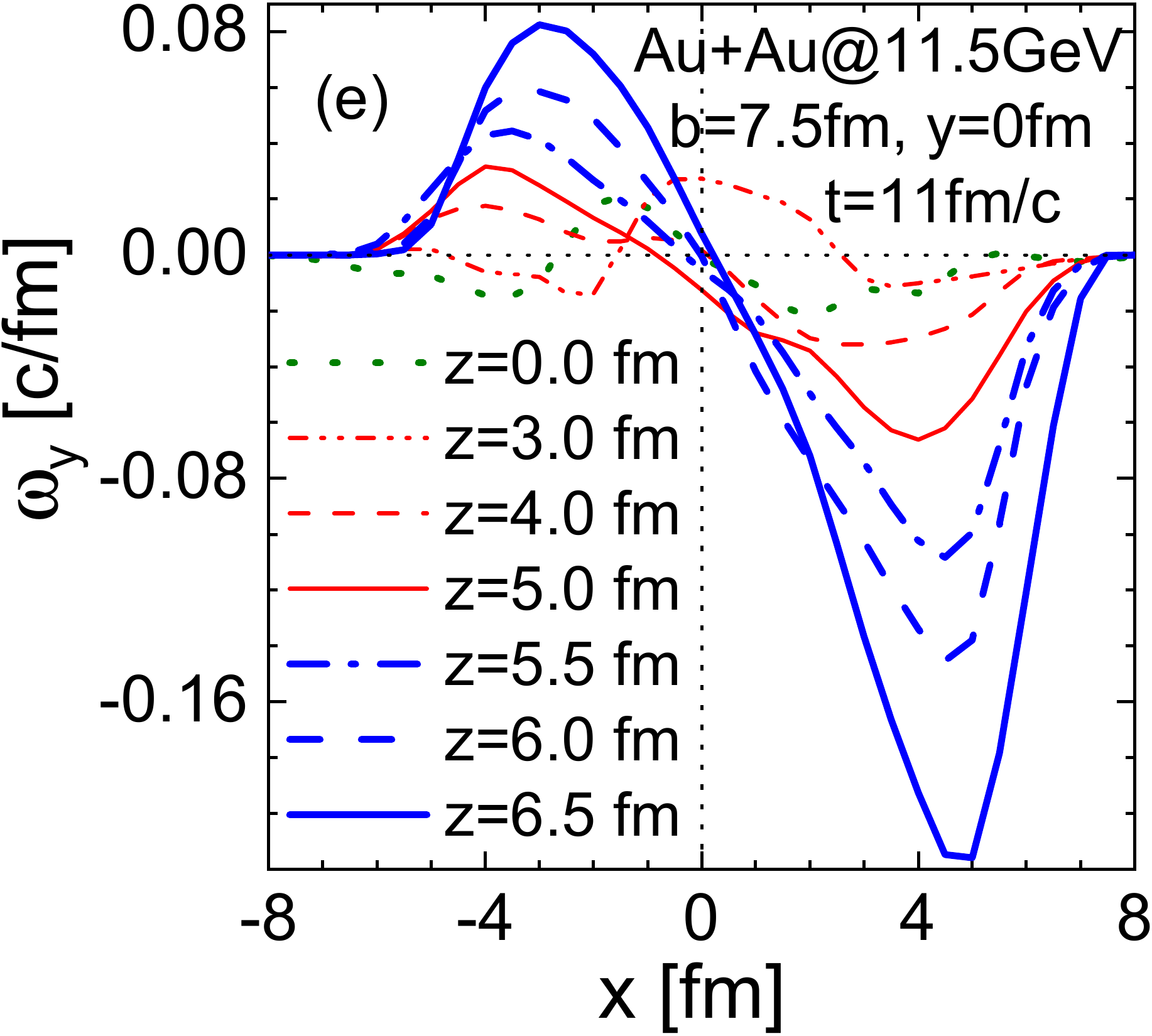}}
		\parbox{4.2cm}{\includegraphics[width=4.2cm]{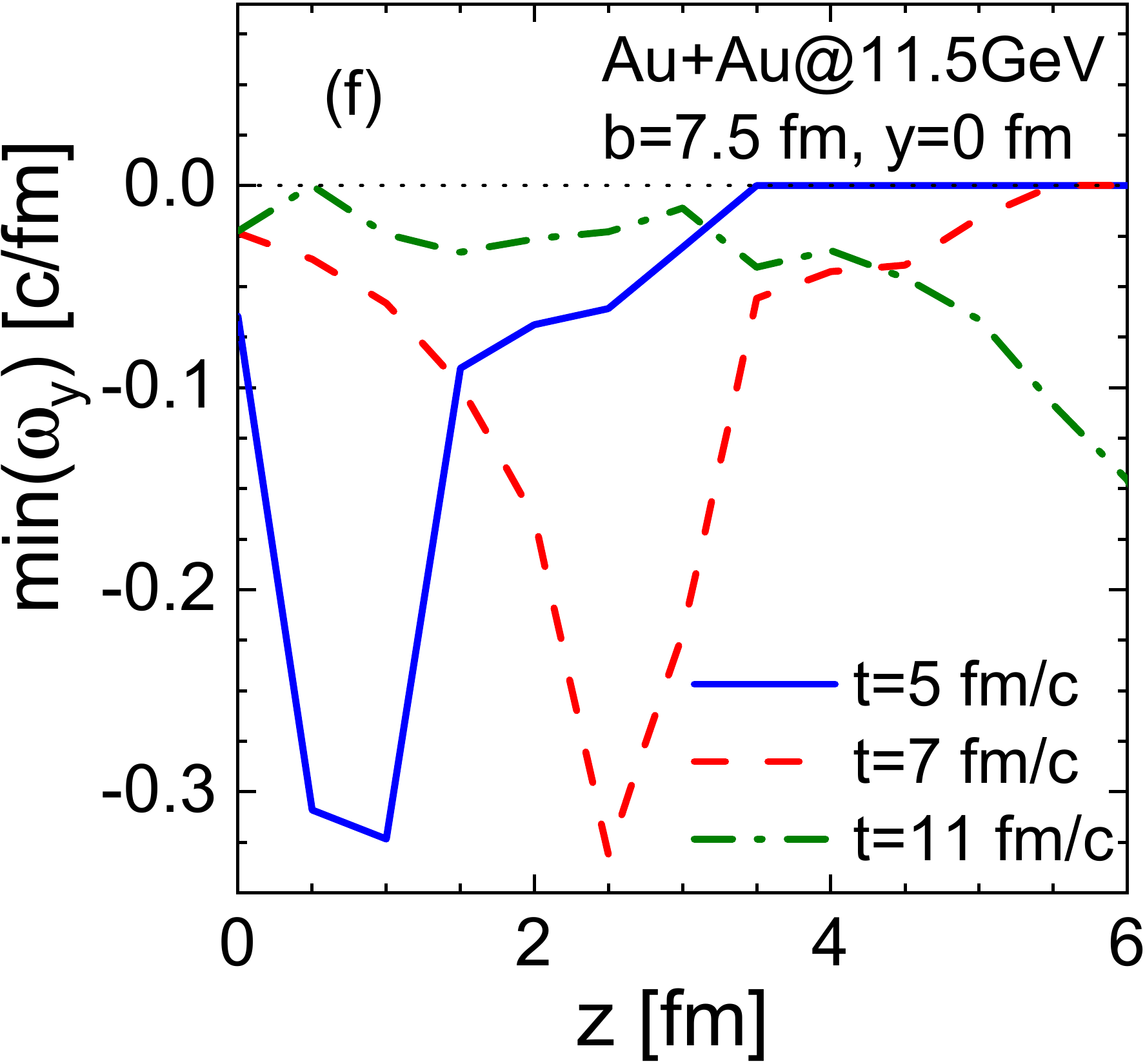}}\\
		\caption{(a,c,e) The $\om_y$ component of the vorticity field as function of $x$ at $y=0$ for various values of $z$.
		(b,d,f) Position of the minimum of the $\om_y$ component as a function of $z$ for various times.
			Calculations are done for various collision energies: $\sqrt{s_{NN}}=7.7$\,GeV shown on panels (a) and (b), $\sqrt{s_{NN}}=4.5$\,GeV shown on panels (c) and (d), and $\sqrt{s_{NN}}=11.5$\,GeV shown on panels (e) and (f). On all plots we apply the cut $\epsilon>0.05\,{\rm GeV/fm^3}$.}
		\label{fig:bagel-cut}
	\end{figure}

	\begin{figure}
		\centering
		\includegraphics[width=0.49\textwidth]{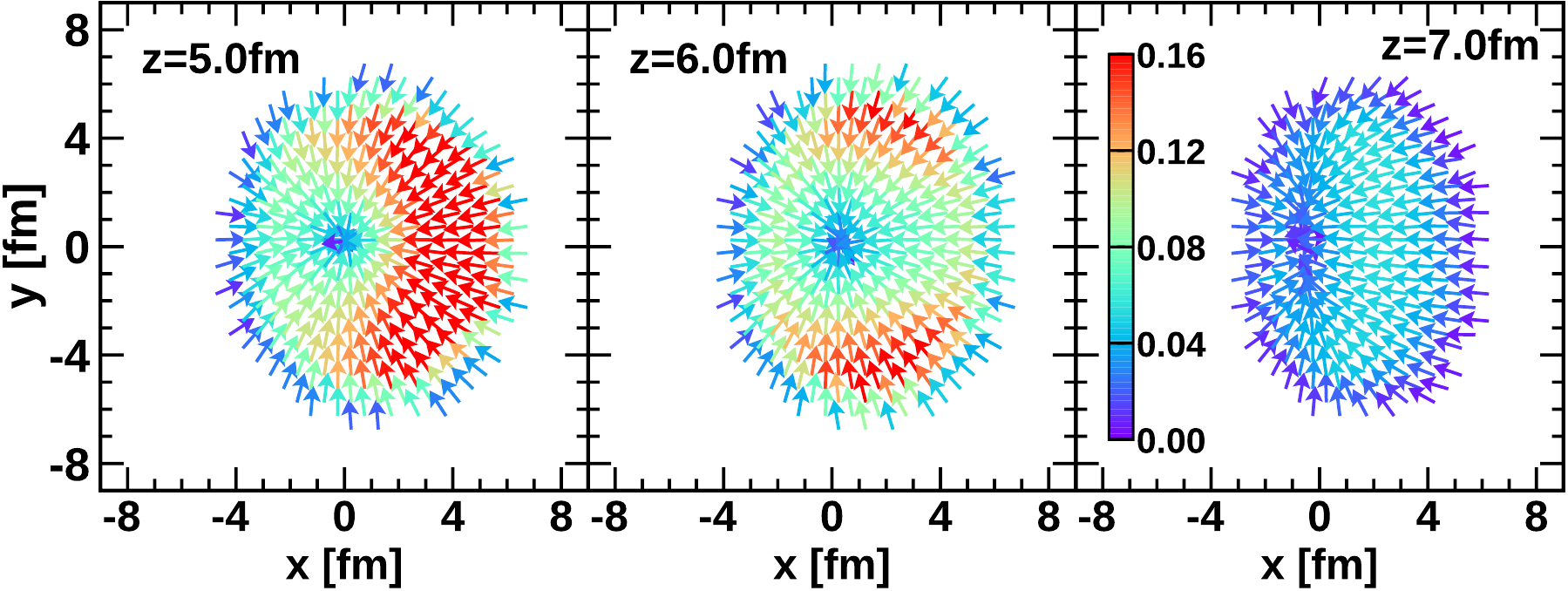}
		\caption{Lamb vector field ($\lambda_{\om_{x}},\lambda_{\om_{y}}$) in the $x$-$y$ plane at various $z$ slices for Au+Au collisions at $\sqrt{s_{NN}}=7.7$\,GeV with the impact parameter $b=7.5$\,fm at time $t=11\,{\rm fm}/c$. The arrows are of equal length. The color scale is given in units $c^2$/fm.}
		\label{fig:Lamb-Au77-b75}
	\end{figure}

	In this section we illustrate the vorticity field created in the heavy-ion collision as modeled by the PHSD transport code. We start with Au+Au collisions at  $\sqrt{s_{NN}}=7.7$\,GeV and the impact parameter $b=7.5$. In Fig.~\ref{fig:W-Au77-b75} we show ($\om_x,\om_y$) components of the vorticity field in the $x$-$y$ plane for various $z$ slices and at various moments of time. The arrows show the local direction of the vorticity field and magnitude is given by color code.
	For time  $t=5$\,fm/$c$ we show slices for both positive and negative values of $z$ to illustrate the symmetry the vorticity field under simultaneous replacements $z\to -z$, $x\to -x$ and $\om_x\to-\om_x$. Taking this symmetry into account, we will show only the positive-$z$ slices on other plots of the vorticity field.
	In Fig.~\ref{fig:W-Au77-b75} we see that at the moment of the maximal overlap of nuclei ($t=5$\,fm/$c$) the maximum vorticity lies in the central slice $z=0$ and is almost homogeneously oriented in the $-y$ direction. The sign is related to the initial relative position of colliding nuclei with respect to impact parameter. With an increase of $z$ the vorticity field becomes weaker (smaller magnitude) and the field orientation gets deformed, building a hole at small $x$ and $y$. At $z \approx 1\mbox{--}1.5$\,fm the vorticity field takes the form of a deformed bublik. For larger $z$ this structure fades away.
	As time passes (see the frame strip for 7\,fm/$c$), the maximum of the vorticity shifts from slice $z=0$ to slices with $z=1$ and 2\,fm where two ringlike structures are distinctly seen. At $t=11$\,fm/$c$ the structure has moved to $z=5\mbox{--}6$\,fm. At the same time the vorticity field in the center at $|z|\lsim 1$\,fm becomes very weak and disoriented. Note that the same bublik exists symmetrically at negative $z$ and propagates in the $-z$ direction. The vorticity field for $z\gsim 1$\,fm is oriented mainly clockwise and for $z\lsim -1$\,fm counterclockwise.  Comparing the obtained results with Fig.~\ref{fig:ENT-contour-Au77-b75}, it can be seen that the vorticity field has an ordered structure around hot matter and reaches the largest value at the outer boundary of the hot clusters.
	
	In the outer layers the vorticity fields become weak (lilac arrows) and disordered. We obtain this behavior because in calculations of vorticities on the fireball border we formally take into account also the cells where the energy density is below the imposed minimal value $\epsilon_c=0.05$\,GeV/fm$^3$. The collected collision statistics is high enough to obtain smooth nonfluctuating values of fluid velocities. As the result we have vanishing vorticity on the fluid boundary $\vec{\om}|_{S}=0$. The cut in $\epsilon$ is applied after velocity gradients are evaluated and only cells with $\epsilon>\epsilon_c$ are shown. If we would first artificially cut away the low-energy-density cells we would obtain large velocity gradients on the boundary, which is defined by the cutoff condition. In this case the boundary condition would be $(\vec{\om}\,\vec{n}_{S})=0$, where $\vec{n}_S$ is the local normal vector to the boundary. Such an enhanced vorticity field on the boundary will look like a vortex sheet/blanket around the fireball as seen in Refs.~\cite{BGST-Hseparation,BGST-Vsheet}.
	
	Thus, we may conclude that within the PHSD code calculations we observe the formation of two (deformed) vortex rings in heavy-ion collisions at $\sqrt{s_{NN}}=7.7$\,GeV. Similarly, two vortex rings were predicted Ref.~\cite{Ivanov-Soldatov-PRC97} for collisions at higher energy, $\sqrt{s_{NN}}=39$\,GeV.
	
	To illustrate the structure of the vortex rings seen in the vorticity distribution in Fig.~\ref{fig:W-Au77-b75}, we show in panel (a) of Fig.~\ref{fig:bagel-cut} the $\om_y$ component of vorticity as a function of $x$ for $y=0$ and various $z$ slices. The results are shown for $t= 11$\,fm/$c$. At this time the significant vorticity is located at $z=5\pm 1$\,fm, as seen in Fig.~\ref{fig:bagel-cut}(a), with a maximum at $z\simeq 5$\,fm and $x\simeq 5$\,fm. The asymmetry of the ring thickness is clearly visible.
	In panel (b) of Fig.~\ref{fig:bagel-cut} we show the minimum of $\om_y$ as function of $z$ for various times. One can see how the vortex ring propagates in the $z$ direction: for $t=5$\,fm/$c$ the ring is at $z \approx 0$\,fm, for $t=7$\,fm/$c$ at $z\simeq 1.5$\,fm, and for $t=11$\,fm/$c$ at $z\simeq 5$\,fm. The ring thickness in the $z$ direction increases with time.
	The same quantities calculated for collisions at lower energy $\sqrt{s_{NN}}=4.5$\,GeV are presented in panels (c) and (d) in Fig.~\ref{fig:bagel-cut}. We see that the ring structure is more diffuse and its evolution in the $z$ direction looks more like diffusion of vorticity to higher $z$ while the peak at $z\simeq 1$ remains but decreases in height. The peak hight at 4.5\,GeV is smaller than that at the collision energy 7.7\,GeV. For higher collision energies, see panels (e) and (f) calculated for $\sqrt{s_{NN}}=11.5$\,GeV, the ring is more pronounced and narrow and move moves fast in the $z$ directions, keeping its narrow structure.

	\begin{figure*}
		\centering
		\includegraphics[width=0.99\textwidth]{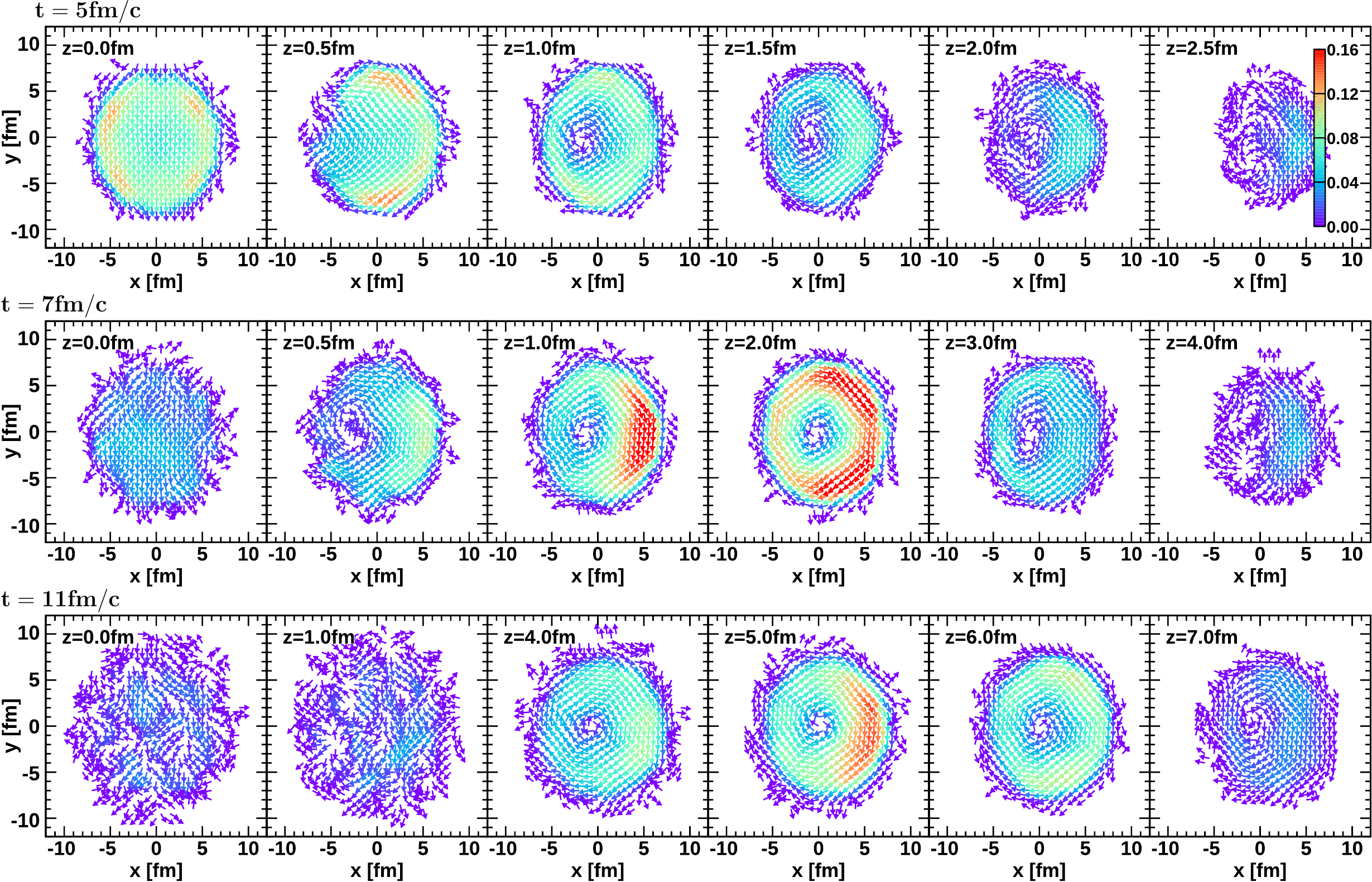}
		\caption{The same as in Fig.~\ref{fig:W-Au77-b75} but for the impact parameter $b=5.0$\,fm.}
		\label{fig:W-Au77-b50}
	\end{figure*}

	It seems interesting to look at the flow pattern created in the collision from the point of view of the Lamb vector. We see in Fig.~\ref{fig:W-Au77-b75} for $t=11$\,fm/$c$ that for $z\simeq 5\pm 1$\,fm the vorticity is mainly oriented as $\vec{\om}=\om_\phi [\vec{e}_T\times \vec{e}_z]$, where $\om_\phi>0$. Note that on each plot in Fig.~\ref{fig:ENT-contour-Au77-b75} the $z$ axis is oriented towards the reader. Using the main components of the velocity $\vec{v}\simeq v_T \vec{e}_T + v_z \vec{e}_z$ we can estimate the direction of the Lamb vector
	\begin{align}
		\vec{\lambda}_\om \simeq \om_\phi(-v_z \vec{e}_T + v_T\vec{e}_z).
		\label{lamb-vec-est}
	\end{align}
	In Fig.~\ref{fig:Lamb-Au77-b75} we depict the projection of the Lamb vector on the $x$-$y$ plane. The Lamb vector as expected~\cite{Rousseaux} points to the center of the vortex.
	
\begin{figure*}
		\centering
		\includegraphics[width=0.99\textwidth]{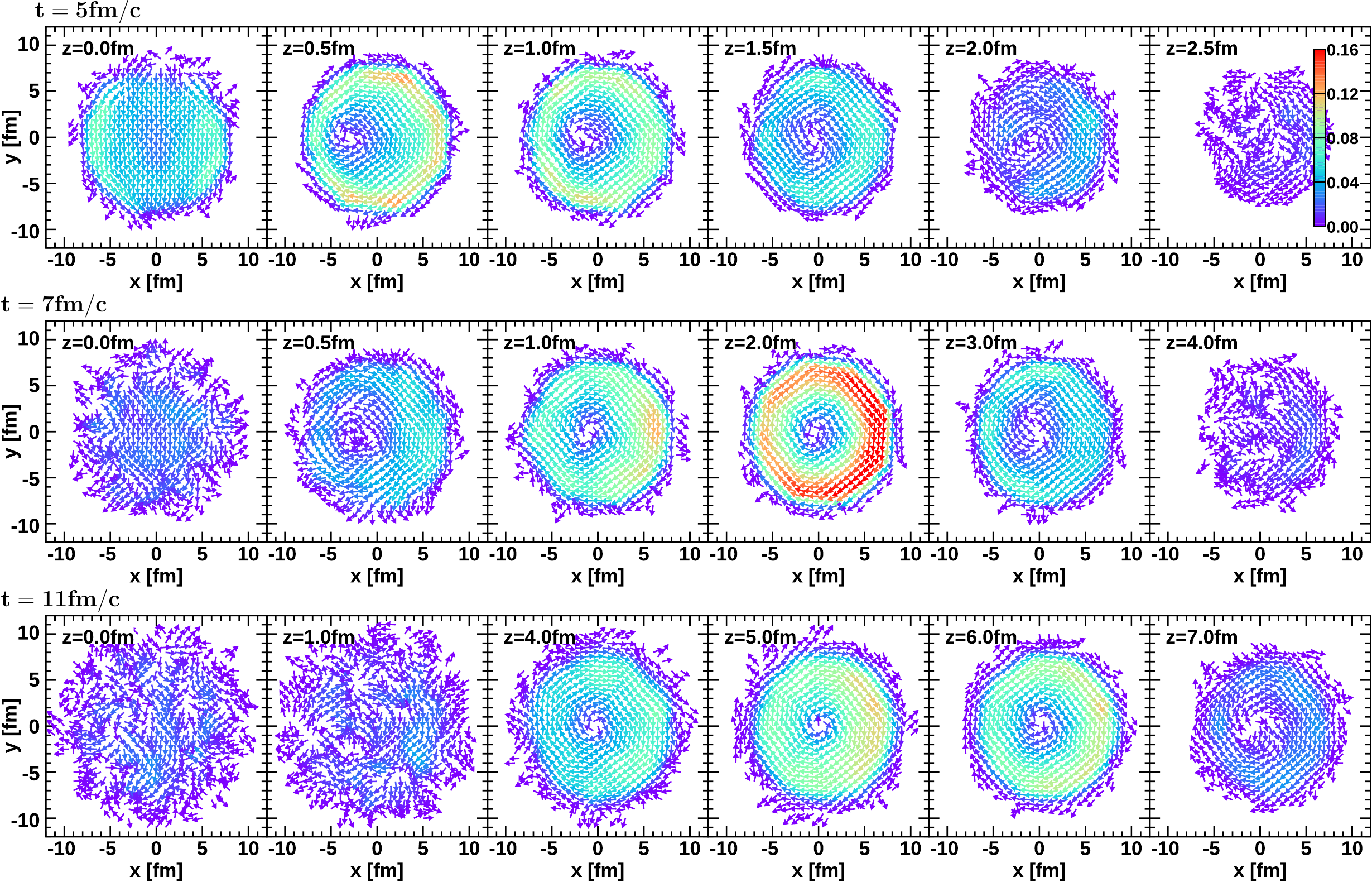}
		\caption{The same as in Fig.~\ref{fig:W-Au77-b75} but for the impact parameter $b=2.5$\,fm.}
		\label{fig:W-Au77-b25}
	\end{figure*}

	In  Figs.~\ref{fig:W-Au77-b50} and \ref{fig:W-Au77-b25} we show the vorticity field for the same collisions as in Fig.~\ref{fig:W-Au77-b75} but for smaller impact parameters, $b=5.0$\,fm and 2.5\,fm, respectively. With a decrease of the impact parameter, the vortex rings become more symmetric and the magnitude of the vorticity field increases.

	\subsection{Measure of rotationality}\label{ssec:Wk}
	
	Significant hyperon polarization in heavy-ion collisions was discovered by the STAR Collaboration~\cite{Adamczyk-Nature}, and the $\sqrt{s_{NN}}$-averaged vorticity of the fluid created in the collision was estimated to be of the order $\langle|\om|\rangle_{\rm exp} \approx 10^{22}$\,s. That leads to the conclusion that the collision created the fastest-spinning fluid ever observed in nature~\cite{Petersen-Nature}.  In more natural units this value would be equal to $\langle|\om|\rangle_{\rm exp} \approx 0.03$\,$c$/fm or in the energy units $\langle|\om|\rangle_{\rm exp} \approx 6$\,MeV/$\hbar$, that is nevertheless a typical temperature of the matter in collisions, $T\gsim 100$\,MeV; see Fig.~\ref{fig:profilesXZ-4.5}. The instantaneous magnitude of the vorticity could be much larger in the course of a collision. So, for Au+Au collisions at $\sqrt{s_{NN}}=7.7$\,GeV at
	the impact parameter $b=7.5\,$fm it reaches the value $|\om|=0.23\,c/{\rm fm}= 47$\,MeV/$\hbar$  in the center slice $z=0$ at $t=5\,$fm$/c$, while at $t=7$\,fm$/c$ the maximum vorticity is $0.34\,c/{\rm fm} =67$\,MeV/$\hbar$ in the slices $|z|=\mbox{1--2}$\,fm.
	
	To compare the vorticity values obtained in the PHSD calculations with the experimental estimations, we calculate the average vorticity of the fireball. For example, at $t=5$\,fm$/c$ and $t=7$\,fm$/c$ we obtain $|\langle\om_{y}\rangle|=9.3$\,MeV/$\hbar$ and $|\langle\om_{y}\rangle|\approx7.3$\,MeV/$\hbar$, respectively. So, the averaged vorticity at times $t=5-13$\,fm$/c$, corresponding to most of the detected $\Lambda$ and $\bar{\Lambda}$ hyperons, is 6.1\,MeV/$\hbar$, which is close to the value $\langle|\vec{\om}|\rangle_{\rm exp}$ quoted in Ref.~\cite{Adamczyk-Nature}.
	
	\begin{figure}
		\includegraphics[width=0.49\textwidth]{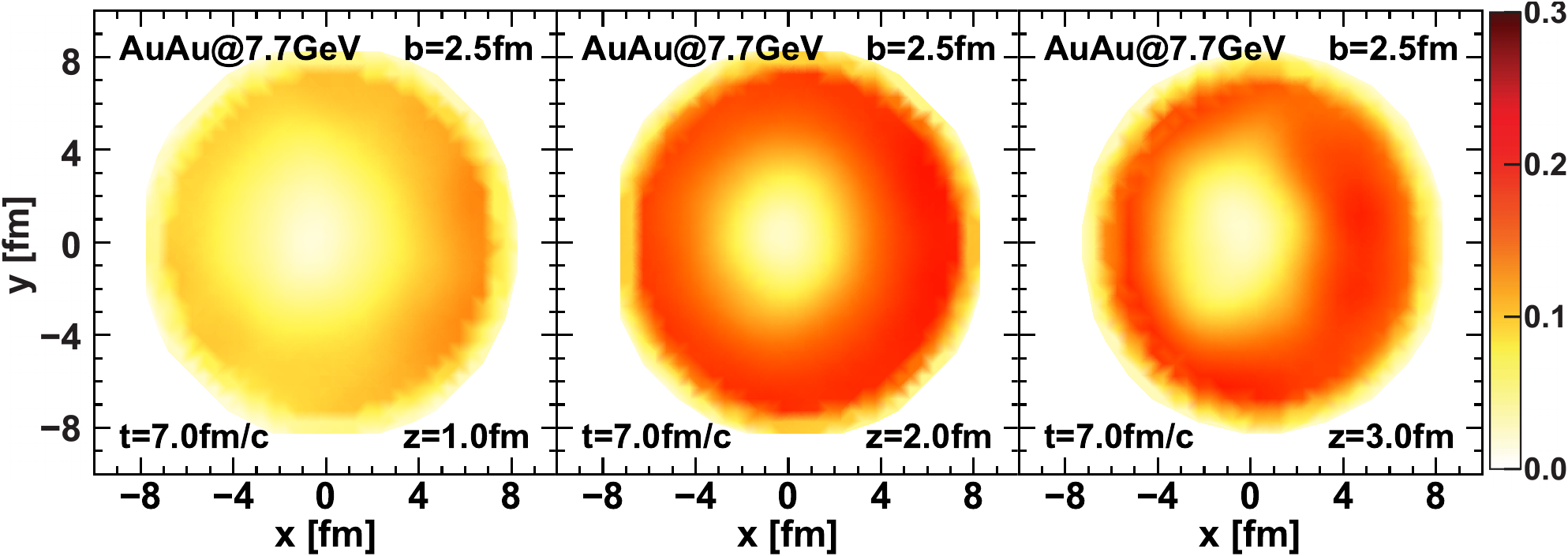}\\
		\includegraphics[width=0.49\textwidth]{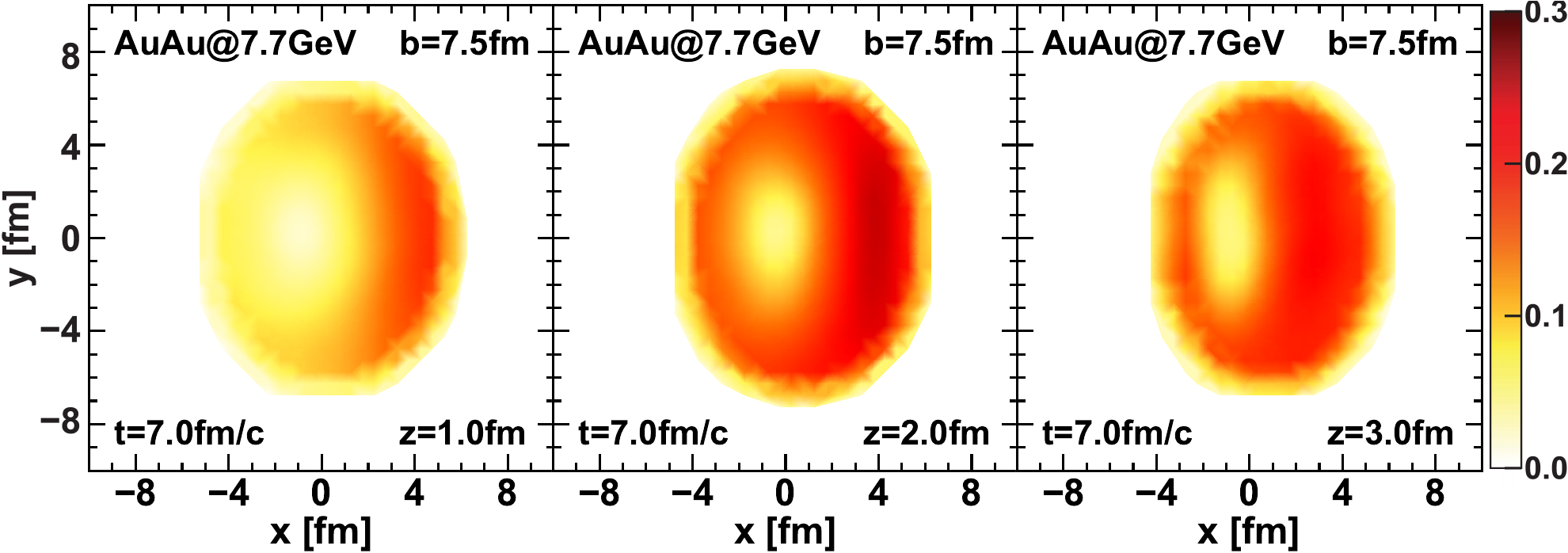}
		\caption{Kinematic vorticity number $\mathfrak{V}_{k}$[ see Eqs.~(\ref{Wk-def}) and (\ref{Vk-def})] for Au+Au collisions at $\sqrt{s_{NN}}=7.7$\,GeV at $t=7$\,fm$/c$ and the impact factors $b=2.5$\,fm (upper row) and 7.5\,fm (lower row).
		}
		\label{fig:KVN-b75-t5}
	\end{figure}

	However, the value of vorticity itself cannot be a basis for conclusion about degree of rotationality of the medium. As pointed out by Truesdell in~\cite{Truesdell53,Truesdell-book} the magnitude of vorticity $\vec{\om}$ as a dimensional quantity is an arbitrary quantity, being dependent on the choice of the unit used. The measure of rotationality, he argued, should indicate not the relative angular speed but the rotational quality or degree. The trivial limiting cases are easy to identify: if $|\vec{\om}|=0$ the motion is not rotational; if $|\vec{\om}|\neq 0$ the motion is rotational. But the desired measure, for instance, should have the same value for all rigid rotations, which are qualitatively identical; i.e., it should be constant in time and independent of the angular speed.
	
	In~\cite{Truesdell53} a convenient dimensionless measure of rotationality was proposed; see also Sec.~55 in~\cite{Truesdell-book}. One starts with the decomposition of $\partial_i v_j=\xi_{ij,+}+\xi_{ij,-}$ into symmetric and antisymmetric tensors, see Eq.~(\ref{xi-decomp}). The symmetric tensor $\xi_{ij,+}$ is the strain rate tensor and constitutes a measure of the rate at which the squared element of arc length is changing. It vanishes if and only if the motion is locally and instantaneously like the rigid body motion. The quantity $\xi_+^2= \xi^{ij}_+ \xi_{ij,+}$, which is called the intensity of deformation and represents the total amount of deformation. It is essentially positive and cannot be zero unless every component of $\xi_{ij,+}$ vanishes.
	Reference~\cite{Truesdell53} suggests comparing the norms of anti-symmetric and symmetric tensors $\|\xi_{ij,-}\|/\|\xi_{ij,+}\|$. Then one defines the kinematic vorticity number
	\begin{align}
		\mathfrak{W}_{\rm k} = \sqrt{\frac{\xi^{ij}_- \xi_{ij,-}}{\xi^{kl}_+ \xi_{kl,+}}} = \frac{|\vec{\om}|}{\sqrt{2}\xi_+},
		\label{Wk-def}
	\end{align}
	where we used here that $\xi^{ij}_- \xi_{ij,-}=\vec{\om}^2/2$.
	For the pure rigid rotation at a given point we have $\xi_{+}=0$ and $\vec{\omega} \neq 0$, what corresponds to $\mathfrak{W}_{\rm k} = \infty$, while an irrotational motion is characterized by $\omega=0$ and $\xi_+\neq 0$ and, consequently, $\mathfrak{W}_{\rm k} = 0$. Thus, all possible motions with the sole exception of rigid translations are assigned a numerical degree of rotationality on a scale from 0 to $\infty$, a rigid motion being the most rotational type of motion possible. To distinguish between almost irrotational motion, $\mathfrak{W}_{\rm k}\ll 1$, and a strong-rotationality case $\mathfrak{W}_{\rm k}\ge 1$, suggest taking in Ref.~\cite{Truesdell53,Truesdell-book} to take the ``dividing'' value $\mathfrak{W}_{\rm k}=\mathfrak{W}^{\rm(div)}_{\rm k}= 1$. Such a value of the kinematic vorticity number corresponds  to a generalized Poiseuille motion, and a simple shearing motion belongs to this class.
	
	The kinematic vorticity number is broadly used in hydrodynamics~\cite{Jeong-Hussain-95},  geology~\cite{Tikoff-Fossen-95}, and meteorology~\cite{Schielicke-16} for identification of vortices and their centers, where $\mathfrak{W}_{\rm k}$ would be maximal.
	
	\begin{figure}
		\includegraphics[width=0.49\textwidth]{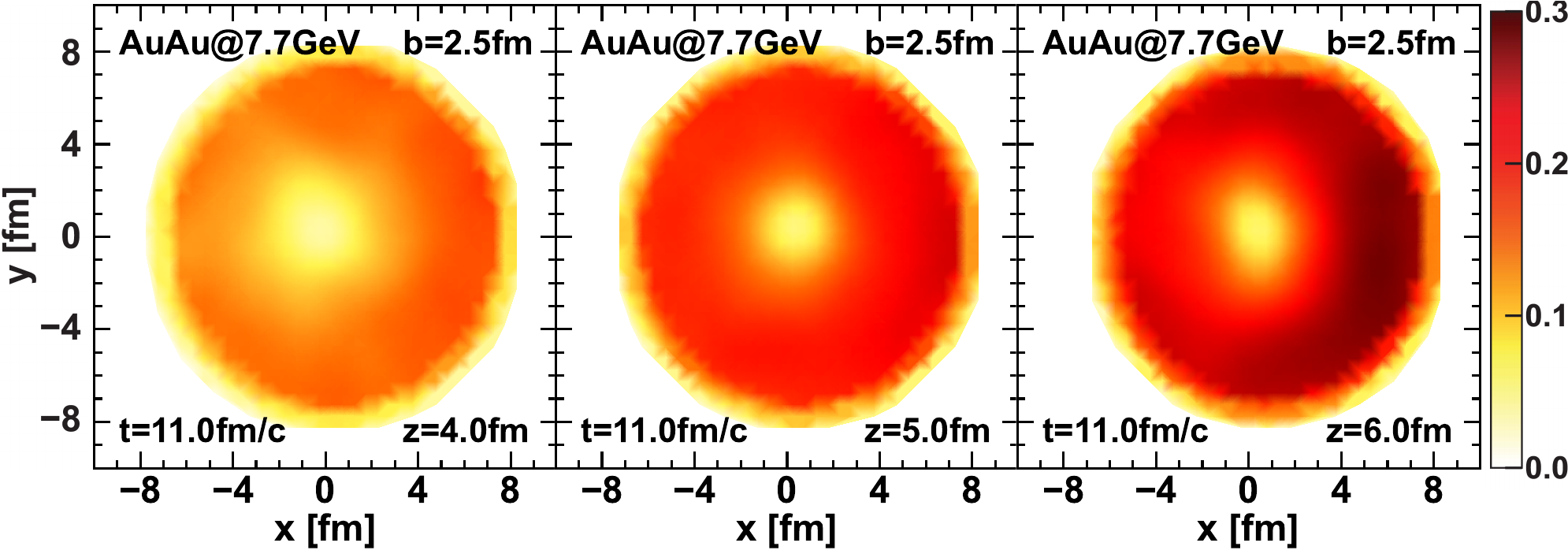}\\
		\includegraphics[width=0.49\textwidth]{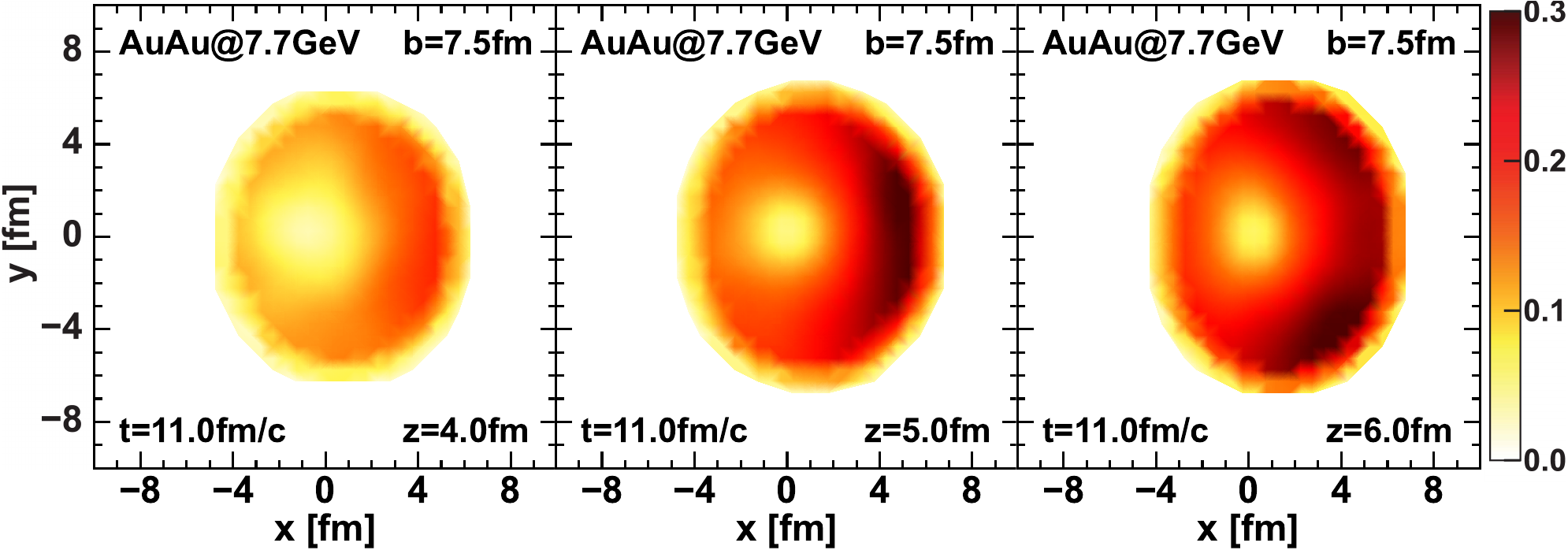}
		\caption{The same as in Fig.~\ref{fig:KVN-b75-t5} but for $t=11$\,fm$/c$. }
		\label{fig:KVN-b75-t11}
	\end{figure}

	In numerical calculations it is convenient to reduce the scale of $\mathfrak{W}_{\rm k}$ variation to a finite interval, and replace $\mathfrak{W}_{\rm k}$ by the quantity
	\begin{equation}
		\mathfrak{V}_{\rm k} = \frac{2}{\pi} \arctan \mathfrak{W}_{\rm k},
		\label{Vk-def}
	\end{equation}
	which takes the value $\mathfrak{V}_{\rm k} = 0$ for the irrotational case and $\mathfrak{V}_{\rm k} = 1$ for the rigid rotation. The dividing value
	$\mathfrak{W}^{\rm(div)}_{\rm k}$ corresponds now to $\mathfrak{V}_{\rm k}^{\rm (div)} =\half$.
	
	It is interesting to quantify the degree of rotationality of the medium created in collisions using the kinematic vorticity number. In Fig.~\ref{fig:KVN-b75-t5} we plot the kinematic vorticity number for the time $t=7$\,fm/$c$ of the Au+Au collisions at $\sqrt{s_{NN}}=7.7$\,GeV for three $z$ slices at $z=1$, 2, and 3\,fm and the impact parameters  $b=2.5$ and 7.5\,fm, the vorticity fields for which are explicitly shown in Fig.~\ref{fig:W-Au77-b75} for $b=2.5$\,fm and Fig.~\ref{fig:W-Au77-b25} for $b=7.5$\,fm. For the chosen $z$ slices, the vortex structure is well developed, and the vorticity magnitudes reach maximal values for each impact parameters. For $b=2.5$\,fm the maximum of kinematic vorticity number forms the nice ring structure both for $z=2$\,fm and for $z=3$\,fm, whereas for the latter slice the vorticity ring in Fig.~\ref{fig:W-Au77-b25} is already less pronounced.
	The maximum values of $\frak{V}_k$ reached in slices with $z=1$, 2, and 3\,fm are 0.14, 0.21, and 0.21, respectively. For impact parameter $b=7.5$, the rings are deformed into ellipses. The maximum values of $\mathfrak{V}_k$ in the same slices are higher, $\mathfrak{V}_k(z=1\,{\rm fm})=0.20$, $\mathfrak{V}_k(z=2\,{\rm fm})=0.25$, and  $\mathfrak{V}_k(z=3\,{\rm fm})=0.23$.
	
	In Fig.~\ref{fig:KVN-b75-t11} we show the kinematic number distribution for a later moment of time, $t=11$\,fm/$c$. We show here slices with $z=4$, 5, and 6\,fm since the maximum of vorticity is shifted now to slices with larger $z$; cf. Figs.~\ref{fig:W-Au77-b75} and \ref{fig:W-Au77-b25}.
	The picture is qualitatively similar to that we see for $t=7$\,fm/$c$, only the kinematic vorticity number reaches higher values: we have $\mathfrak{V}_k(z=4\,{\rm fm})=0.14$,
	$\mathfrak{V}_k(z=5\,{\rm fm})=0.24$ for $b=2.5$\,fm, and $\mathfrak{V}_k(z=6\,{\rm fm})=0.29$ for $b=2.5$\,fm, and
	$\mathfrak{V}_k(z=4\,{\rm fm})=0.20$, $\mathfrak{V}_k(z=5\,{\rm fm})=0.31$, and $\mathfrak{V}_k(z=6\,{\rm fm})=0.31$ for $b=7.5$\,fm.
	
	\begin{figure*}
		\includegraphics[height=8cm]{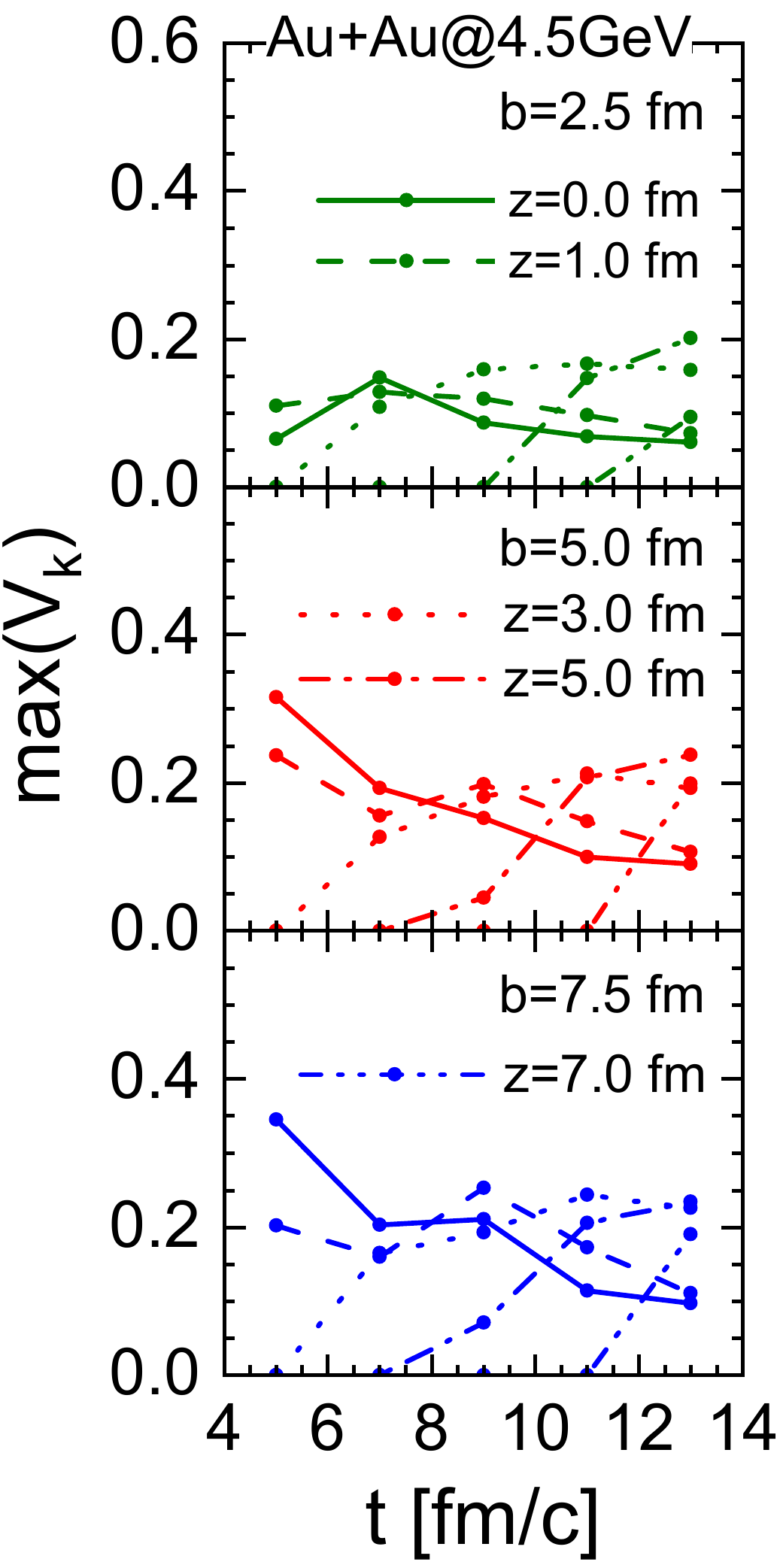}\includegraphics[height=8cm]{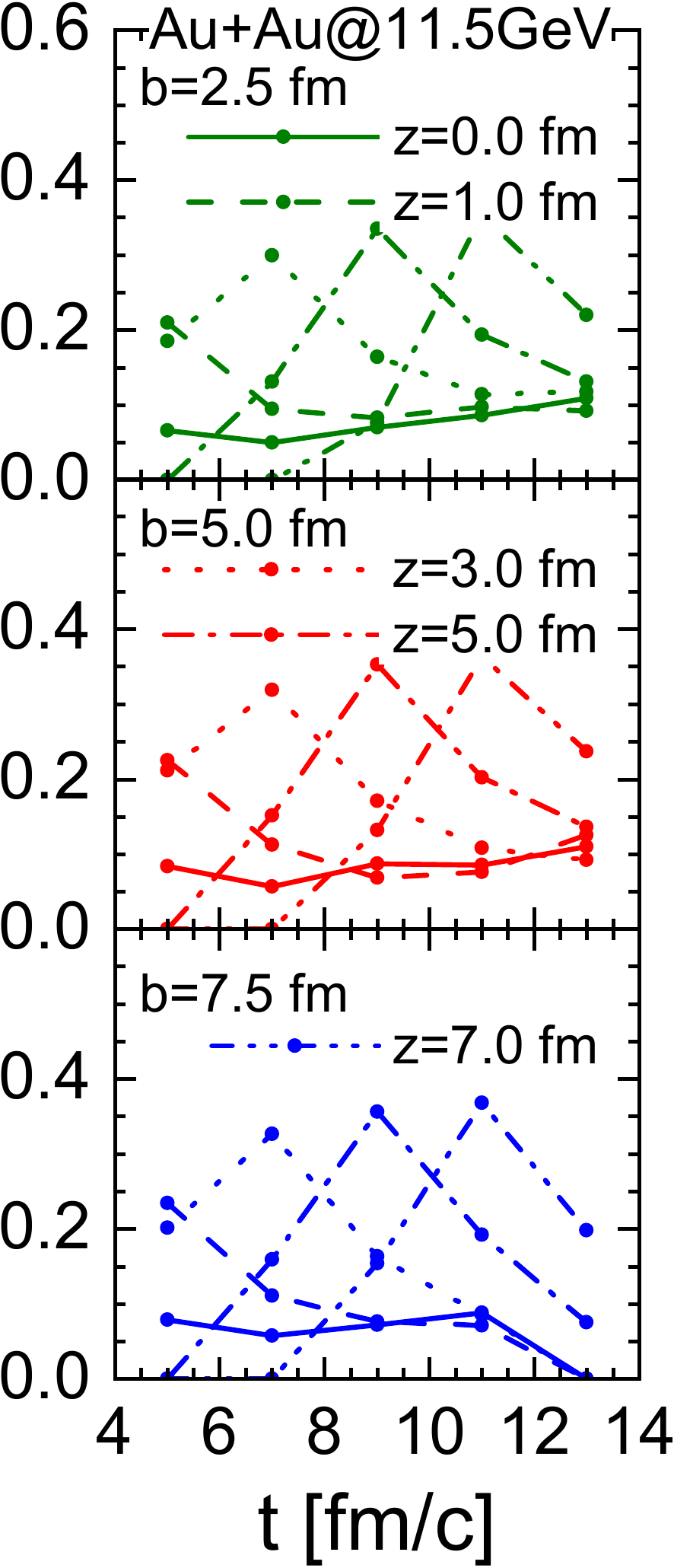}\includegraphics[height=8cm]{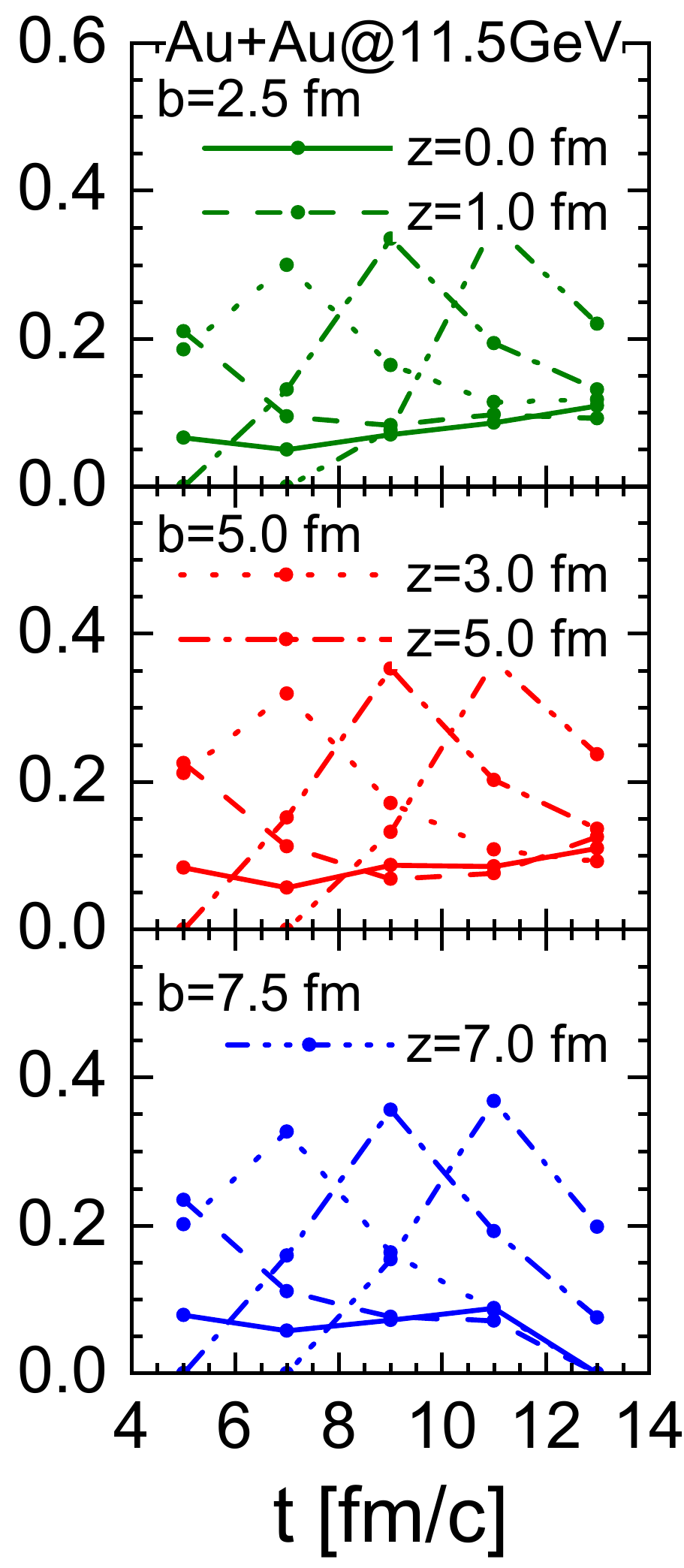}
		\caption{Maximum kinematic vorticity number $\mathfrak{V}_{k}$ as a function of time for various $z$ slices for Au+Au collisions at energies $\sqrt{s_{NN}}=4.5$, 7.7, and 11.5\,GeV with impact parameters $b=2.5$, 5.0, and 7.5\,fm.}
		\label{fig:Vk-max}
	\end{figure*}
	
	The maximum values of the kinematic vorticity number for various collision energies are illustrated in Fig.~\ref{fig:Vk-max} as functions of time for various impact parameters and $z$ slices. The common picture for all three energies is that $\max\{\mathfrak{V}_k\}$ moves with time from slices with smaller $z$ to those with larger $z$  and increases in magnitude. The largest values of $\max\{\mathfrak{V}_k\}$ are realized for largest $z$, except for collisions with $\sqrt{s_{NN}}=4.5$\,GeV where for impact parameters $b=5.0$ and 7.5\,fm the maxima correspond to $z=0$ and smallest times. From Fig.~\ref{fig:Vk-max} we conclude that for $\sqrt{s_{NN}}=4.5$\,GeV $\max\{\mathfrak{V}_k\}<0.35$. For $\sqrt{s_{NN}}=7.7$\,GeV we have $\max\{\frak{V}_k\}<0.32$ and for $\sqrt{s_{NN}}=11.5$\,GeV we find
	$\max\{\mathfrak{V}_k\}<0.38$. The maxima correspond to collisions with the impact parameter $b=7.5$\,fm
	
	We see that the fireball medium created in the collisions in the energy range 4.5--11.5\,GeV has rather mediocre degree of rationality, $\max\{\mathfrak{V}_{\rm k}\} < \mathfrak{V}_{\rm k}^{\rm (div)}=1/2$, which smaller than for the Poiseuille flow and is close to the pure shear deformation corresponding to just a flattening of fluid cells.

	\section{Helicity separation}\label{sec:helicity}
	
	\begin{figure*}
		\parbox{10cm}{\centering  
			\includegraphics[height=4cm]{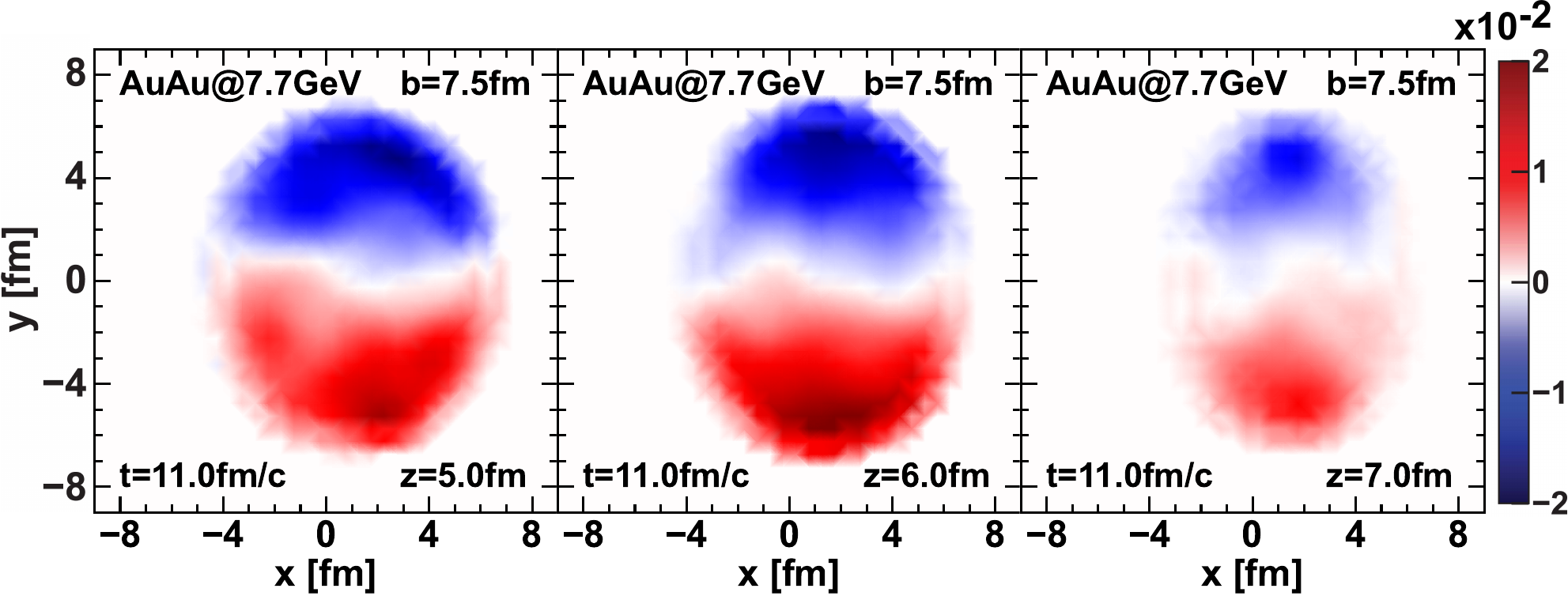}\\
			\caption{Hydrodynamic helicity field (\ref{h-def}) created in Au+Au at $\sqrt{s_{NN}}=7.7$\,GeV and the impact factor $b=7.5$\,fm at the moment $t=11$\,fm/$c$ for various values of $z$. The colore scale is in units of $c^2/{\rm fm}$.
				\label{fig:helicityXY-sep-z}} \vfill \phantom{x}
		}
		\hspace{1cm}
		\parbox{6.5cm}{
			\centering
			\includegraphics[height=4cm]{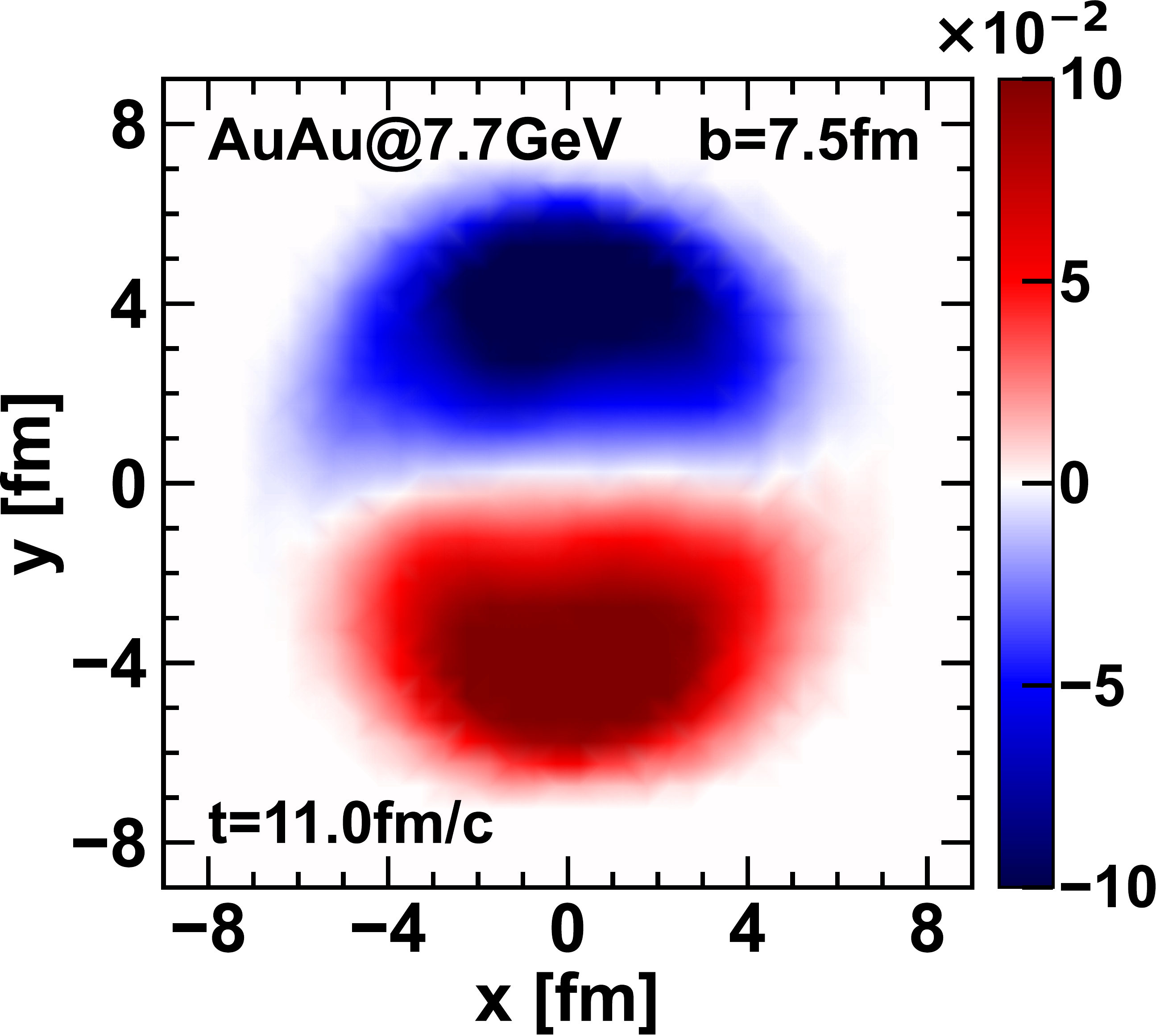}
			\caption{Hydrodynamic helicity field integrated over $z$ for Au+Au collisions at $\sqrt{s_{NN}}=7.7$\,GeV, with the impact factor $b=7.5$\,fm at $t=11$\,fm/$c$. The color scale is in units of $c^2$.	
				\label{fig:helicityXY-sep-int}}
		}
	\end{figure*}

	The integral hydrodynamic helicity (\ref{H-int}) was suggested in Ref.~\cite{BGST-Hseparation} to be a source of a nonvanishing strange chiral charge, which would be carried by the strange quarks and antiquarks determining the finite average spin orientation of $\Lambda$ and $\overline{\Lambda}$ hyperons. Therefore, it is interesting to look at the helicity field generated in heavy-ion collisions within the PHSD transport approach.

	Using Eqs.~(\ref{Vtot}), (\ref{V-Hubble}), (\ref{dV-sym}), and (\ref{dV-asym}) for the velocity and corresponding Eq.~(\ref{vort-exp})
	for the vorticity, and dropping subleading terms responsible for the axial symmetry violation, $\delta\alpha_{T,{\rm as}}$,
	we obtain the following expression for the helicity:
	\begin{align}
		h &=-\alpha_T\frac{y}{r_T}\Big[ \Big(\alpha_T\,r_T + \delta\alpha_{T}(r_T,z)\Big) \frac{\partial x_0(z)}{\partial z}
		\nonumber\\
		&+ x_0(z)\Big(\frac{\partial \delta\alpha_{\|}}{\partial r_T} z
		-\frac{\partial\delta\alpha_{T}}{\partial z} r_T\Big)
		\Big]\,.
		\label{h-field}
	\end{align}
	We see that the main source of the helicity is the offset of the Hubble expansion field of the velocity, i.e., terms $\propto x_0(z)$ in (\ref{dV-asym}).
	The expression in the square brackets is axially symmetric and the dependence on the azimuthal angle is determined solely by the prefactor $y$. Hence, for $y>0$ the helicity is negative, whereas for $y<0$ it is positive. This expectation is confirmed by our calculation shown in Fig.~\ref{fig:helicityXY-sep-z}, where we can see the clear separation of the helicity. This separation is stronger, the smaller the $z$ value is. So for slices with $z\ge 7$\,fm it is almost washed out. In Fig.~\ref{fig:helicityXY-sep-int} we present the $z$-integrated helicity field, which shows the sharp separation of the regions of positive and negative helicity. A similar pattern was found in~Ref.~\cite{Deng-Huang-PRC93} for heavy-ion collisions at LHC energies.

	The offset of the transversal velocity field $x_0(z)$ determines not only the hydrodynamic helicity field (\ref{h-field}) but also the directed hydrodynamic flow (\ref{hydro-flow}). Thus, one can expect that the helicity field will change sign at the energy where the directed flow for the fluid $v_1^{\rm (hydro)}$ changes sign.
	
	\begin{figure*}
		{\includegraphics[height=4cm]{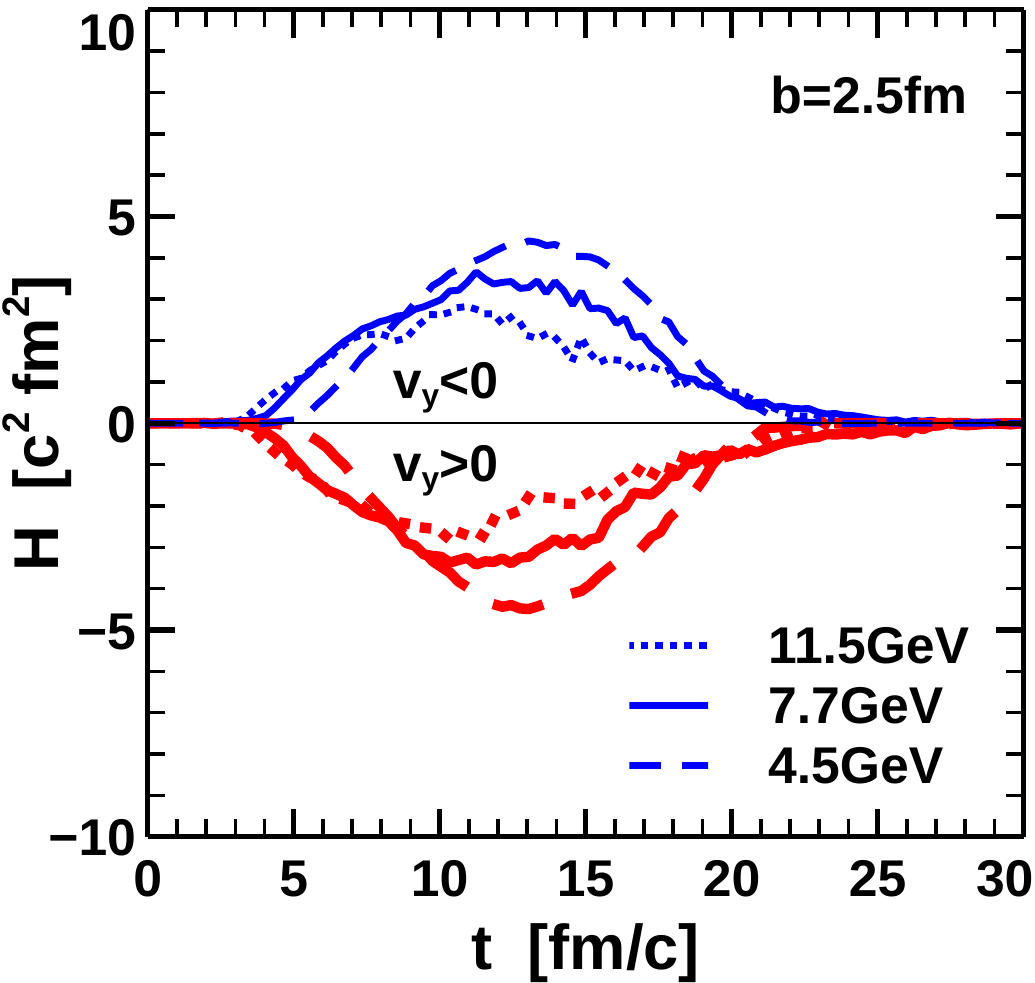}\,\includegraphics[height=4cm]{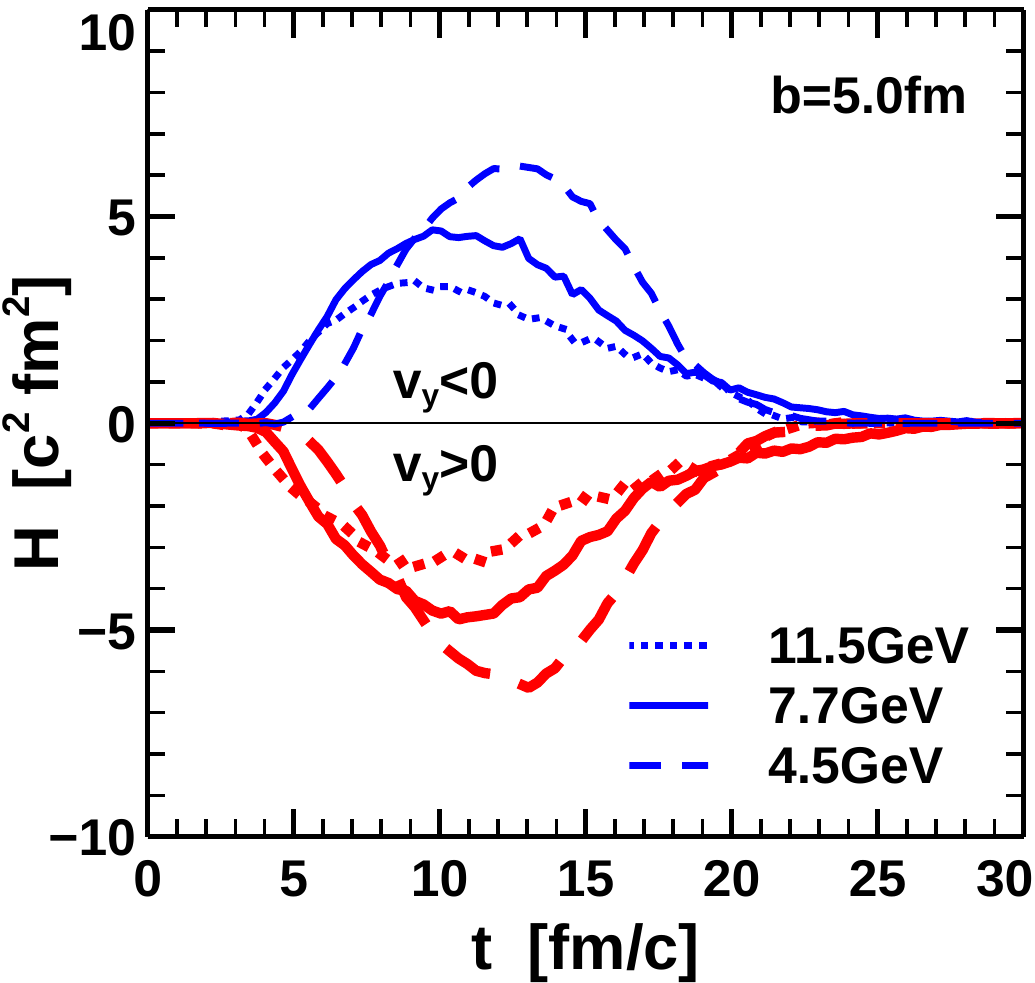}\,\includegraphics[height=4cm]{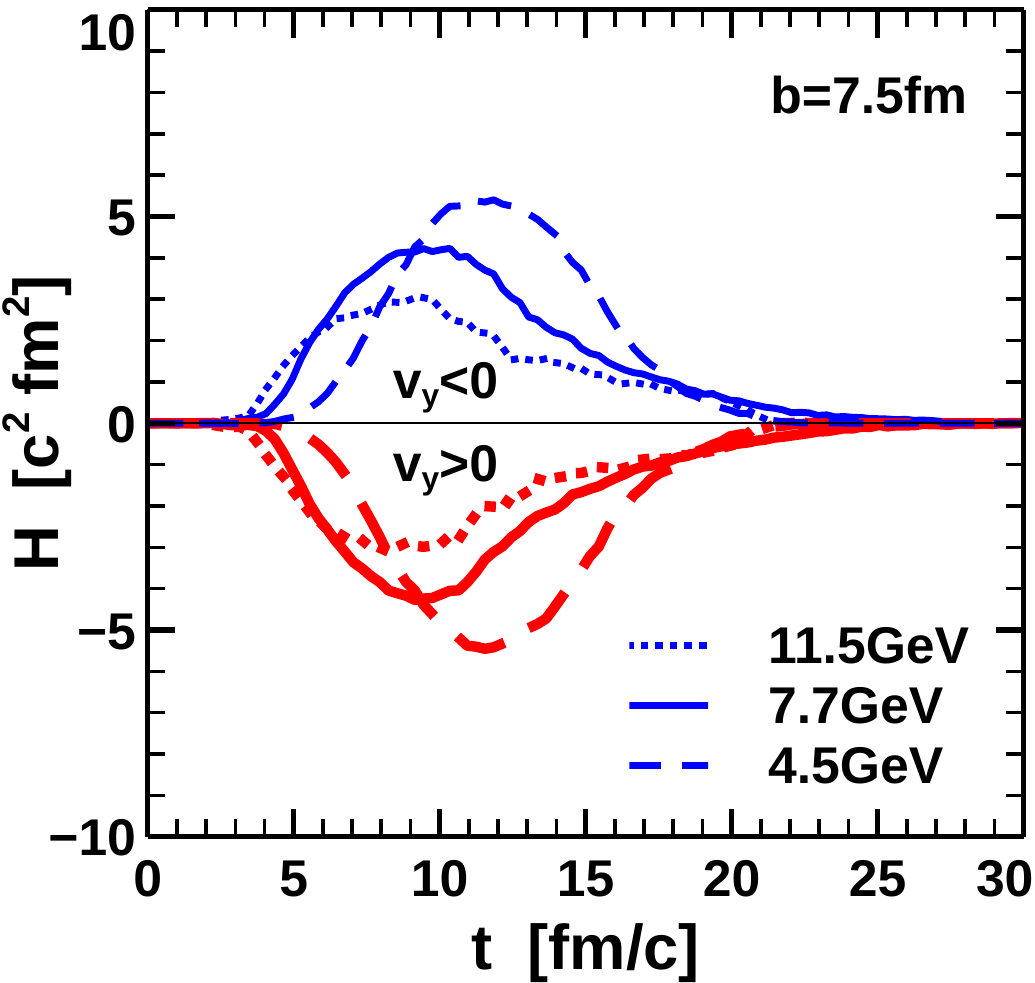}}
		\caption{Evolution of the hydrodynamic helicity integrated over the regions with $v_y<0$ (thin lines) and $v_y>0$ (thick lines) for Au+Au collision at energies $\sqrt{s_{NN}}=4.5$, 7.7, and 11.5\,GeV and different impact parameters. The cut $\epsilon>0.05\,{\rm GeV/fm^3}$ is applied.}
		\label{fig:helicityPy-sep}
	\end{figure*}

	Although, in the course of the collision, large parts of the fireball develop a nonvanishing helicity, the integral helicity of the whole fireball remains strictly zero, since no nontrivial topological structures are expected to develop in the course of heavy-ion collisions. At least they cannot be formed within the transport code operating on a finite mesh. Figures~\ref{fig:helicityXY-sep-z} and \ref{fig:helicityXY-sep-int} support this expectation.
	To illustrate the dynamical evolution of local helicity fields of various signs, Ref.~\cite{BGST-Hseparation} suggested plotting separately the integral helicity
	from the areas with positive and negative values of $v_y$. The results obtained within our approach are shown in Fig.~\ref{fig:helicityPy-sep}. Thin lines show the integral helicity for regions with $v_y<0$ as a function of time for various impact parameters and collision energies, and thick lines show it for regions with $v_y>0$. Thin and thick lines are specular symmetric with respect to the $x$ axis, so that their sum is zero. For $\sqrt{s_{NN}}=7.7$\,GeV, the magnitudes of the integral helicity in both regions increase with time over the first 9-13\,fm/$c$ (depending on the impact parameters) and drop then to zero at 27\,fm/$c$; see the solid thin and thick lines. The growth time becomes shorter for collisions with the larger impact parameter. The maximum value of the integrated vorticity depends also on the impact parameter, and is maximal for $b\simeq 5$\,fm with $H(v_y<0)=-H(v_y>0)\simeq 4.7\,c^2{\rm fm}^2$. This behavior can be confronted against the results of Ref.~\cite{BGST-Hseparation} obtained with QGSM~\cite{QGSM-1,QGSM-2,QGSM-3}; see Fig.~3 there. The time interval shown there is short, so that only those part are visible where the integral helicities are increasing.
	
	For the smaller collision energy $\sqrt{s_{NN}}=4.5$\,GeV (see dashed lines in Fig.~\ref{fig:helicityPy-sep}), the increase times of $|H(v_y<0)|$ and $|H(v_y>0)|$ becomes shorter, the maximal values are smaller, and the $H$ decay time longer than for collisions with $\sqrt{s_{NN}}=7.7$\,GeV. So, the maxima of $|H|$ are reached at earlier times. The maximal value is $H(v_y<0,b\simeq 5)\simeq 6.2\,c^2{\rm fm}^2$. For larger collision energy $\sqrt{s_{NN}}=11.5$\,GeV, the overall evolution time of $H$ is shorter but the maximum value is reached at later times than for collisions with $\sqrt{s_{NN}}=7.7$\,GeV. The maximum is $|H(v_y<0,b\simeq 5)|\simeq 3.5\,c^2{\rm fm}^2$.
	Thus the maximum helicity increases with a decrease of collision energy.
	
\begin{figure}[!ht]
		\centering
		\includegraphics[width=6cm]{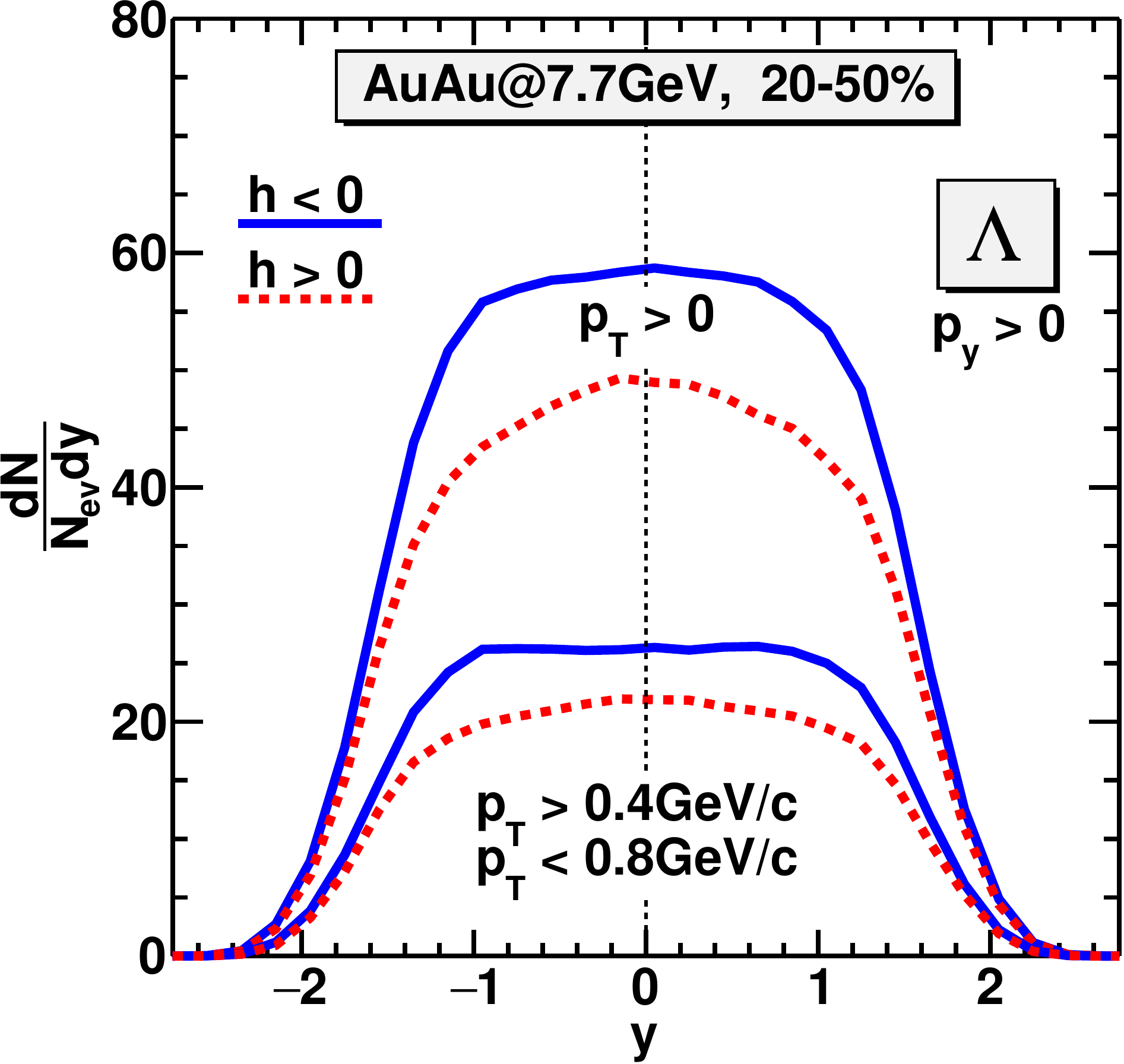}
		\includegraphics[width=6cm]{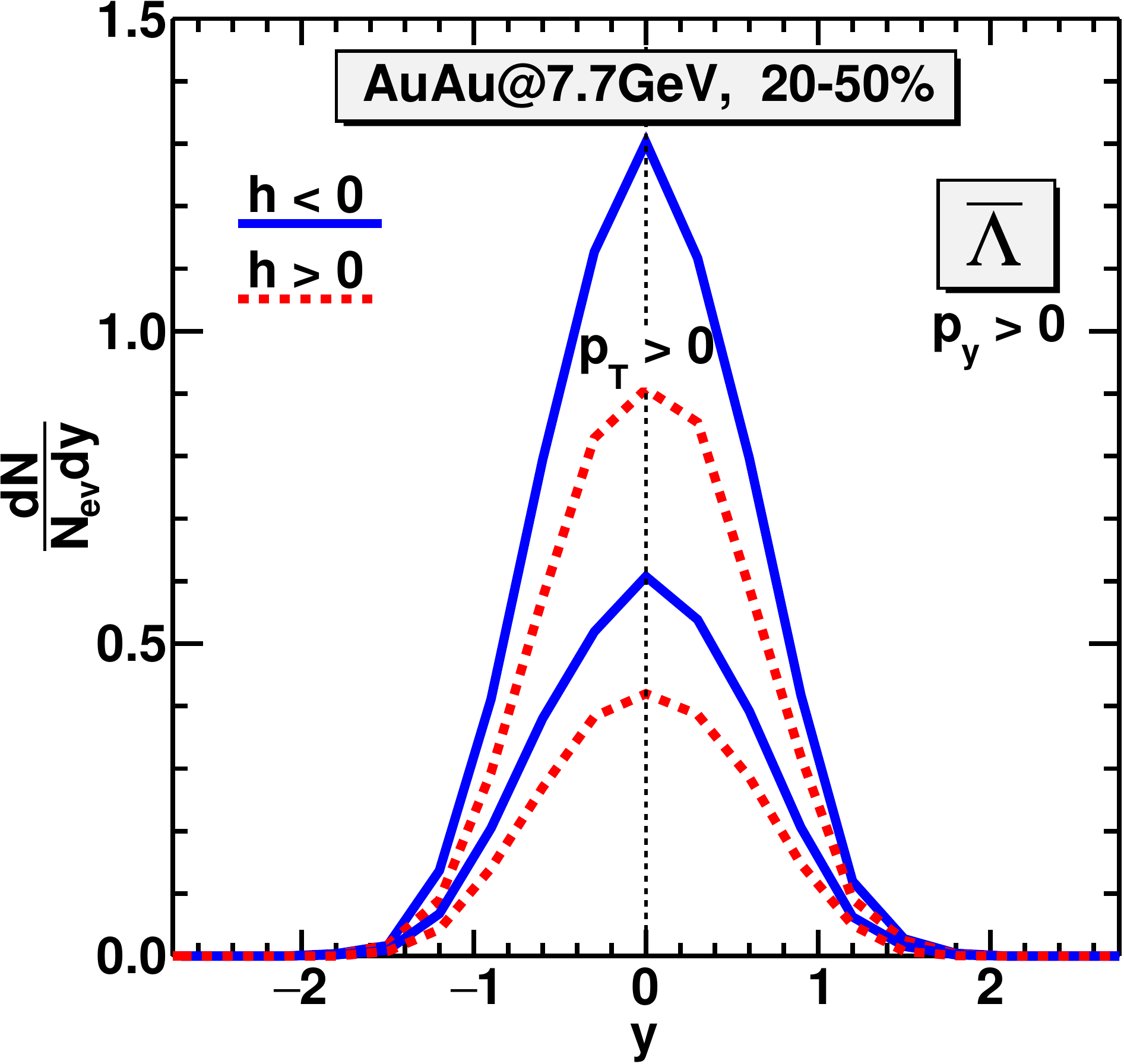}
		\caption{The rapidity spectrum of $\Lambda$ and $\bar{\Lambda}$ for different signs of hydrodynamic helicity density $h>0\,(h<0)$ and $p_{T}$-spectrum cuts in Au+Au collisions at $\sqrt{s_{NN}}=7.7$\,GeV with the centrality range 20--50\%.}
		\label{fig:dndy}
	\end{figure}

	The question remains if one can get some experimental access to the fireball regions with various helicities. From the hydrodynamics point of view, particles from the fluid cells involved in the transversal Hubble-like motion will be predominantly emitted in the positive $y$ direction, i.e., $p_y>0$ if they originate from the fluids moving in the positive $y$ directions and therefore having negative helicity. Oppositely, particles with $p_y<0$ will more probably stem from fluids with positive helicity. In Fig.~\ref{fig:dndy} we illustrate this by direct calculations. We show rapidity distributions of  $\Lambda$ (upper plane) and  $\overline{\Lambda}$ (lower plane) selected by the condition $p_y>0$. Solid lines corresponding to the emission from fluids with $h<0$ lie above the dotted lines corresponding to $h>0$. The relative strength of the enhancement is about 20\% percent for $\Lambda$ and 40\% for $\overline{\Lambda}$.
	The enhancement gets reduced when one imposes cuts in the hyperon transverse momentum $p_T$; see the difference between the dashed and short-dotted lines.
	
	Thus we conclude that selecting hyperons with positive and negative projections of the momentum, $p_y$, one can enhance the polarization signal of hyperons if it is related to the axial vortex effect as proposed in Refs.~\cite{Sorin-Teryaev-2017} and implemented  in Refs.~\cite{BGST-Hseparation,BGST-PRC97,BGST-Vsheet,Ivanov-PRC102-AVE}.
	
	\section{Conclusions}\label{sec:conclusion}
	
	We applied the PHSD transport code~\cite{PHSD,PHSD-contin} to the analysis of the formation and evolution of the vorticity and hydrodynamic helicity fields in Au+Au collisions at NICA energies $\sqrt{s_{NN}}=4.5\mbox{ -- }11.5$\,GeV. First, in Sec.~\ref{sec:Spec-separ} we argued that it is necessary to separate properly the spectator nucleons, which should not be involved in the determinations of the hydrodynamic parameters of the fluid as they experienced no interaction and cannot be equilibrated with the medium. The spectators are selected as the particles whose rapidities do not differ from the beam rapidity by more than $\Delta y_{\rm b} =0.27$, the rapidity uncertainty due to the Fermi motion of a nucleon inside the colliding nucleus. Applying this criterion, we studied the transfer of the angular momentum to the fireball created in the collision. It lasts for about 10\,fm/$c$ and the maximum fraction of the angular momentum is transferred in collisions with impact parameter $b\simeq 5$\,fm, independently of the energy. We showed that the collisions with the highest transferred angular momentum can be selected by choosing a sufficiently narrow centrality window, $C=10\pm 5\%$; see Fig.~\ref{fig:L-aver}.
	
	The method of fluidization of particle distributions generated by the transport code is presented Sec.~\ref{sec:Fluid}. We used the cloud-in-cell method with a parabolic smearing function, with the help of which we identify contributions of every particle to the energy-momentum tensor and the baryon current in grid points. Then, we smoothly interpolated them to any point of the fluid from the neighboring 27 cells. The hydrodynamic velocity was determined as the velocity of the energy transfer (the Landau frame). The resulting temperature baryon density were presented in Sec.~\ref{sec:Tne}, and velocity fields were presented in Sec.~\ref{sec:V-field}. The velocity field, has to a large extent the Hubble-like structure in transverse and longitudinal directions. The non-Hubble corrections are relatively small, but these corrections are the source of hydrodynamic vorticity. Parameters of the Hubble-like expansion were determined and their time evolutions for various collision energies were investigated. Evolution of transverse and longitudinal parameters for collisions at $\sqrt{s_{NN}}=4.5$\,GeV is found to be quite different from that for higher collision energies, 7.7 and 11.5\,GeV.
	
	The vorticity field is studied in Sec.~\ref{sec:vorticity}. We demonstrated that in collisions two asymmetric vortex rings are formed, which are moving along the $z$ axis in opposite directions. For smaller impact parameter the rings become more symmetric. For small energy, 4.5\,GeV, the ring is more diffuse, and becomes more pronounced at higher energies, and the vorticity magnitude increases also. Also, we demonstrated that the vortex ring center can be also identified with the help of the Lamb vector distribution.

	In Sec.~\ref{ssec:Wk}, the degree of the rotationality of the fluid is evaluated with the help of the hydrodynamic invariant proposed by Truesdell in~\cite{Truesdell53}: the kinematic vorticity number. Variation of the spatial distribution of this number over the collision time (see Figs.~\ref{fig:KVN-b75-t5} and \ref{fig:KVN-b75-t11}) indicates that the degree of vorticity is rather moderate and does not reach even the rotationality of the Poiseuille flow.
	The $z$ and time dependences of the maximum vorticity number shown in Fig.~\ref{fig:Vk-max} confirm the structure of the vortex rings seen in the vorticity field.
	
	In Sec.~\ref{sec:helicity} we study the hydrodynamic helicity distribution. We support the conclusion drawn in Refs.~\cite{Deng-Huang-PRC93,BGST-Hseparation,BGST-Vsheet} about the separation of the positive and negative helicity fields on different side of the reaction plane (the $xz$ plane), which can be selected according to the sign of the $y$ component of the fluid velocity.
	We showed that selecting particles with the particular sign of the $y$ projection of the momentum, say $p_y>0$, one would detect more particles from the area with negative helicity than from those with positive helicity. Thereby one could enhance a signal of the axial vortical effect.
	
	Finally we conclude that the proposed scheme of the spectator separation, fluidization, and the determination of thermodynamic variables provides a smooth weakly fluctuating velocity field which can be used for calculations of hyperon polarization in heavy-ion collisions.
	

	\acknowledgments
	We thank D.~N.~Voskresensky and Yu.~B.~Ivanov for discussions.
	The calculations were performed on the ``Govorun'' computational cluster provided by the Laboratory of Information Technologies of JINR, Dubna. The work was supported in part by the Grant No. VEGA~1/0521/22.
	
	\appendix

	\section{Equation for the vorticity}\label{app:vorticity-eq}
	
	The evolution of the nonrelativistic fluid is described by the set of two equations. One is the continuity (Euler) equation
	\begin{align}
		\frac{\partial \rho}{\partial t} + (\vec{v}\cdot \vec{\nabla}) \rho + \rho \, \theta = 0,
		\label{eq:hydro:euler}
	\end{align}
	where $\vec{v}$ is the velocity field and $\rho$ stands for the matter density, $\rho=\varepsilon/c^2$ with $\varepsilon$ being the energy density, and $\theta=\Div\vec{v}$ is the dilation.
	The second equation is the Navier-Stokes equation
	\begin{align}
		\vec{a}=\frac{\partial \vec{v}}{\partial t} & + (\vec{v}\cdot \vec{\nabla}) \vec{v} =
		-\frac{\vec{\nabla}p}{\rho} +\vec{\tau},
		\label{NS-eq}
	\end{align}
	where the acceleration of the fluid is determined by the gradient of the pressure, $p$, and the viscosity force
	\begin{align}
		(\rho\,\vec{\tau})_i &= \partial_j \Big(
		\eta(\partial_i v_j + \partial_j v_i) +\delta_{ij}\big(\zeta-\frac23\eta\big)\theta\Big).
	\end{align}
	Here $\eta$ and $\zeta$ are the shear and bulk viscosities, respectively.
	Using the relations
	\begin{align}
		\vec{\nabla}\theta &= \Delta\vec{v}+\rot\vec{\om},
		\label{v-om}\\
		(\partial_j\eta)(\partial_i v_j) &
		=\partial_i (  v_j (\partial_j\eta)) - v_j \partial_j \partial_i \eta,
	\end{align}
	we can write the viscous force in the vector form
	\begin{align}
		\rho\vec{\tau} &=  \Big(\frac43\eta +\zeta\Big) \Delta \vec{v} +
		\Big(\frac13\eta +\zeta\Big) \rot\vec{\om}
		\nonumber\\
		&+ \vec{\nabla} (  \vec{v}\cdot\vec{\nabla}\eta) - (\vec{v}\cdot \vec{\nabla}) \vec{\nabla} \eta + ((\vec{\nabla}\eta)\cdot\vec{\nabla}) \vec{v}
		- \theta\vec{\nabla}\eta
		\nonumber\\
		&+ \theta \,\vec{\nabla}\Big(\frac13\eta +\zeta\Big).
	\end{align}
	Using the relation
	\begin{align}
		\vec{\nabla} (  \vec{v}\cdot\vec{\nabla}\eta) &=
		(\vec{v}\cdot \vec{\nabla}) \vec{\nabla}\eta +
		((\vec{\nabla}\eta)\cdot \vec{\nabla}) \vec{v}
		+ [(\vec{\nabla}\eta)\times \vec{\om}]
		\label{vec-rel-1}
	\end{align}
	we can cast
	\begin{align}
		\rho\vec{\tau} &=  \Big(\frac43\eta +\zeta\Big) \Delta \vec{v} +
		\Big(\frac13\eta +\zeta\Big) \rot\vec{\om} + [(\vec{\nabla}\eta)\times \vec{\om}]
		\nonumber\\
		& + 2[(\vec{\nabla}\eta)\cdot\vec{\nabla}] \vec{v}
		- \theta\vec{\nabla}\eta
		+ \theta \,\vec{\nabla}\Big(\frac13\eta +\zeta\Big).
	\end{align}
	To eliminate $\rho$ on the left-hand side of this equation one introduces kinematic viscosities
	\begin{align}
		\nu =\frac{\eta}{\rho}\,,\quad
		\bar\nu=\frac{1}{\rho}\Big(\frac13\eta +\zeta\Big)
		\label{kin-viscos}
	\end{align}
	and obtains
	\begin{align}
		\vec{\tau} &= (\nu+\bar\nu) \Delta \vec{v} +
		\bar\nu \rot\vec{\om} +\tau_{\nu\rho}.
	\end{align}
	Here we separated terms depending on the gradients of kinetic viscosities and density,
	\begin{align}
		\tau_{\nu\rho} &= \theta(\bar\nu-\nu) \vec{\nabla}\ln\big((\bar\nu- \nu)\rho\big)
		+ \nu[\big(\vec{\nabla}\ln(\nu\rho)\big)\times \vec{\om}]
		\nonumber\\
		& + 2\nu\big(\big(\vec{\nabla}\ln(\nu\rho)\big)\cdot\vec{\nabla}\big) \vec{v} .
		\label{tau_nr-1}
	\end{align}

	Finally, using the relation
	\begin{align}
		\frac12\vec{\nabla} \vec{v}^{2} = (\vec{v} \cdot \vec{\nabla} ) \vec{v} + [\vec{v} \times \vec{\omega}],
		\label{eq:hydro:id-1}
	\end{align}
	the Navier-Stokes equation (\ref{NS-eq})
	can be written in the following from:
	\begin{align}
		\frac{\partial \vec{v}}{\partial t} +[\vec{\om}\times \vec{v}]
		&=-\vec{\nabla}\Big(\frac{p}{\rho}+\frac{ \vec{v}^2}{2}\Big)
		+(\nu+\bar\nu) \Delta \vec{v} + \bar\nu \rot\vec{\om}
		\nonumber\\
		&-\frac{p}{\rho^2}\vec{\nabla}\rho + \tau_{\nu\rho}\,.
		\label{NS-eq-2}
	\end{align}
	For the case of incompressible fluid, $\rho=const$ and $\Div\vec{v}=0$, this equation turns into Eq.~(\ref{NS-eq-text}), taking into account Eq.~(\ref{v-om}).
	
	To obtain the equation for the viscosity it is convenient to use Eq.~(\ref{NS-eq-2}). Taking circulation from both sides of this equation, we find
	\begin{align}
		&\frac{\partial \vec{\om}}{\partial t} +\rot \vec{\lambda}_\om
		-\nu\Delta \vec{\om} =
		+\frac{1}{\rho^2}[(\vec{\nabla}\rho) \times (\vec{\nabla} p)] +\vec{f}_{\nu\rho},
		\nonumber\\
		&\quad \vec{f}_{\nu\rho} = \rot \tau_{\nu\rho}.
		\label{NS-eq-rot}
	\end{align}
	Without two terms on the right-hand side which disappear in the case of barotropic fluid and the constant (density and temperature independent) kinematic viscosities, we would recover the Helmholtz equation (\ref{HH-eq}) for the vorticity.
	The new term on the right-hand side induced by density and viscosity gradients can be written in the following form:
	\begin{align}
		(\vec{f}_{\nu\rho})_i=-\nu\Gamma^{[\nu\rho]}_{ij}\om_j - \nu D^{[\nu\rho]}_{ijk}\nabla_j \om_k +S^{[\nu\rho]}_{i}\,,
	\end{align}
	where
	\begin{align}
		\Gamma^{[\nu\rho]}_{ij} &=\big(\nabla_i\ln(\nu\rho)\big) (\nabla_j\ln\nu)
		+\nabla_j\nabla_i\ln(\nu\rho)
		\nonumber\\
		& - \delta_{ij}\big\{((\vec{\nabla}\ln\nu)\cdot\big(\vec{\nabla}\ln(\nu\rho)\big))
		+ \Delta \ln(\nu\rho)\big\},
		\label{app:Gamma}
	\end{align}
	\begin{align}
		D^{[\nu\rho]}_{ijk} &= -\delta_{ik}\nabla_j\ln(\nu\rho) +
		\delta_{ij} \nabla_k\ln\nu  - \delta_{ik}\nabla_j\ln\nu,
		\label{app:D}
	\end{align}
	\begin{align}
		\vec{S}^{[\nu\rho]}&=
		2[(\vec{\nabla}\nu)\times \vec{\nabla}\theta]
		+ (\bar\nu-\nu)[(\vec{\nabla}\theta)\times \vec{\nabla}\ln\rho]
		\nonumber\\
		&+ \theta[(\vec{\nabla}(\bar\nu+\nu))\times \vec{\nabla}\ln\rho]
		+2\big[(\vec{\nabla}\nu)\times\vec{v}\big]\Delta\ln(\nu\rho)
		\nonumber\\
		& - 2\big[(\vec{\nabla}\nu)\times (\vec{v}\cdot \vec{\nabla})\big] \vec{\nabla}\ln(\nu\rho)
		\nonumber\\
		& - 2\nu\big[\vec{\nabla}\times (\vec{v}\cdot \vec{\nabla})\big] \vec{\nabla}\ln(\nu\rho).
		\label{app:S}
	\end{align}

\end{document}